\documentclass[longauth]{aaEC}

\usepackage{graphicx}
\usepackage{natbib}
\usepackage{scalerel}
\usepackage[table]{xcolor}
\usepackage[varg]{txfonts}
\usepackage{natbib}
\usepackage{siunitx}
\usepackage{listings}
\usepackage{float}
\usepackage{enumitem}
\usepackage[outercaption]{sidecap}

\DeclareSIUnit\jansky{Jy}

\newcommand{\computed}[1]{\textcolor{black}{#1}}

\newcommand{\myfigurewidth}[0]{.9}

\usepackage{txfonts}
\usepackage[pdfencoding=auto,psdextra]{hyperref}
\usepackage{rotating}

\hypersetup{
    colorlinks=true,
    linkcolor=blue,
    filecolor=magenta,
    urlcolor=blue,
    citecolor=blue
}
\makeatletter
\renewcommand*\aa@pageof{, page \thepage{} of \pageref*{LastPage}}
\makeatother

\usepackage[utf8]{inputenc}

\newcommand{\linenumbers}[0]{}

\usepackage{euclid}

\begin{document}
\newcommand{\orcid}[1]{} 
\author{M.~Fabricius\orcid{0000-0002-7025-6058}\thanks{\email{mxhf@mpe.mpg.de}}\inst{\ref{aff1},\ref{aff2}}
\and R.~Saglia\orcid{0000-0003-0378-7032}\inst{\ref{aff2},\ref{aff1}}
\and F.~Balzer\orcid{0009-0005-6733-5432}\inst{\ref{aff1}}
\and L.~R.~Ecker\orcid{0009-0005-3508-2469}\inst{\ref{aff2},\ref{aff1}}
\and J.~Thomas\orcid{0000-0003-2868-9244}\inst{\ref{aff1},\ref{aff2}}
\and R.~Bender\orcid{0000-0001-7179-0626}\inst{\ref{aff1},\ref{aff2}}
\and J.~Gracia-Carpio\inst{\ref{aff1}}
\and M.~Magliocchetti\orcid{0000-0001-9158-4838}\inst{\ref{aff3}}
\and O.~Marggraf\orcid{0000-0001-7242-3852}\inst{\ref{aff4}}
\and A.~Rawlings\inst{\ref{aff5}}
\and J.~G.~Sorce\orcid{0000-0002-2307-2432}\inst{\ref{aff6},\ref{aff7}}
\and K.~Voggel\orcid{0000-0001-6215-0950}\inst{\ref{aff8}}
\and L.~Wang\orcid{0000-0002-6736-9158}\inst{\ref{aff9},\ref{aff10}}
\and A.~van~der~Wel\inst{\ref{aff11}}
\and B.~Altieri\orcid{0000-0003-3936-0284}\inst{\ref{aff12}}
\and A.~Amara\inst{\ref{aff13}}
\and S.~Andreon\orcid{0000-0002-2041-8784}\inst{\ref{aff14}}
\and N.~Auricchio\orcid{0000-0003-4444-8651}\inst{\ref{aff15}}
\and C.~Baccigalupi\orcid{0000-0002-8211-1630}\inst{\ref{aff16},\ref{aff17},\ref{aff18},\ref{aff19}}
\and M.~Baldi\orcid{0000-0003-4145-1943}\inst{\ref{aff20},\ref{aff15},\ref{aff21}}
\and A.~Balestra\orcid{0000-0002-6967-261X}\inst{\ref{aff22}}
\and S.~Bardelli\orcid{0000-0002-8900-0298}\inst{\ref{aff15}}
\and A.~Biviano\orcid{0000-0002-0857-0732}\inst{\ref{aff17},\ref{aff16}}
\and E.~Branchini\orcid{0000-0002-0808-6908}\inst{\ref{aff23},\ref{aff24},\ref{aff14}}
\and M.~Brescia\orcid{0000-0001-9506-5680}\inst{\ref{aff25},\ref{aff26}}
\and J.~Brinchmann\orcid{0000-0003-4359-8797}\inst{\ref{aff27},\ref{aff28},\ref{aff29}}
\and S.~Camera\orcid{0000-0003-3399-3574}\inst{\ref{aff30},\ref{aff31},\ref{aff32}}
\and G.~Ca\~nas-Herrera\orcid{0000-0003-2796-2149}\inst{\ref{aff33},\ref{aff34}}
\and V.~Capobianco\orcid{0000-0002-3309-7692}\inst{\ref{aff32}}
\and C.~Carbone\orcid{0000-0003-0125-3563}\inst{\ref{aff35}}
\and J.~Carretero\orcid{0000-0002-3130-0204}\inst{\ref{aff36},\ref{aff37}}
\and M.~Castellano\orcid{0000-0001-9875-8263}\inst{\ref{aff38}}
\and G.~Castignani\orcid{0000-0001-6831-0687}\inst{\ref{aff15}}
\and S.~Cavuoti\orcid{0000-0002-3787-4196}\inst{\ref{aff26},\ref{aff39}}
\and K.~C.~Chambers\orcid{0000-0001-6965-7789}\inst{\ref{aff40}}
\and A.~Cimatti\inst{\ref{aff41}}
\and C.~Colodro-Conde\inst{\ref{aff42}}
\and G.~Congedo\orcid{0000-0003-2508-0046}\inst{\ref{aff43}}
\and C.~J.~Conselice\orcid{0000-0003-1949-7638}\inst{\ref{aff44}}
\and L.~Conversi\orcid{0000-0002-6710-8476}\inst{\ref{aff45},\ref{aff12}}
\and Y.~Copin\orcid{0000-0002-5317-7518}\inst{\ref{aff46}}
\and F.~Courbin\orcid{0000-0003-0758-6510}\inst{\ref{aff47},\ref{aff48}}
\and H.~M.~Courtois\orcid{0000-0003-0509-1776}\inst{\ref{aff49}}
\and M.~Cropper\orcid{0000-0003-4571-9468}\inst{\ref{aff50}}
\and H.~Degaudenzi\orcid{0000-0002-5887-6799}\inst{\ref{aff51}}
\and G.~De~Lucia\orcid{0000-0002-6220-9104}\inst{\ref{aff17}}
\and C.~Dolding\orcid{0009-0003-7199-6108}\inst{\ref{aff50}}
\and H.~Dole\orcid{0000-0002-9767-3839}\inst{\ref{aff7}}
\and F.~Dubath\orcid{0000-0002-6533-2810}\inst{\ref{aff51}}
\and C.~A.~J.~Duncan\orcid{0009-0003-3573-0791}\inst{\ref{aff43}}
\and X.~Dupac\inst{\ref{aff12}}
\and S.~Dusini\orcid{0000-0002-1128-0664}\inst{\ref{aff52}}
\and S.~Escoffier\orcid{0000-0002-2847-7498}\inst{\ref{aff53}}
\and M.~Farina\orcid{0000-0002-3089-7846}\inst{\ref{aff3}}
\and R.~Farinelli\inst{\ref{aff15}}
\and S.~Ferriol\inst{\ref{aff46}}
\and F.~Finelli\orcid{0000-0002-6694-3269}\inst{\ref{aff15},\ref{aff54}}
\and M.~Frailis\orcid{0000-0002-7400-2135}\inst{\ref{aff17}}
\and E.~Franceschi\orcid{0000-0002-0585-6591}\inst{\ref{aff15}}
\and M.~Fumana\orcid{0000-0001-6787-5950}\inst{\ref{aff35}}
\and S.~Galeotta\orcid{0000-0002-3748-5115}\inst{\ref{aff17}}
\and B.~Gillis\orcid{0000-0002-4478-1270}\inst{\ref{aff43}}
\and C.~Giocoli\orcid{0000-0002-9590-7961}\inst{\ref{aff15},\ref{aff21}}
\and A.~Grazian\orcid{0000-0002-5688-0663}\inst{\ref{aff22}}
\and F.~Grupp\inst{\ref{aff1},\ref{aff2}}
\and S.~V.~H.~Haugan\orcid{0000-0001-9648-7260}\inst{\ref{aff55}}
\and J.~Hoar\inst{\ref{aff12}}
\and H.~Hoekstra\orcid{0000-0002-0641-3231}\inst{\ref{aff34}}
\and W.~Holmes\inst{\ref{aff56}}
\and I.~M.~Hook\orcid{0000-0002-2960-978X}\inst{\ref{aff57}}
\and F.~Hormuth\inst{\ref{aff58}}
\and A.~Hornstrup\orcid{0000-0002-3363-0936}\inst{\ref{aff59},\ref{aff60}}
\and K.~Jahnke\orcid{0000-0003-3804-2137}\inst{\ref{aff61}}
\and M.~Jhabvala\inst{\ref{aff62}}
\and B.~Joachimi\orcid{0000-0001-7494-1303}\inst{\ref{aff63}}
\and E.~Keih\"anen\orcid{0000-0003-1804-7715}\inst{\ref{aff64}}
\and S.~Kermiche\orcid{0000-0002-0302-5735}\inst{\ref{aff53}}
\and A.~Kiessling\orcid{0000-0002-2590-1273}\inst{\ref{aff56}}
\and B.~Kubik\orcid{0009-0006-5823-4880}\inst{\ref{aff46}}
\and K.~Kuijken\orcid{0000-0002-3827-0175}\inst{\ref{aff34}}
\and M.~K\"ummel\orcid{0000-0003-2791-2117}\inst{\ref{aff2}}
\and M.~Kunz\orcid{0000-0002-3052-7394}\inst{\ref{aff65}}
\and H.~Kurki-Suonio\orcid{0000-0002-4618-3063}\inst{\ref{aff5},\ref{aff66}}
\and A.~M.~C.~Le~Brun\orcid{0000-0002-0936-4594}\inst{\ref{aff67}}
\and S.~Ligori\orcid{0000-0003-4172-4606}\inst{\ref{aff32}}
\and P.~B.~Lilje\orcid{0000-0003-4324-7794}\inst{\ref{aff55}}
\and V.~Lindholm\orcid{0000-0003-2317-5471}\inst{\ref{aff5},\ref{aff66}}
\and I.~Lloro\orcid{0000-0001-5966-1434}\inst{\ref{aff68}}
\and G.~Mainetti\orcid{0000-0003-2384-2377}\inst{\ref{aff69}}
\and D.~Maino\inst{\ref{aff70},\ref{aff35},\ref{aff71}}
\and E.~Maiorano\orcid{0000-0003-2593-4355}\inst{\ref{aff15}}
\and O.~Mansutti\orcid{0000-0001-5758-4658}\inst{\ref{aff17}}
\and M.~Martinelli\orcid{0000-0002-6943-7732}\inst{\ref{aff38},\ref{aff72}}
\and N.~Martinet\orcid{0000-0003-2786-7790}\inst{\ref{aff73}}
\and F.~Marulli\orcid{0000-0002-8850-0303}\inst{\ref{aff74},\ref{aff15},\ref{aff21}}
\and R.~J.~Massey\orcid{0000-0002-6085-3780}\inst{\ref{aff75}}
\and E.~Medinaceli\orcid{0000-0002-4040-7783}\inst{\ref{aff15}}
\and S.~Mei\orcid{0000-0002-2849-559X}\inst{\ref{aff76},\ref{aff77}}
\and Y.~Mellier\inst{\ref{aff78},\ref{aff79}}
\and M.~Meneghetti\orcid{0000-0003-1225-7084}\inst{\ref{aff15},\ref{aff21}}
\and E.~Merlin\orcid{0000-0001-6870-8900}\inst{\ref{aff38}}
\and G.~Meylan\inst{\ref{aff80}}
\and A.~Mora\orcid{0000-0002-1922-8529}\inst{\ref{aff81}}
\and M.~Moresco\orcid{0000-0002-7616-7136}\inst{\ref{aff74},\ref{aff15}}
\and L.~Moscardini\orcid{0000-0002-3473-6716}\inst{\ref{aff74},\ref{aff15},\ref{aff21}}
\and R.~Nakajima\orcid{0009-0009-1213-7040}\inst{\ref{aff4}}
\and C.~Neissner\orcid{0000-0001-8524-4968}\inst{\ref{aff82},\ref{aff37}}
\and S.-M.~Niemi\orcid{0009-0005-0247-0086}\inst{\ref{aff33}}
\and C.~Padilla\orcid{0000-0001-7951-0166}\inst{\ref{aff82}}
\and S.~Paltani\orcid{0000-0002-8108-9179}\inst{\ref{aff51}}
\and F.~Pasian\orcid{0000-0002-4869-3227}\inst{\ref{aff17}}
\and K.~Pedersen\inst{\ref{aff83}}
\and W.~J.~Percival\orcid{0000-0002-0644-5727}\inst{\ref{aff84},\ref{aff85},\ref{aff86}}
\and V.~Pettorino\orcid{0000-0002-4203-9320}\inst{\ref{aff33}}
\and S.~Pires\orcid{0000-0002-0249-2104}\inst{\ref{aff87}}
\and G.~Polenta\orcid{0000-0003-4067-9196}\inst{\ref{aff88}}
\and M.~Poncet\inst{\ref{aff89}}
\and L.~A.~Popa\inst{\ref{aff90}}
\and L.~Pozzetti\orcid{0000-0001-7085-0412}\inst{\ref{aff15}}
\and F.~Raison\orcid{0000-0002-7819-6918}\inst{\ref{aff1}}
\and A.~Renzi\orcid{0000-0001-9856-1970}\inst{\ref{aff91},\ref{aff52}}
\and J.~Rhodes\orcid{0000-0002-4485-8549}\inst{\ref{aff56}}
\and G.~Riccio\inst{\ref{aff26}}
\and E.~Romelli\orcid{0000-0003-3069-9222}\inst{\ref{aff17}}
\and M.~Roncarelli\orcid{0000-0001-9587-7822}\inst{\ref{aff15}}
\and H.~J.~A.~Rottgering\orcid{0000-0001-8887-2257}\inst{\ref{aff34}}
\and Z.~Sakr\orcid{0000-0002-4823-3757}\inst{\ref{aff92},\ref{aff93},\ref{aff94}}
\and A.~G.~S\'anchez\orcid{0000-0003-1198-831X}\inst{\ref{aff1}}
\and D.~Sapone\orcid{0000-0001-7089-4503}\inst{\ref{aff95}}
\and B.~Sartoris\orcid{0000-0003-1337-5269}\inst{\ref{aff2},\ref{aff17}}
\and M.~Schirmer\orcid{0000-0003-2568-9994}\inst{\ref{aff61}}
\and P.~Schneider\orcid{0000-0001-8561-2679}\inst{\ref{aff4}}
\and T.~Schrabback\orcid{0000-0002-6987-7834}\inst{\ref{aff96}}
\and A.~Secroun\orcid{0000-0003-0505-3710}\inst{\ref{aff53}}
\and G.~Seidel\orcid{0000-0003-2907-353X}\inst{\ref{aff61}}
\and S.~Serrano\orcid{0000-0002-0211-2861}\inst{\ref{aff97},\ref{aff98},\ref{aff99}}
\and P.~Simon\inst{\ref{aff4}}
\and C.~Sirignano\orcid{0000-0002-0995-7146}\inst{\ref{aff91},\ref{aff52}}
\and G.~Sirri\orcid{0000-0003-2626-2853}\inst{\ref{aff21}}
\and J.~Skottfelt\orcid{0000-0003-1310-8283}\inst{\ref{aff100}}
\and L.~Stanco\orcid{0000-0002-9706-5104}\inst{\ref{aff52}}
\and J.-L.~Starck\orcid{0000-0003-2177-7794}\inst{\ref{aff87}}
\and J.~Steinwagner\orcid{0000-0001-7443-1047}\inst{\ref{aff1}}
\and P.~Tallada-Cresp\'{i}\orcid{0000-0002-1336-8328}\inst{\ref{aff36},\ref{aff37}}
\and A.~N.~Taylor\inst{\ref{aff43}}
\and H.~I.~Teplitz\orcid{0000-0002-7064-5424}\inst{\ref{aff101}}
\and I.~Tereno\orcid{0000-0002-4537-6218}\inst{\ref{aff102},\ref{aff103}}
\and N.~Tessore\orcid{0000-0002-9696-7931}\inst{\ref{aff63}}
\and S.~Toft\orcid{0000-0003-3631-7176}\inst{\ref{aff104},\ref{aff105}}
\and R.~Toledo-Moreo\orcid{0000-0002-2997-4859}\inst{\ref{aff106}}
\and F.~Torradeflot\orcid{0000-0003-1160-1517}\inst{\ref{aff37},\ref{aff36}}
\and I.~Tutusaus\orcid{0000-0002-3199-0399}\inst{\ref{aff99},\ref{aff97},\ref{aff93}}
\and L.~Valenziano\orcid{0000-0002-1170-0104}\inst{\ref{aff15},\ref{aff54}}
\and J.~Valiviita\orcid{0000-0001-6225-3693}\inst{\ref{aff5},\ref{aff66}}
\and T.~Vassallo\orcid{0000-0001-6512-6358}\inst{\ref{aff17}}
\and G.~Verdoes~Kleijn\orcid{0000-0001-5803-2580}\inst{\ref{aff10}}
\and A.~Veropalumbo\orcid{0000-0003-2387-1194}\inst{\ref{aff14},\ref{aff24},\ref{aff23}}
\and Y.~Wang\orcid{0000-0002-4749-2984}\inst{\ref{aff101}}
\and J.~Weller\orcid{0000-0002-8282-2010}\inst{\ref{aff2},\ref{aff1}}
\and M.~Wetzstein\inst{\ref{aff1}}
\and A.~Zacchei\orcid{0000-0003-0396-1192}\inst{\ref{aff17},\ref{aff16}}
\and G.~Zamorani\orcid{0000-0002-2318-301X}\inst{\ref{aff15}}
\and I.~A.~Zinchenko\orcid{0000-0002-2944-2449}\inst{\ref{aff107}}
\and E.~Zucca\orcid{0000-0002-5845-8132}\inst{\ref{aff15}}
\and M.~Huertas-Company\orcid{0000-0002-1416-8483}\inst{\ref{aff42},\ref{aff108},\ref{aff109},\ref{aff110}}
\and V.~Scottez\orcid{0009-0008-3864-940X}\inst{\ref{aff78},\ref{aff111}}
\and D.~Scott\orcid{0000-0002-6878-9840}\inst{\ref{aff112}}
\and M.~Siudek\orcid{0000-0002-2949-2155}\inst{\ref{aff108},\ref{aff99}}}
										   
\institute{Max Planck Institute for Extraterrestrial Physics, Giessenbachstr. 1, 85748 Garching, Germany\label{aff1}
\and
Universit\"ats-Sternwarte M\"unchen, Fakult\"at f\"ur Physik, Ludwig-Maximilians-Universit\"at M\"unchen, Scheinerstr.~1, 81679 M\"unchen, Germany\label{aff2}
\and
INAF-Istituto di Astrofisica e Planetologia Spaziali, via del Fosso del Cavaliere, 100, 00100 Roma, Italy\label{aff3}
\and
Universit\"at Bonn, Argelander-Institut f\"ur Astronomie, Auf dem H\"ugel 71, 53121 Bonn, Germany\label{aff4}
\and
Department of Physics, P.O. Box 64, University of Helsinki, 00014 Helsinki, Finland\label{aff5}
\and
Univ. Lille, CNRS, Centrale Lille, UMR 9189 CRIStAL, 59000 Lille, France\label{aff6}
\and
Universit\'e Paris-Saclay, CNRS, Institut d'astrophysique spatiale, 91405, Orsay, France\label{aff7}
\and
Universit\'e de Strasbourg, CNRS, Observatoire astronomique de Strasbourg, UMR 7550, 67000 Strasbourg, France\label{aff8}
\and
SRON Netherlands Institute for Space Research, Landleven 12, 9747 AD, Groningen, The Netherlands\label{aff9}
\and
Kapteyn Astronomical Institute, University of Groningen, PO Box 800, 9700 AV Groningen, The Netherlands\label{aff10}
\and
Sterrenkundig Observatorium, Universiteit Gent, Krijgslaan 281 S9, 9000 Gent, Belgium\label{aff11}
\and
ESAC/ESA, Camino Bajo del Castillo, s/n., Urb. Villafranca del Castillo, 28692 Villanueva de la Ca\~nada, Madrid, Spain\label{aff12}
\and
School of Mathematics and Physics, University of Surrey, Guildford, Surrey, GU2 7XH, UK\label{aff13}
\and
INAF-Osservatorio Astronomico di Brera, Via Brera 28, 20122 Milano, Italy\label{aff14}
\and
INAF-Osservatorio di Astrofisica e Scienza dello Spazio di Bologna, Via Piero Gobetti 93/3, 40129 Bologna, Italy\label{aff15}
\and
IFPU, Institute for Fundamental Physics of the Universe, via Beirut 2, 34151 Trieste, Italy\label{aff16}
\and
INAF-Osservatorio Astronomico di Trieste, Via G. B. Tiepolo 11, 34143 Trieste, Italy\label{aff17}
\and
INFN, Sezione di Trieste, Via Valerio 2, 34127 Trieste TS, Italy\label{aff18}
\and
SISSA, International School for Advanced Studies, Via Bonomea 265, 34136 Trieste TS, Italy\label{aff19}
\and
Dipartimento di Fisica e Astronomia, Universit\`a di Bologna, Via Gobetti 93/2, 40129 Bologna, Italy\label{aff20}
\and
INFN-Sezione di Bologna, Viale Berti Pichat 6/2, 40127 Bologna, Italy\label{aff21}
\and
INAF-Osservatorio Astronomico di Padova, Via dell'Osservatorio 5, 35122 Padova, Italy\label{aff22}
\and
Dipartimento di Fisica, Universit\`a di Genova, Via Dodecaneso 33, 16146, Genova, Italy\label{aff23}
\and
INFN-Sezione di Genova, Via Dodecaneso 33, 16146, Genova, Italy\label{aff24}
\and
Department of Physics "E. Pancini", University Federico II, Via Cinthia 6, 80126, Napoli, Italy\label{aff25}
\and
INAF-Osservatorio Astronomico di Capodimonte, Via Moiariello 16, 80131 Napoli, Italy\label{aff26}
\and
Instituto de Astrof\'isica e Ci\^encias do Espa\c{c}o, Universidade do Porto, CAUP, Rua das Estrelas, PT4150-762 Porto, Portugal\label{aff27}
\and
Faculdade de Ci\^encias da Universidade do Porto, Rua do Campo de Alegre, 4150-007 Porto, Portugal\label{aff28}
\and
European Southern Observatory, Karl-Schwarzschild-Str.~2, 85748 Garching, Germany\label{aff29}
\and
Dipartimento di Fisica, Universit\`a degli Studi di Torino, Via P. Giuria 1, 10125 Torino, Italy\label{aff30}
\and
INFN-Sezione di Torino, Via P. Giuria 1, 10125 Torino, Italy\label{aff31}
\and
INAF-Osservatorio Astrofisico di Torino, Via Osservatorio 20, 10025 Pino Torinese (TO), Italy\label{aff32}
\and
European Space Agency/ESTEC, Keplerlaan 1, 2201 AZ Noordwijk, The Netherlands\label{aff33}
\and
Leiden Observatory, Leiden University, Einsteinweg 55, 2333 CC Leiden, The Netherlands\label{aff34}
\and
INAF-IASF Milano, Via Alfonso Corti 12, 20133 Milano, Italy\label{aff35}
\and
Centro de Investigaciones Energ\'eticas, Medioambientales y Tecnol\'ogicas (CIEMAT), Avenida Complutense 40, 28040 Madrid, Spain\label{aff36}
\and
Port d'Informaci\'{o} Cient\'{i}fica, Campus UAB, C. Albareda s/n, 08193 Bellaterra (Barcelona), Spain\label{aff37}
\and
INAF-Osservatorio Astronomico di Roma, Via Frascati 33, 00078 Monteporzio Catone, Italy\label{aff38}
\and
INFN section of Naples, Via Cinthia 6, 80126, Napoli, Italy\label{aff39}
\and
Institute for Astronomy, University of Hawaii, 2680 Woodlawn Drive, Honolulu, HI 96822, USA\label{aff40}
\and
Dipartimento di Fisica e Astronomia "Augusto Righi" - Alma Mater Studiorum Universit\`a di Bologna, Viale Berti Pichat 6/2, 40127 Bologna, Italy\label{aff41}
\and
Instituto de Astrof\'{\i}sica de Canarias, E-38205 La Laguna, Tenerife, Spain\label{aff42}
\and
Institute for Astronomy, University of Edinburgh, Royal Observatory, Blackford Hill, Edinburgh EH9 3HJ, UK\label{aff43}
\and
Jodrell Bank Centre for Astrophysics, Department of Physics and Astronomy, University of Manchester, Oxford Road, Manchester M13 9PL, UK\label{aff44}
\and
European Space Agency/ESRIN, Largo Galileo Galilei 1, 00044 Frascati, Roma, Italy\label{aff45}
\and
Universit\'e Claude Bernard Lyon 1, CNRS/IN2P3, IP2I Lyon, UMR 5822, Villeurbanne, F-69100, France\label{aff46}
\and
Institut de Ci\`{e}ncies del Cosmos (ICCUB), Universitat de Barcelona (IEEC-UB), Mart\'{i} i Franqu\`{e}s 1, 08028 Barcelona, Spain\label{aff47}
\and
Instituci\'o Catalana de Recerca i Estudis Avan\c{c}ats (ICREA), Passeig de Llu\'{\i}s Companys 23, 08010 Barcelona, Spain\label{aff48}
\and
UCB Lyon 1, CNRS/IN2P3, IUF, IP2I Lyon, 4 rue Enrico Fermi, 69622 Villeurbanne, France\label{aff49}
\and
Mullard Space Science Laboratory, University College London, Holmbury St Mary, Dorking, Surrey RH5 6NT, UK\label{aff50}
\and
Department of Astronomy, University of Geneva, ch. d'Ecogia 16, 1290 Versoix, Switzerland\label{aff51}
\and
INFN-Padova, Via Marzolo 8, 35131 Padova, Italy\label{aff52}
\and
Aix-Marseille Universit\'e, CNRS/IN2P3, CPPM, Marseille, France\label{aff53}
\and
INFN-Bologna, Via Irnerio 46, 40126 Bologna, Italy\label{aff54}
\and
Institute of Theoretical Astrophysics, University of Oslo, P.O. Box 1029 Blindern, 0315 Oslo, Norway\label{aff55}
\and
Jet Propulsion Laboratory, California Institute of Technology, 4800 Oak Grove Drive, Pasadena, CA, 91109, USA\label{aff56}
\and
Department of Physics, Lancaster University, Lancaster, LA1 4YB, UK\label{aff57}
\and
Felix Hormuth Engineering, Goethestr. 17, 69181 Leimen, Germany\label{aff58}
\and
Technical University of Denmark, Elektrovej 327, 2800 Kgs. Lyngby, Denmark\label{aff59}
\and
Cosmic Dawn Center (DAWN), Denmark\label{aff60}
\and
Max-Planck-Institut f\"ur Astronomie, K\"onigstuhl 17, 69117 Heidelberg, Germany\label{aff61}
\and
NASA Goddard Space Flight Center, Greenbelt, MD 20771, USA\label{aff62}
\and
Department of Physics and Astronomy, University College London, Gower Street, London WC1E 6BT, UK\label{aff63}
\and
Department of Physics and Helsinki Institute of Physics, Gustaf H\"allstr\"omin katu 2, University of Helsinki, 00014 Helsinki, Finland\label{aff64}
\and
Universit\'e de Gen\`eve, D\'epartement de Physique Th\'eorique and Centre for Astroparticle Physics, 24 quai Ernest-Ansermet, CH-1211 Gen\`eve 4, Switzerland\label{aff65}
\and
Helsinki Institute of Physics, Gustaf H{\"a}llstr{\"o}min katu 2, University of Helsinki, 00014 Helsinki, Finland\label{aff66}
\and
Laboratoire d'etude de l'Univers et des phenomenes eXtremes, Observatoire de Paris, Universit\'e PSL, Sorbonne Universit\'e, CNRS, 92190 Meudon, France\label{aff67}
\and
SKAO, Jodrell Bank, Lower Withington, Macclesfield SK11 9FT, UK\label{aff68}
\and
Centre de Calcul de l'IN2P3/CNRS, 21 avenue Pierre de Coubertin 69627 Villeurbanne Cedex, France\label{aff69}
\and
Dipartimento di Fisica "Aldo Pontremoli", Universit\`a degli Studi di Milano, Via Celoria 16, 20133 Milano, Italy\label{aff70}
\and
INFN-Sezione di Milano, Via Celoria 16, 20133 Milano, Italy\label{aff71}
\and
INFN-Sezione di Roma, Piazzale Aldo Moro, 2 - c/o Dipartimento di Fisica, Edificio G. Marconi, 00185 Roma, Italy\label{aff72}
\and
Aix-Marseille Universit\'e, CNRS, CNES, LAM, Marseille, France\label{aff73}
\and
Dipartimento di Fisica e Astronomia "Augusto Righi" - Alma Mater Studiorum Universit\`a di Bologna, via Piero Gobetti 93/2, 40129 Bologna, Italy\label{aff74}
\and
Department of Physics, Institute for Computational Cosmology, Durham University, South Road, Durham, DH1 3LE, UK\label{aff75}
\and
Universit\'e Paris Cit\'e, CNRS, Astroparticule et Cosmologie, 75013 Paris, France\label{aff76}
\and
CNRS-UCB International Research Laboratory, Centre Pierre Bin\'etruy, IRL2007, CPB-IN2P3, Berkeley, USA\label{aff77}
\and
Institut d'Astrophysique de Paris, 98bis Boulevard Arago, 75014, Paris, France\label{aff78}
\and
Institut d'Astrophysique de Paris, UMR 7095, CNRS, and Sorbonne Universit\'e, 98 bis boulevard Arago, 75014 Paris, France\label{aff79}
\and
Institute of Physics, Laboratory of Astrophysics, Ecole Polytechnique F\'ed\'erale de Lausanne (EPFL), Observatoire de Sauverny, 1290 Versoix, Switzerland\label{aff80}
\and
Telespazio UK S.L. for European Space Agency (ESA), Camino bajo del Castillo, s/n, Urbanizacion Villafranca del Castillo, Villanueva de la Ca\~nada, 28692 Madrid, Spain\label{aff81}
\and
Institut de F\'{i}sica d'Altes Energies (IFAE), The Barcelona Institute of Science and Technology, Campus UAB, 08193 Bellaterra (Barcelona), Spain\label{aff82}
\and
DARK, Niels Bohr Institute, University of Copenhagen, Jagtvej 155, 2200 Copenhagen, Denmark\label{aff83}
\and
Waterloo Centre for Astrophysics, University of Waterloo, Waterloo, Ontario N2L 3G1, Canada\label{aff84}
\and
Department of Physics and Astronomy, University of Waterloo, Waterloo, Ontario N2L 3G1, Canada\label{aff85}
\and
Perimeter Institute for Theoretical Physics, Waterloo, Ontario N2L 2Y5, Canada\label{aff86}
\and
Universit\'e Paris-Saclay, Universit\'e Paris Cit\'e, CEA, CNRS, AIM, 91191, Gif-sur-Yvette, France\label{aff87}
\and
Space Science Data Center, Italian Space Agency, via del Politecnico snc, 00133 Roma, Italy\label{aff88}
\and
Centre National d'Etudes Spatiales -- Centre spatial de Toulouse, 18 avenue Edouard Belin, 31401 Toulouse Cedex 9, France\label{aff89}
\and
Institute of Space Science, Str. Atomistilor, nr. 409 M\u{a}gurele, Ilfov, 077125, Romania\label{aff90}
\and
Dipartimento di Fisica e Astronomia "G. Galilei", Universit\`a di Padova, Via Marzolo 8, 35131 Padova, Italy\label{aff91}
\and
Institut f\"ur Theoretische Physik, University of Heidelberg, Philosophenweg 16, 69120 Heidelberg, Germany\label{aff92}
\and
Institut de Recherche en Astrophysique et Plan\'etologie (IRAP), Universit\'e de Toulouse, CNRS, UPS, CNES, 14 Av. Edouard Belin, 31400 Toulouse, France\label{aff93}
\and
Universit\'e St Joseph; Faculty of Sciences, Beirut, Lebanon\label{aff94}
\and
Departamento de F\'isica, FCFM, Universidad de Chile, Blanco Encalada 2008, Santiago, Chile\label{aff95}
\and
Universit\"at Innsbruck, Institut f\"ur Astro- und Teilchenphysik, Technikerstr. 25/8, 6020 Innsbruck, Austria\label{aff96}
\and
Institut d'Estudis Espacials de Catalunya (IEEC),  Edifici RDIT, Campus UPC, 08860 Castelldefels, Barcelona, Spain\label{aff97}
\and
Satlantis, University Science Park, Sede Bld 48940, Leioa-Bilbao, Spain\label{aff98}
\and
Institute of Space Sciences (ICE, CSIC), Campus UAB, Carrer de Can Magrans, s/n, 08193 Barcelona, Spain\label{aff99}
\and
Centre for Electronic Imaging, Open University, Walton Hall, Milton Keynes, MK7~6AA, UK\label{aff100}
\and
Infrared Processing and Analysis Center, California Institute of Technology, Pasadena, CA 91125, USA\label{aff101}
\and
Departamento de F\'isica, Faculdade de Ci\^encias, Universidade de Lisboa, Edif\'icio C8, Campo Grande, PT1749-016 Lisboa, Portugal\label{aff102}
\and
Instituto de Astrof\'isica e Ci\^encias do Espa\c{c}o, Faculdade de Ci\^encias, Universidade de Lisboa, Tapada da Ajuda, 1349-018 Lisboa, Portugal\label{aff103}
\and
Cosmic Dawn Center (DAWN)\label{aff104}
\and
Niels Bohr Institute, University of Copenhagen, Jagtvej 128, 2200 Copenhagen, Denmark\label{aff105}
\and
Universidad Polit\'ecnica de Cartagena, Departamento de Electr\'onica y Tecnolog\'ia de Computadoras,  Plaza del Hospital 1, 30202 Cartagena, Spain\label{aff106}
\and
Astronomisches Rechen-Institut, Zentrum f\"ur Astronomie der Universit\"at Heidelberg, M\"onchhofstr. 12-14, 69120 Heidelberg, Germany\label{aff107}
\and
 Instituto de Astrof\'{\i}sica de Canarias, E-38205 La Laguna; Universidad de La Laguna, Dpto. Astrof\'\i sica, E-38206 La Laguna, Tenerife, Spain\label{aff108}
\and
Universit\'e PSL, Observatoire de Paris, Sorbonne Universit\'e, CNRS, LERMA, 75014, Paris, France\label{aff109}
\and
Universit\'e Paris-Cit\'e, 5 Rue Thomas Mann, 75013, Paris, France\label{aff110}
\and
ICL, Junia, Universit\'e Catholique de Lille, LITL, 59000 Lille, France\label{aff111}
\and
Department of Physics and Astronomy, University of British Columbia, Vancouver, BC V6T 1Z1, Canada\label{aff112}}    

\newcommand{\numhavemultiple}[0]{\computed{129}}
\newcommand{\numone}[0]{\computed{316}}
\newcommand{\numtwo}[0]{\computed{86}}
\newcommand{\numthree}[0]{\computed{25}}
\newcommand{\morethanthree}[0]{\computed{18}}
\newcommand{\numcandhostswithphot}[0]{\computed{449}}
\newcommand{\numearlytypes}[0]{\computed{4420}}
\newcommand{\numinitialsampleobjects}[0]{\computed{15\,172}}
\newcommand{\numetnodnuclei}[0]{\computed{3916}}
\newcommand{\numetwithdnuclei}[0]{\computed{504}}
\newcommand{\numcanddnuclei}[0]{\computed{666}}
\newcommand{\bwlmagtostmass}[0]{\computed{-0.50}}
\newcommand{\numcanddnucleiwithphotandz}[0]{\computed{648}}
\newcommand{\numcandlttenkpc}[0]{\computed{440}}
\newcommand{\numcandlttwokpc}[0]{\computed{44}}
\newcommand{\numcandlttenkpcwithspecz}[0]{\computed{120}}
\newcommand{\numcandlttwokpcwithspecz}[0]{\computed{14}}
\newcommand{\numltohpointeight}[0]{\computed{639}}
\title{\vspace{-0.25cm}\Euclid: Quick Data Release (Q1) --
Secondary nuclei in early-type galaxies\thanks{This paper is published on behalf of the Euclid Consortium.}}
\titlerunning{\Euclid Q1: Secondary nuclei in early-type galaxies}
\authorrunning{Fabricius, M., et al.}
\abstract{Massive early-type galaxies (ETGs;  $M > 10^{11} M_\odot$ ) are believed to form primarily through
mergers of less massive progenitors, leaving behind numerous traces of violent
formation histories, such as stellar streams and shells. A particularly striking
signature of these mergers is the formation of supermassive black hole (SMBH)
binaries, which can create depleted stellar cores through interactions with
stars on radial orbits -- a process known as core scouring.
The secondary SMBH in such systems may still carry a dense stellar envelope and
thereby remain observable for some time as a secondary nucleus, while it is sinking
towards the shared gravitational potential of the merged galaxy.
Direct observations of
secondary nuclei on sub-kiloparsec scales remain rare, with only a few notable
cases, such as NGC\,5419. Investigating such features and building up statistics
requires both high spatial resolution and wide-field coverage, a capability
uniquely provided by \Euclid{}.
In this study, we leverage \Euclid{}’s Q1 Early Release data to systematically
search for secondary nuclei in ETGs. We present a preliminary sample of
\numcanddnuclei{} candidate systems distributed over \numetwithdnuclei{} hosts
(some of which contain multiple secondary nuclei). The vast majority of these
fall at separations of \SI{3}{\kilo\parsec} to \SI{15}{\kilo\parsec}, indicative
of normal mergers. \numcandlttwokpc{} fall at projected separations of less than
\SI{2}{\kilo\parsec}.
In the discussion, we argue that this most
interesting subset of secondary nucleus candidates -- those at very close angular
separations -- are unlikely to be a consequence of chance alignments. We show that
their stellar masses are mostly too large for them to be globular clusters and
that a significant subset are unresolved even at \Euclid's spatial resolution,
rendering them too small to be dwarf galaxies. These may represent
the highest-density nuclei of a previously merged galaxy, currently sinking into
the centre of the new, common gravitational potential and thus likely to host a
secondary SMBH. We then demonstrate that convolutional neural networks offer a
viable avenue to detect multiple nuclei in the thirty-times larger sky coverage
of the future \Euclid{} DR1.
Finally, we argue that our method
could detect the remnants of a recoil event from two merged SMBHs: Two of our
secondary nuclei candidates are unresolved at
the \Euclid{} spatial resolution, occur
at projected physical separations of less than 2\,kpc, and occur in hosts
of $M > 10^{11} M_\odot$, which makes them viable candidates.}
\keywords{
Surveys --
Galaxies: nuclei --
Galaxy: evolution --
Galaxy: center --
Galaxies: supermassive black holes
}
\maketitle
\section{Introduction} \label{sec:intro}
Early-type galaxies (ETGs) are traditionally characterised by their smooth,
featureless light distributions and the absence of significant star formation
(e.g.\ \citealp{Hoessel1980, Schneider1983, BenderMoellenhoff1987}).
However, high-resolution imaging, particularly from the \HST{} (HST), has revealed that
many of these galaxies possess complex internal structures, including central
cores, dust lanes, nuclear disks, and multiple nuclei \citep{Carollo1997},
and kinematically decoupled cores (e.g.\ \citealp{Bender1988}). These
discoveries challenge the simplistic view of ETGs as dynamically relaxed
systems.

Observational studies have demonstrated that a significant fraction of ETGs may
exhibit multiple nuclei. For example, \citet{Goullaud2018} reported that
approximately 37\% of massive ETGs display isophotal position angle rotations,
often attributable to the presence of multiple nuclei or nearby companions.
These features are frequently indicative of past merger events, where two or
more galaxies have coalesced, leaving behind remnants of their original cores.

Theoretical models predict that the presence of multiple nuclei in ETGs is a
natural outcome of hierarchical galaxy formation \citep{Volonteri2003a,
Springel2005}. In this framework, galaxies evolve through successive mergers and
accretion episodes. During such interactions, the central regions of the
progenitor galaxies can persist for
several Gyrs (e.g. \citealp{Tremmel2015}) before
ultimately merging into a single system. This process is particularly prevalent
in dense environments such as galaxy clusters, where galaxy interactions occur
frequently.

The study of multiple nuclei in ETGs offers insight into the role of
supermassive black holes (SMBHs) in galaxy evolution. In some cases, each
nucleus may host its own black hole, leading to the formation of binary or even
multiple SMBH systems \citep{Valtonen1996}. Observational evidence for such
scenarios includes the presence of central cores in massive elliptical galaxies,
which exhibit tangential anisotropies in their stellar orbit distributions
\citep{Valtonen1996, Rusli2013, Thomas2014, Thomas2016, Rantala2018}. These
features are understood to result from the late stages of galaxy mergers, in
which SMBHs settle into the centre of the newly formed gravitational potential.
The resulting binary SMBH interacts preferentially with stars on radial orbits,
ejecting them from the central region and thereby forming a depleted core
\citep{Volonteri2003b}. The core size may then get even further increased as a
consequence of the gravitational wave recoil of the merged SMBH
\citep{Khonji2024}. In fact, cored galaxies have already been identified in
\Euclid{} data: \cite{Saglia24} detected a 0.4\,kpc core in NGC\,1272 and show
that it hosts a $5 \times 10^9 M_\odot$ SMBH.

\sidecaptionvpos{figure}{c}
\begin{SCfigure*}[20][h]
\includegraphics[scale=.38]{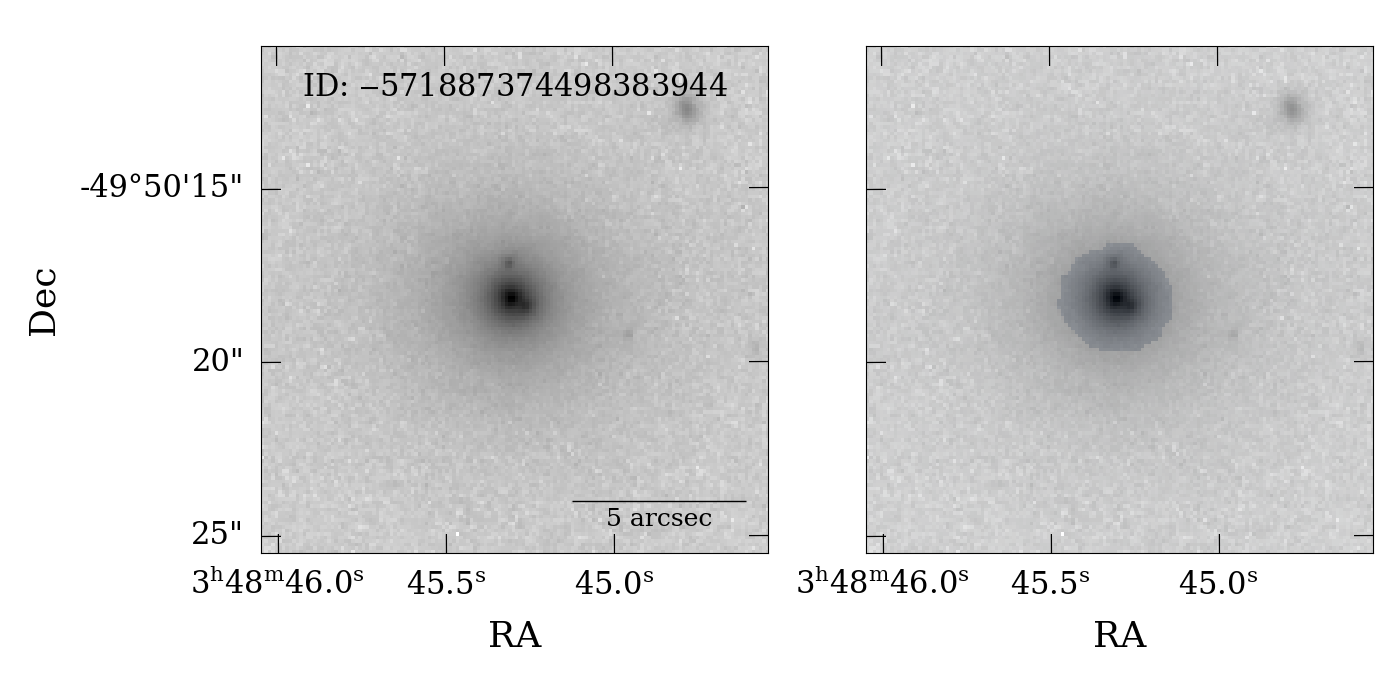}
\caption[]{VIS cutout. \emph{Left panel}: Basis for all our searches are VIS
    cutouts that we generate on the ESA Datalabs service (see
    Sect.~\ref{sect:fitting-models} for details). The image size is adjusted
    dynamically to be 1.23 times the Kron radius as determined by MER.
    \emph{Right Panel}: We initially compute a segmentation map (dark shaded
    region) that loosely selects the central high-surface brightness region of a
    galaxy.}
\label{fig:central-segmentation}
\end{SCfigure*}

This merger-driven scenario is further supported by the recent detection of a
low-frequency gravitational wave background via pulsar timing arrays (PTAs). The
amplitudes reported by various PTA collaborations \citep{Agazie2023,
EPTA_InPTA2023, Reardon2023} are consistent with merger rates derived from the
high-mass end of the galaxy mass function \citep{Liepold2024}.

A natural precursor to such a core formation stage is the presence of a
secondary nucleus -- the remnant core of an accreted galaxy stripped of its
outer stellar envelope. Many ultra-compact dwarf galaxies (UCDs) are now
understood to be the remnants of larger galaxies that have undergone tidal
stripping in dense environments such as galaxy clusters. In particular,
\cite{Neumeyer2012} and \cite{Mieske2013} show that many UCDs have elevated
dynamical mass-to-light ratios, which cannot be explained by normal stellar
populations. These high ratios suggest the presence of additional mass
components, such as central massive black holes -- a hallmark of former
galactic nuclei (see also \citealp{Seth2014}). The occurrence of multiple nuclei
in ETGs therefore carries significant implications for our understanding of
galaxy assembly. Their frequency should be consistent with PTA amplitudes and
with the redshift-dependent growth function of SMBHs as a function of host mass
and environment.

Despite their relevance, systems with clearly identifiable secondary nuclei have
remained relatively rare until recently. Generally, a large spatial coverage is
required to study the statistical properties of these systems.
\cite{Bhattacharya2023} have detected 159 dual AGN candidates in the Sloan
Digital Sky Survey, using automated searches (see also \citealp{Q1-SP072}).
But core formation is expected to
set in at SMBH separations of a few kiloparsecs. The detection of such systems is
difficult to achieve from the ground due to the limited spatial resolution of ground
based surveys. An alternate possibility is to search for dual AGNs
spectroscopically, and to confirm their dual nature through the identification
of multiple redshift manifestations of the same emission line
\citep{Goulding2019}. Using high spatial resolution observations from
HST, research has started to uncover a more massive population
of SMBH pairs with kiloparsec-scale separations by detecting multiple distinct nuclear
cores associated with unresolved AGN emissions \citep{Xu2009, Fu2012}.

A prominent example for a close separation system is NGC\,5419, a cored ETG
hosting a secondary nucleus at a projected separation of approximately
\SI{70}{\parsec} from the galactic centre \citep{Capetti2005, Lauer2005}. This
galaxy hosts a central SMBH with a mass of $10^{10}{M_\odot}$
\citep{Mazzalay2016, Neureiter2023}.

Recent searches for dual AGN using \textit{Gaia} have been successful, uncovering several
binary AGN through a novel approach known as the \textit{Gaia} multi-peak method
\citep{Mannucci2022}. A key advantage of focusing the search on AGN, as
highlighted, is the ability to examine emission lines in spectroscopic
follow-up observations \citep{Mannucci2023}, which enables the determination of relative
redshifts and confirmation of their nature through the detection of multiple
velocity emission lines. Additionally, the point-like nature of AGN allows for
the identification of systems with the smallest spatial separations. However, a
significant portion of passive secondary nuclei remains undetected, which
complicates efforts to compare their frequency of occurrence with theoretical
models of core formation.

While we do not select against AGN in this study, we focus on the detection of
passive secondary nuclei in \Euclid{} VIS images. We analyse cutout images
from the VIS instrument \citep{Cropper16, EuclidSkyVIS, Q1-TP002} included in
the \Euclid{} \citep{EuclidSkyOverview} Q1 data release \citep{Q1cite, Q1-TP001}, to
investigate the occurrence of multiple nuclei in early-type galaxies.
For the resulting sample, we compute the frequency of occurrence, derive their
luminosities, and examine the redshift distributions.
We conclude by discussing
the prospects for extending this analysis to the upcoming \Euclid{} Data
Release~1 (DR1), and consider the potential to identify recoiling SMBH candidates
within this framework.

The structure of this paper is as follows: In Sect.~\ref{sec:data},
we describe the data, specifically the \Euclid{} Q1 data release.
The methodology for sample selection, structural modelling, and
visual inspection is detailed in Sect.~\ref{sec:methods}. Our
results, including the frequency, luminosity, and redshift distribution
of multiple nuclei, are presented in Sect.~\ref{sec:results}. We
discuss the implications of our findings and potential avenues for
future work, including extensions to the forthcoming DR1 and the
identification of recoiling supermassive black holes, in
Sect.~\ref{sec:discussion}. We summarise our conclusions in
Sect.~\ref{sec:conclusions}.
We assume a \citet{Planck2018} cosmology with  $H_0\approx 67.66\,{\rm
km\,s^{-1}\,Mpc^{-1}}$, $\Omega_{\text{m}} = 0.31$, and  $\Omega_{\Lambda} =
0.69$ for the computation of luminosities and to translate observed angular to
physical separations throughout this paper.

\section{Data}
\label{sec:data}
The \Euclid{} Q1 data release \citep{Q1cite} represents
the first significant public dataset from the \Euclid{} mission
\citep{EuclidSkyOverview}. The release is described in detail
in \citet{Q1-TP001}.
Q1 covers 63.1\,\si{\deg\squared} of sky area across the three
Euclid Deep Fields, to the depth of the nominal Euclid Wide
Survey \citep{Scaramella-EP1}.  The release provides imaging and
photometric data, with coverage in both the visible and near-infrared
bands.

A key component of this release is the data from the Visible Imaging Channel
(VIS; \citealp{EuclidSkyVIS}). The VIS imaging data consist of
deep optical observations obtained in a broad visible band, ranging
approximately from \SI{550}{\nano\meter} to \SI{900}{\nano\meter}
at a spatial resolution of \ang{;;0.18}. It provides
high-resolution visible imaging with a field of view of 0.57\,deg$^2$ and a
pixel scale of \ang{;;0.1}. \Euclid{} and the VIS channel are optimised to
suppress stray light and to maintain near diffraction-limited image quality to
detect faint galaxies and measure their shapes with exceptional precision,
enabling weak lensing analyses. The Q1 release includes both reduced images and
associated detection catalogues, featuring detections and photometric
measurements and segmentation maps.

In addition to VIS, the Near-Infrared Spectrometer and Photometer (NISP;
\citealp{EuclidSkyNISP}) operates in the 950\,nm--2020\,nm range,
enabling observations in the near-infrared spectrum. NISP complements VIS by
providing photometric and spectroscopic data, which are essential for measuring
photometric redshifts and exploring galaxy clustering.

The Q1 data release provides an essential resource for studies of galaxy
morphology, structure, and clustering, as well as for the detection of
substructures such as multiple nuclei in early-type galaxies. It offers a unique
opportunity to explore the detailed central morphologies of galaxies at
space-based spatial resolution in an unprecedented large sky area, observed in
unbiased rather than pointed observations. Q1 serves as a benchmark for the
development and validation of methodologies in preparation for the later, more
extensive \Euclid{} mission data releases.

\section{Methods}
\label{sec:methods}
\subsection{Sample selection}
All objects discussed here are part of the MER detection catalogue
\citep{Q1-TP004} as provided in the Q1 data release of the \Euclid
project \citep{Q1-TP001}.
Based on some early experiments on workable image sizes
and object brightnesses we apply an initial selection based on
two criteria only:
The segmentation area of the objects must exceed
5500 pixels, and the flux within the one times full width at half
maximum (FWHM) aperture must be greater than \SI{20}{\micro\jansky},
amounting to a magnitude of 20.5\,AB in VIS. We give the
exact ADQL statement in Appendix~\ref{sec:cut_study}.
This effectively
excludes objects from subsequent analysis that are too faint or have
too small projected sizes. The segmentation cut also excludes
objects in crowded regions, where there is significant spatial
overlap between neighbouring objects.
This results in 90\,502 MER detections. In
Appendix~\ref{sec:cut_study} we look in more detail at how
our selection criteria affects the redshift space and stellar mass distribution compared to the full MER catalogue.

Following this initial sample selection, we obtain the \textit{Gaia} DR3
catalogue \citep{BabusiauxFabricius2023, Gaia-CollaborationBrown2021,
Gaia-CollaborationPrusti2016, Gaia-CollaborationVallenari2023,
RowellDavidson2021} covering the Euclid Deep Field North (EDF-N), Euclid Deep
Field South (EDF-S), and Euclid Deep Field Fornax (EDF-F) regions. We cross-match
the \textit{Gaia} catalogue (all sources with $G < 20$) with our initial sample.
To exclude foreground stellar contamination, we reject all detections located
within \ang{;;5} of any entry in the \textit{Gaia} catalogue,
which removes 83\% of all detections.  We also require
all sources in our sample to have a Kron radius \citep{Kron1980} measurement
from MER.\footnote{\url{http://st-dm.pages.euclid-sgs.uk/data-product-doc/dmq1/}}
This leaves \numinitialsampleobjects~sources for further investigation.

\subsection{Fitting of analytic models of stellar light distribution}
\label{sect:fitting-models}
We generate VIS image cutouts using the Datalabs service
\citep{Navarro2024} provided by ESA. The cutouts are created with
dimensions set to 1.2 times the Kron radius in diameter.\footnote{This number is somewhat arbitrary, and ultimately arose from an example given
by ESA. The exact choice does not affect the results.} Prior to
initiating the fitting process, we compute a segmentation that
focuses on the central light distribution. This is done using the
{\tt skimage.filters.threshold-triangle
} method from {\tt scikit},\footnote{\url{https://scikit-image.org/docs/stable/}}
applied to a median-filtered version of the image, which consistently
identifies a reasonably bright region around the brightest central
light distribution of each object (Fig.~\ref{fig:central-segmentation}).

We then proceed to model the light distribution of each object in a two-step
process: Initially, we fit a multi-Gaussian expansion
(MGE; \citealp{Cappellari2002}) to the pixels contained in our central
segmentation area. This initial modelling captures the broader underlying light
distribution of the host galaxy. Following this, we identify and flag central
features that deviate significantly from the modelled profile, indicating
potential substructures or additional nuclei. For this, we employ a simple
$\kappa$-$\sigma$ clipping scheme\footnote{The \(\kappa\)-\(\sigma\) clipping
scheme iteratively excludes data points that deviate more than \(\kappa\)
standard deviations from the mean, recalculating the mean and standard deviation
after each iteration until convergence is achieved. This method is commonly used
to mitigate the impact of outliers in astronomical datasets.} with $\kappa$ set
to 2. Where fewer than 200 pixels survive this clipping, we successively
increase $\kappa$ by 50\% until at least 200 pixels remain to leave a sufficient
number of data points for the model fitting.

The MGE modelling process is very stable and finds reasonable representations
of the light distribution in virtually all cases, even when nearby foreground or
background galaxies overlap the segmentation area. However, in some cases, an
inner Gaussian component can be assigned to a secondary nucleus even if it is
offset. For this reason, and also motivated by the desire to construct a more
physically meaningful model, we turn to a Sérsic model for the next step.

\begin{figure*}
    \centering
    \includegraphics[width=.9\linewidth]{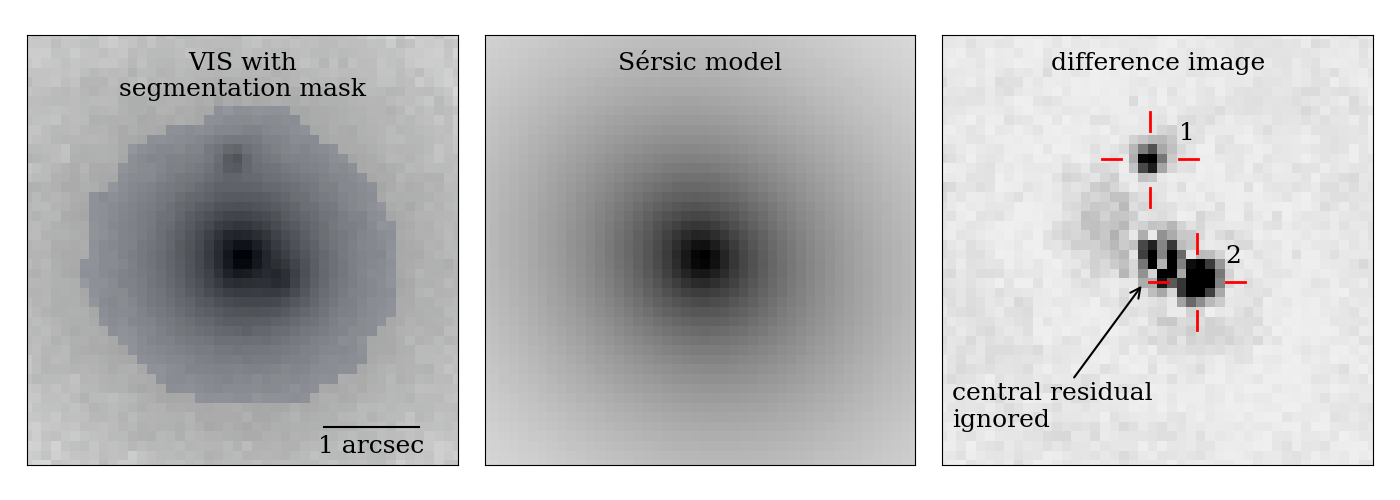}
    \caption{Modelling process. The object is the same as in
    Fig.~\ref{fig:central-segmentation} (MER ID: $-571887374498383944$). To all
    the pixels that fall within our segmentation map (\emph{left}), we fit a
    two-dimensional Sérsic light distribution using {\tt imfit} (\emph{centre}).
    We often observe central residuals after the Sérsic model subtraction, but
    spatially offset potential secondary nuclei become readily visible in the
    difference image on the \emph{right}; there are two in this particular example.}
    \label{fig:image_modelling_part2_01}
\end{figure*}

Next, we refine the light distribution modelling using the
\texttt{imfit} tool \citep{Erwin2015}. This second fit is applied
to the previously defined segmentation area, excluding the flagged
image regions corresponding to potential central structures. We
employ a single Sérsic model, allowing the position, the central
intensity, Sérsic index, effective radius, ellipticity, and position
angle to vary freely during the fitting process. We incorporate the
provided VIS point spread function (PSF) specific to the image
location of each object, allowing \texttt{imfit} to convolve the
model with the PSF during the fit.

\subsection{Visual inspections}
To analyse the central substructures, we begin by subtracting the best-fitting
\texttt{imfit} model from the original object's light distribution. This
difference image emphasises any residual central structures that might be
present.

Figures~\ref{fig:central-segmentation} and \ref{fig:image_modelling_part2_01}
show an example of this for object $-571887374498383944$ from the MER
catalogue. The reader may notice a central residual in the difference image.
This may be due to several factors, such as imperfections in the VIS PSF model,
an actual central component like a nuclear star cluster that we do not include
in our model, or a general deviation from a Sérsic light profile in the centre.
Such central residual features are very common; in fact, the majority of our
difference images exhibit them. We ignore those in our analysis. Of particular
interest for this work, however, are centrally offset structures. In the right
panel of Fig.~\ref{fig:image_modelling_part2_01}, one can see two additional,
marginally resolved features, offset to the bottom right and to the top of the
centre of the host.

\begin{figure}
    \centering
    \includegraphics[width=1.\linewidth]{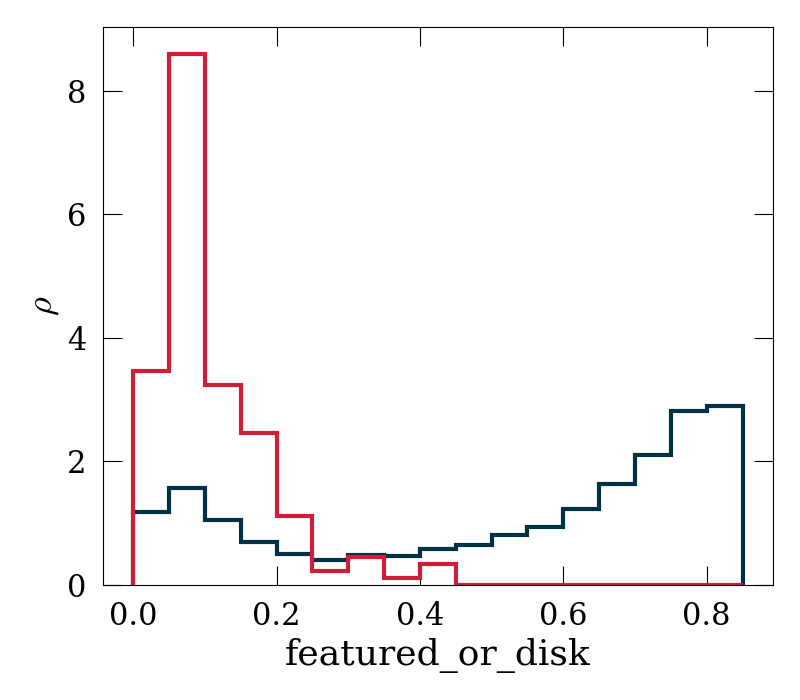}
    \caption{
    Normalized density distributions of the \texttt{Zoobot} computed
    \texttt{featured\_or\_disk} parameter. The blue histogram shows all
    \numinitialsampleobjects~objects that result from our initial sample
    selection. The red histogram shows those
    \numearlytypes~objects that we labelled `early type'.
    }
    \label{fig:featured_or_disk}
\end{figure}

\begin{figure}
    \centering
    \includegraphics[width=\myfigurewidth\linewidth]{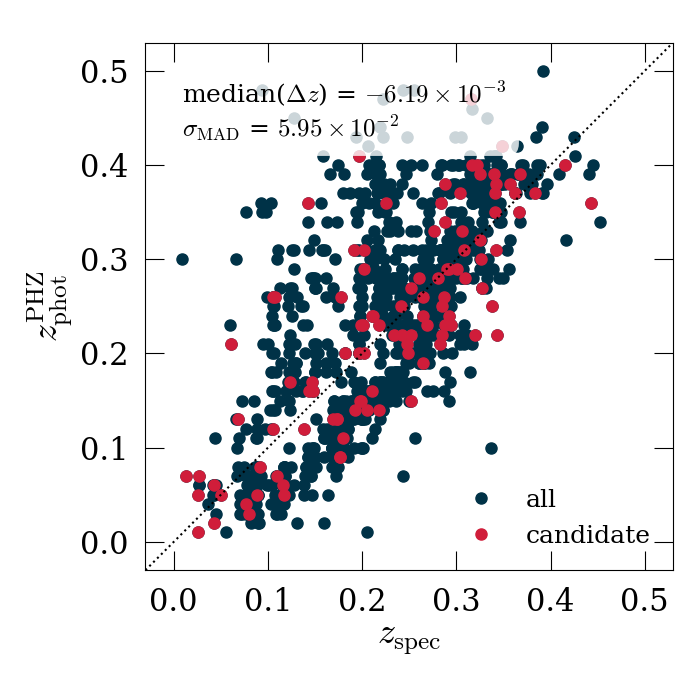}
    \caption{Comparison of photometric redshift estimates by PHZ to
    spectroscopic measurements. We show all objects in our early-type subsample
    where spectroscopic data are available. The limited precision in the various
    space- and ground-based imaging channels still limits the accuracy of
    photometric redshifts in Q1. Any quantity that is derived from the
    photometric redshift thus needs to be taken with care.
     }
    \label{fig:specz_photz}
\end{figure}

\begin{figure}
    \centering
    \includegraphics[width=\myfigurewidth\linewidth]{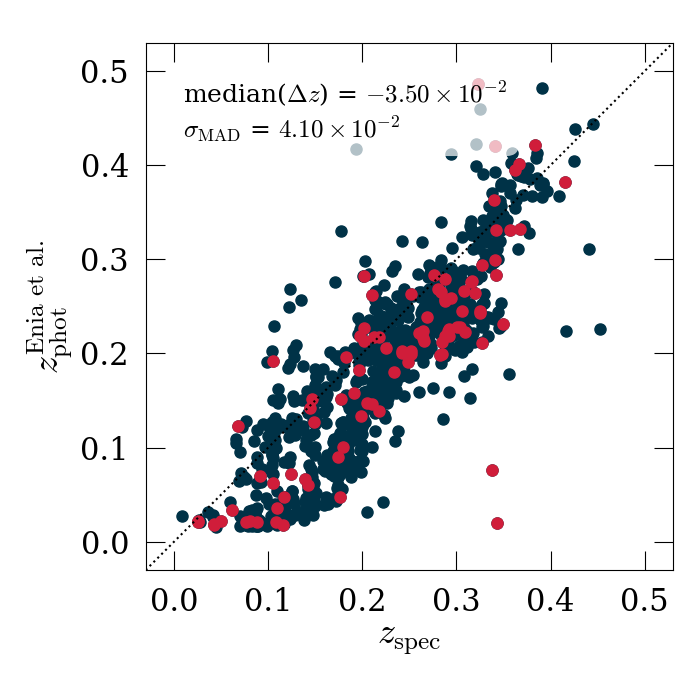}
    \caption{
    Same as Fig.\,\ref{fig:specz_photz}
    but showing the
    comparison to the redshifts by ENIA25.
     }
    \label{fig:specz_photz_enia}
\end{figure}

We visually inspect all cutouts and the corresponding difference images. We
attribute the following labels to each object:

\begin{itemize}
    \item {\bf Early type/smooth light distribution:} We reject
    the presence of dust, bars, spiral structure, but also edge-on cases
    that complicate the interpretation of the central morphology.
    The criterium here really is: Would we detect a secondary
    nucleus if it was present?
    If so, we label the respective object as having
    a smooth and early-type like central light distribution. This
    will include ellipticals but also S0 and dwarf ellipticals.
    \item {\bf Secondary object visible:} In the VIS image, a
    compact secondary object is visible by eye,
    either moderately resolved or unresolved,
    within the footprint of the primary
    host. No distinction is made here with respect to, for
    example, spatial separation, compactness, or brightness.  Such
    secondary objects might include:
    \begin{itemize}
        \item genuine secondary nuclei, potential merger remnant,
        \item globular clusters,
        \item fore- or background objects,
        \item dwarf galaxies,
        \item merger pairs.
    \end{itemize}
    \item {\bf Gravitational lensing images:} During the first
    iterations of visual inspections we realised that  in a number
    of cases, the residual images also make central strong lensing
    images and arcs easily visible. We label such objects separately.
    Over 120 bright lensing candidates are found this way, which
    will be presented in \cite{Q1-SP082}.
    \item {\bf Star:} Not all stars get successfully rejected by
    matching against the \textit{Gaia} catalogue. Also, extended
    diffraction spikes contaminate the image sometimes. We label these images
    as {\it star} and reject them from any further analysis.
    \item {\bf Artefact:} We reject obvious image artefacts such as
    ghosts, diffraction spikes and incomplete cosmic removal.
\end{itemize}

Generally our visual host classification shows agreement with other,
machine-learning based approaches within \Euclid{}: We compare classifications of
systems as early type (or not early type) to labels obtained using
\texttt{Zoobot} \citep{Q1-SP047}. The classifications are part of the \Euclid{}
Q1 data release and are the result of fine-tuning the \texttt{Zoobot} galaxy
foundation models on annotations from an intensive one-month campaign by Galaxy
Zoo volunteers. \citet{Q1-SP040} already explored the correlation of their
Sérsic indices with the \texttt{featured\_or\_disk} and \texttt{smooth}
parameters from the \texttt{Zoobot} team and found that these labels generally
do well selecting systems with high Sérsic index (early type) or low Sérsic
index (late type) systems. In Fig.~\ref{fig:featured_or_disk}, we compare our
classifications of galaxies as `early type' (or not) to the
\texttt{featured\_or\_disk} label. Generally we find that our early type
systems all have low values of \texttt{featured\_or\_disk}.

Our sample must be strictly understood as a candidate sample, as we have no way
to cleanly select against any of the discussed contaminants. We will discuss the
probabilities of chance alignments and contamination by globular clusters later
(see Sect.~\ref{sec:discussion}).
During the inspection, we became increasingly aware of the presence of strong
gravitational lenses that had not been identified in previous studies \citep{Q1-SP082},
which prompted a reinspection of the entire sample. We also
recognised the necessity of explicitly flagging whether or not an object is
considered to exhibit early-type morphology. As a result,
every image is inspected at least three times
in random order, but in the context of this work, we make no attempt to apply
multi-expert classifications or citizen science approaches. Hence, we cannot
claim completeness or a fully unbiased selection of candidates (see
Sect.~\ref{sec:towards_dr1_automated_detections} for a discussion on automated
detection methods).

\subsection{Redshift estimates}
Redshift estimates are required to turn angular sizes or separations into physical
quantities. The scatter between the PHZ-determined photometric redshifts and
spectroscopic redshifts is, however, large. In Fig.~\ref{fig:specz_photz} we
show the comparison between spectroscopic and photometric redshift, for all
early-types in our sample for the PHZ derived photometric redshifts and
Fig.~\ref{fig:specz_photz_enia} for the redshifts by \citet[ENIA25 in
the following]{Q1-SP031}. In this plot, we generally observe a bias in both methods.

We compute the median of the residual $\Delta z = (z_{\rm{phot}} - z_{\rm{spec}})/(1 + z_{\rm{spec}})$ as measure of the respective bias and also compute the
normalized median absolute deviation as measure of the precision
$\sigma_{\rm{MAD}} = 1.4826\times\rm{median}\left(|\Delta z - \rm{median}(\Delta z)|\right)$ as measure of the precision \citep{Hoaglin1983}.
We obtain a bias value of $-6.19 \times 10^{-3}$ for PHZ with $\sigma_{\rm{MAD}} = 5.95 \times 10^{-2}$. For ENIA25 we obtain a bias of $-3.50 \times 10^{-2}$ with
$\sigma_{\rm{MAD}} = 6.30 \times 10^{-2}$.

We apply our computed bias corrections to the
PHZ redshifts and the ENIA25 redshifts wherever we use them.
In general, we give preference to the latter as they show somewhat
smaller intrinsic scatter. Throughout this work, we pick the best available
value for the redshift following the order:

\begin{figure}
    \centering
    \includegraphics[width=\myfigurewidth\linewidth]{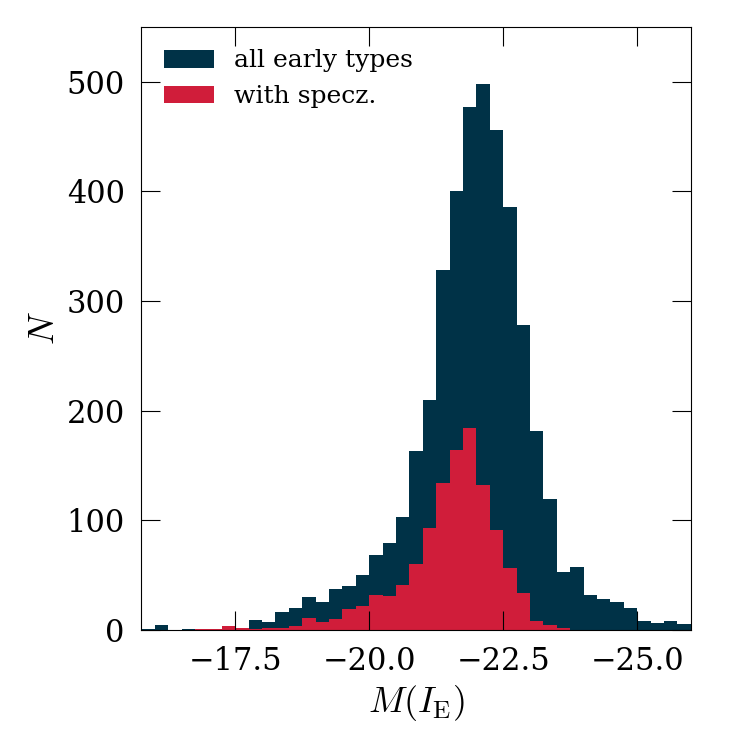}
    \caption{Distribution of absolute
    magnitudes, corrected for Galactic extinction.
    The blue histogram shows all early-type
    systems in our sample, the red histogram shows those that have spectroscopic
    redshifts. We truncate the plot at $M(\IE) = -26$
    as such luminosities are probably a consequence of failures
    in the photometric redshift estimates.} \label{fig:sample_abs_mags}
\end{figure}

\begin{itemize}
    \item We use external spectroscopic redshifts where available
    (see the caption of Table~\ref{tab:catalog} for a list of literature sources).
    \item We use spectroscopic redshifts that we take
    from the Northern Ecliptic Pole Survey (TESLA) data \citep{Ortiz2023},
    which were obtained at the Hobby--Eberly Telescope \citep{Ramsey1998}
    using the VIRUS \citep{Hill2021} instrument in connection with
    the HETDEX survey \citep{Gebhardt2021}. Details of the spectral
    extraction and derivation of redshifts are given in Balzer (in prep.).
    All our candidates fall within $z < 1$.
    \item We use the ENIA25 photometric redshift where available.
    \item As last fallback we use the \Euclid PHZ photometric redshift
    \citep{Q1-TP005}.
\end{itemize}

In Fig.~\ref{fig:sample_abs_mags}, we show the distribution of absolute VIS
magnitudes of all the ETGs in our sample. In this plot, the observed magnitudes
are corrected for Galactic extinction using the dust maps provided by
\cite{Schlafly2011} through the \textsf{dustmaps}
tool.\footnote{\url{https://dustmaps.readthedocs.io/en/latest/maps.html}} The
absolute magnitude distribution of objects that do have spectroscopic redshifts
(in red) does not extend beyond $M(\IE) = -23$,
as expected from the luminosity function of galaxies \citep[for
instance][]{Cuillandre2025}.

In the VIS images, we manually draw circular apertures around compact,
moderately, or unresolved secondary objects.
We then use these to firstly assess their projected separation from the galaxy
centre, and secondly, to measure the brightness of the secondary object.
Attempts to fit the secondary objects with dedicated Sérsic models generally
fail due to their relatively low brightness (compared to the host) and because
they typically cover very few pixels in the image. We thus restrict ourselves to
simple aperture photometry. For this, we rerun the Sérsic model fit after
masking any secondary object, in case it was not already sufficiently masked
during the automatic masking procedure. We then recompute a residual image by
subtracting the model from the actual image and integrate the residual flux
inside the manually drawn aperture. In all cases, the aperture is at least
\ang{;;2.0} in diameter.
Using the VIS PSF, we compute that, within this aperture, virtually no flux
of a secondary object is lost, provided the object was a point source:
As a test, we
model a potential secondary nucleus as a Gaussian with a sigma of \ang{;;0.5}
(as we will see, this is larger than most of our detected candidate
secondary nuclei; Sect.~\ref{sec:chance_alignments}).
We then integrate the signal over a $r =$ \ang{;;2.0} aperture and we find that we
still recover 84\% of the flux.
The on-sky positions of the candidate
nuclei are measured using a simple centre-of-mass method on the residual image.
Using the center-of-mass estimate for the position we finally compute a Kron
radius (\citealp{Kron1980}; see also Appendix~\ref{sec:kron_to_re}) for all
candidate secondary nuclei in our sample that are located within the circular
aperture on the residual image.

\section{Results}
\label{sec:results}
In total, we label \numearlytypes~objects as early-type galaxies or as having a
sufficiently smooth light distribution to confidently detect secondary nuclei.
A total of \numetwithdnuclei~galaxies exhibit one or more secondary objects,
visible within their main footprint. As a galaxy might host more than one
candidate for a secondary nucleus, our list of candidates is larger than the
list of hosts:
\numone~galaxies host only one additional nucleus, \numtwo~two secondary nuclei,
\numthree~three, and \morethanthree~systems host more than three secondary
candidates.
For \numcanddnucleiwithphotandz~candidates, we have redshifts of
their hosts (either photometric or spectroscopic) and measurements of their
luminosity and angular separation from the host centre.

\begin{figure*}
\centering
\begin{tabular}{ccc}
    \includegraphics[width=0.29\textwidth]{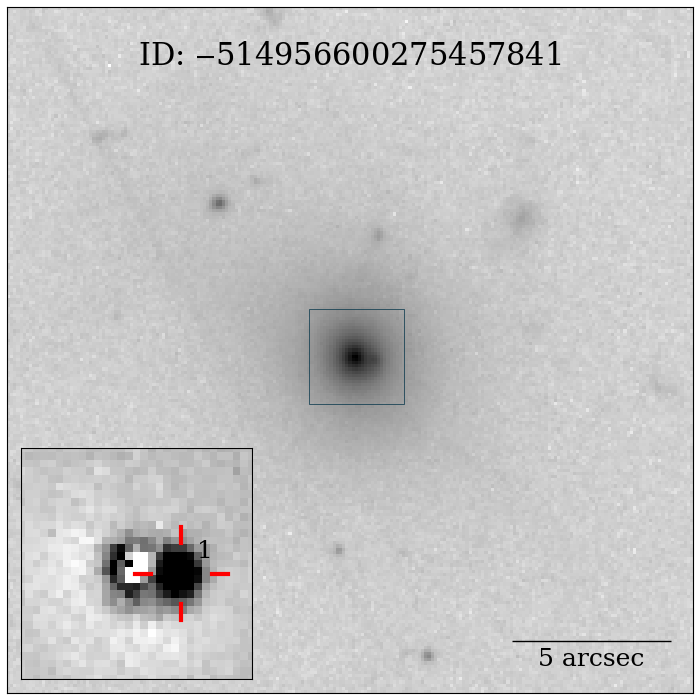} &
    \includegraphics[width=0.29\textwidth]{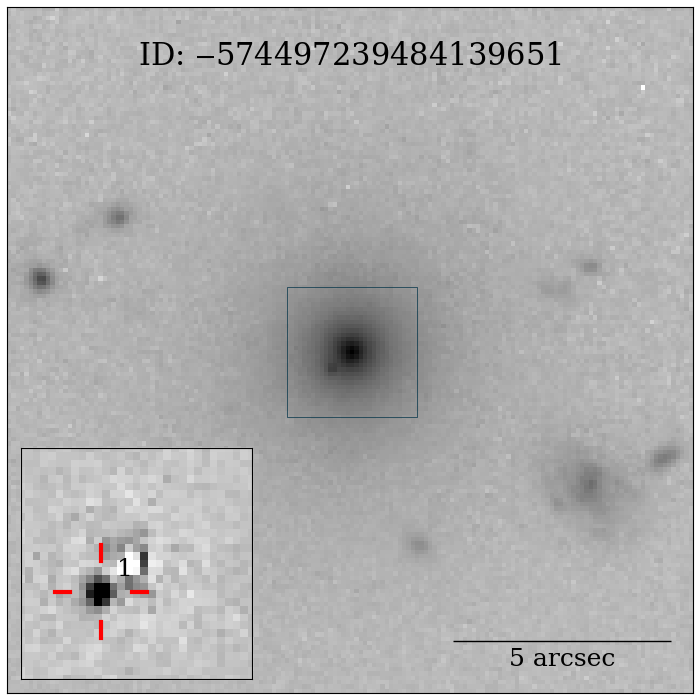} &
    \includegraphics[width=0.29\textwidth]{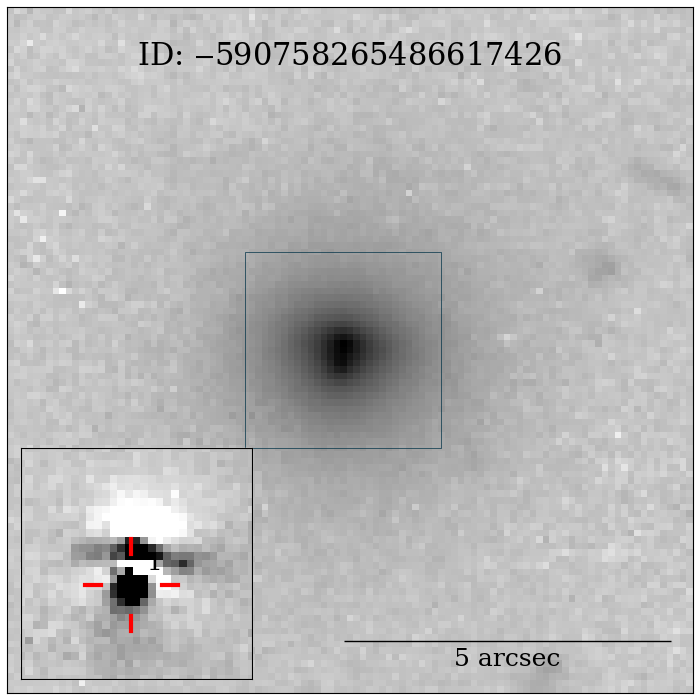} \\
    \includegraphics[width=0.29\textwidth]{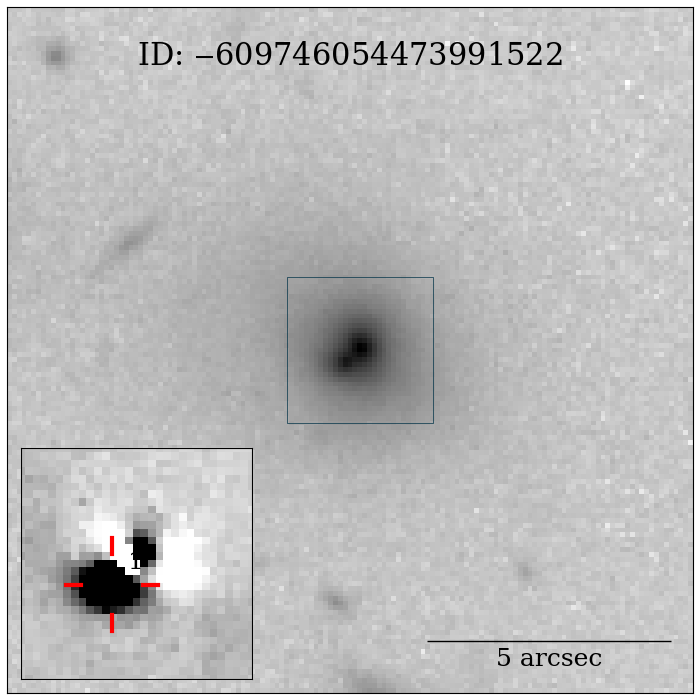} &
    \includegraphics[width=0.29\textwidth]{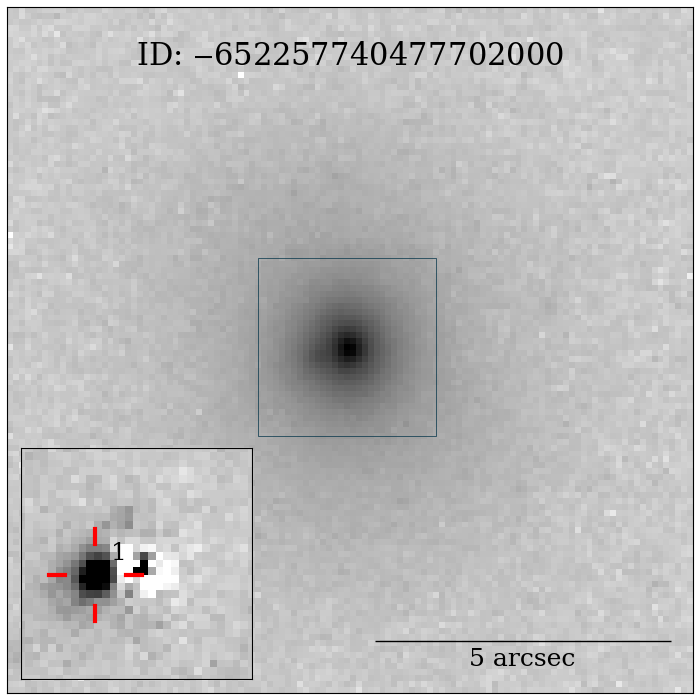} &
    \includegraphics[width=0.29\textwidth]{figures/obj-514956600275457841.png} \\
\end{tabular}
    \caption{Examples of objects hosting a candidate secondary nucleus. The main panels
    display the VIS cutout of the host in logarithmic stretch. The inset
    shows a $3'' \times 3''$ zoom-in on the centre of the host
    after subtraction of the \texttt{imfit} model. The red crosshair indicates
    the location of the secondary nucleus. }
    \label{fig:dnuclei_examples}
\end{figure*}

\begin{figure}[h]
    \centering
    \includegraphics[width=\myfigurewidth\linewidth]{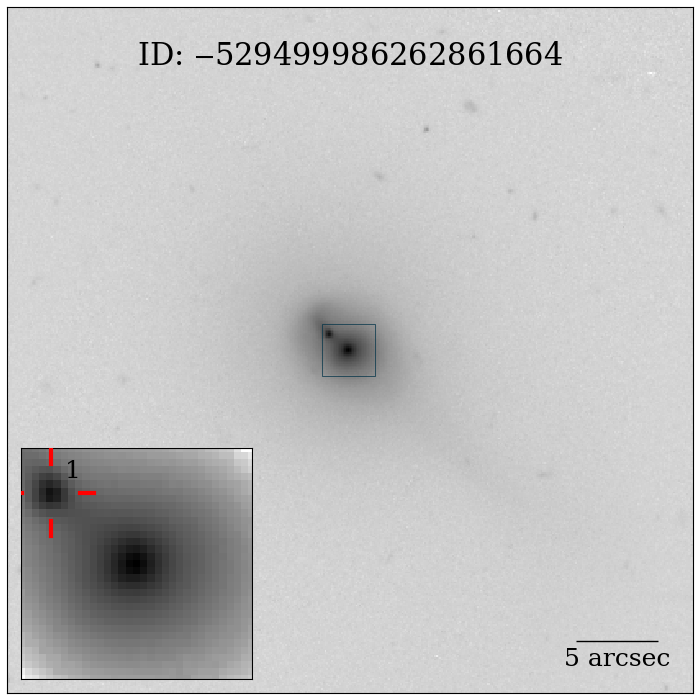}\\
    \includegraphics[width=\myfigurewidth\linewidth]{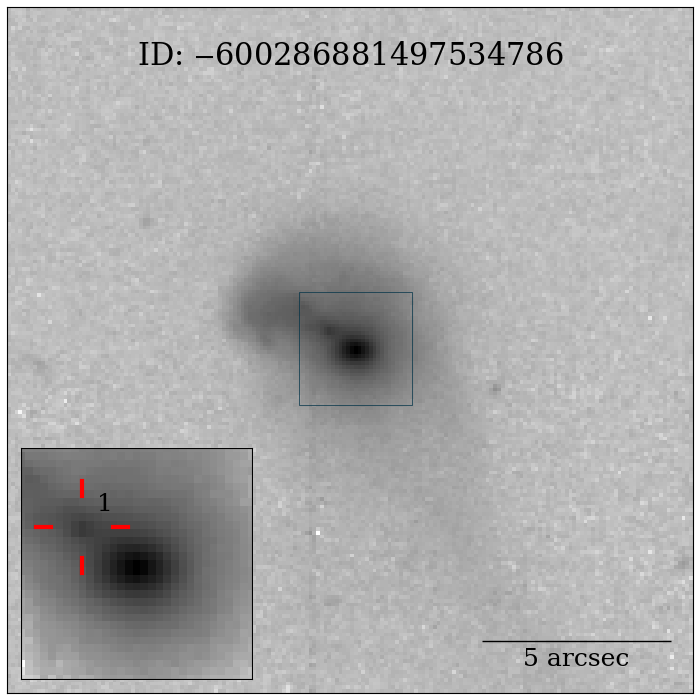}
    \caption{Candidate mergers with secondary nuclei. The significantly
    disturbed light profile of the hosts indicates that these galaxies have
    experienced a recent merger. Note that here the inset shows the original VIS
    image rather than the result of a subtraction of the \texttt{imfit} model.}
    \label{fig:punch_through}
\end{figure}

Figure~\ref{fig:dnuclei_examples} shows a few examples of such visually detected
cases. We again stress that we list candidates only, and that without further
data, we cannot determine which of these are actual nuclei rather than, for
instance, chance alignments or globular clusters. Figure~\ref{fig:punch_through}
shows two particularly interesting objects, where the disturbed morphology of
the host clearly indicates the recent merger out of which the secondary
nucleus results.

\begin{figure}
    \centering
    \includegraphics[width=\myfigurewidth\linewidth]{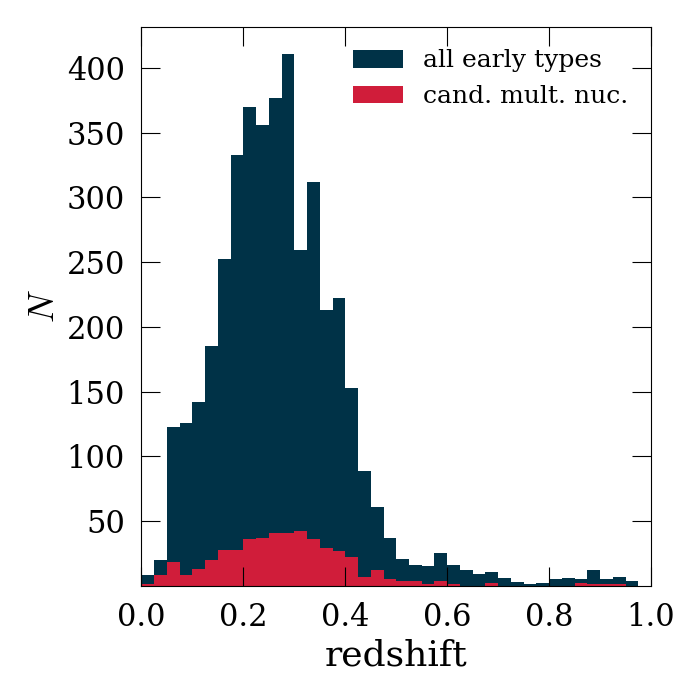}
    \caption{Redshift distributions. The blue histogram shows all early-type
    galaxies. The red histogram shows only hosts of candidate secondary nuclei.
    Where available, we use spectroscopic redshifts. The majority of our
    \numltohpointeight~candidates fall at $z < 0.4$.
    }
    \label{fig:redshift_histogram}
\end{figure}

Table~\ref{tab:catalog} shows an example of the catalogue that
lists all of our visually detected candidates.
The catalogue contains \numcanddnuclei~candidates for secondary nuclei
(these include five objects for which none of our sources listed a redshift).
Figure~\ref{fig:redshift_histogram} shows a histogram of the host
galaxies in redshift space. The catalogue will be made available in
electronic form.

In the context of this work, it is of particular interest to examine
at what physical separations from the host centre we observe possible
secondary nuclei. At separations that fall close to or within typical
core sizes ($\sim 1$\,kpc), we may expect to see an initial onset
of flattening of the central light profile. We measure angular
separations by computing the centre of mass of the residual light
distribution of a secondary nucleus candidate after subtracting the
host galaxy model light. We then translate the observed angular
separations (see Fig.~\ref{fig:angular_separation}) to projected
physical separations, using
spectroscopic redshifts where they are available and photometric
redshifts otherwise.

\begin{figure}
    \centering
    \includegraphics[width=\myfigurewidth\linewidth]{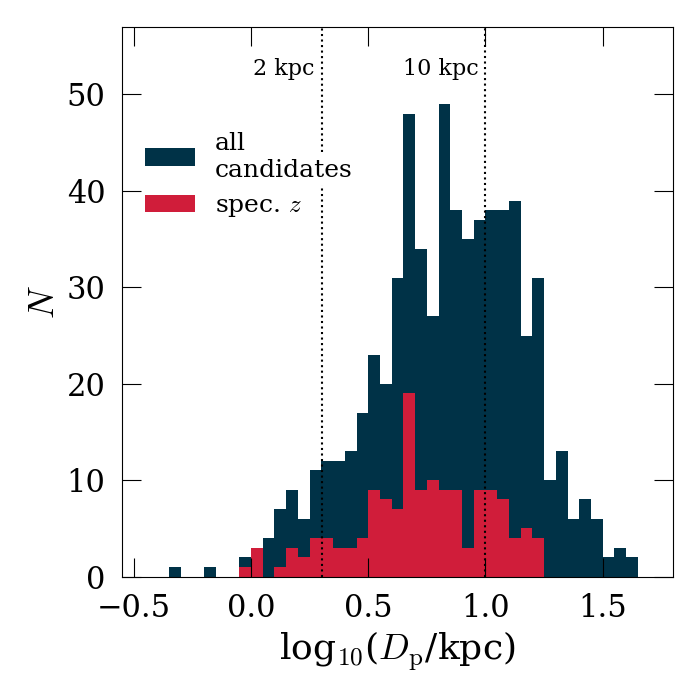}
    \caption{
    Projected physical separations of secondary nucleus
    candidate from host centre.
     }
    \label{fig:physical_separations}
\end{figure}

Figure~\ref{fig:physical_separations} shows the histogram of the projected
physical separations. The peak of the distribution falls around 10\,kpc. The
cut-off towards larger separations is entirely set by our
methodology (image cutout size, requirement to fall within the perceived
footprint of the host, etc.). More interesting is the shape of the distribution
below approximately 5\,kpc, with the smallest value falling around 0.5\,kpc.
A total of \numcandlttwokpc~candidates, of which \numcandlttwokpcwithspecz~have
spectroscopic redshifts, fall at a projected physical separation of less than
2\,kpc, which is comparable to some of the largest observed core sizes in
elliptical galaxies (\citealp{Lauer2007, Rusli2013,Khonji2024}; but see
\citealp{Mehrgan2019} for an extreme case: here, the core break radius $r_{\rm{b}}$
measures some 3~kpc). At separations this small, dynamical effects of the
secondary nucleus are expected to start flattening the core of the host galaxy.

\begin{figure}
    \centering
    \includegraphics[width=\myfigurewidth\linewidth]{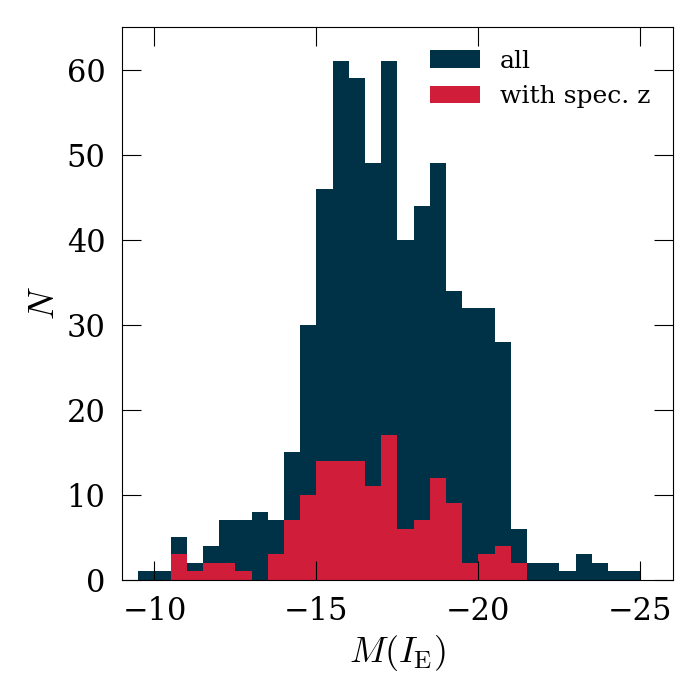}
    \caption{Distribution of absolute magnitudes of our secondary nuclear
    candidates. The faint tail at $M(\IE) \gtrsim -11$
    may be caused by bright globular clusters. The magnitudes have been
    corrected for Galactic extinction.
    }
    \label{fig:cand_photometry}
\end{figure}

Figure~\ref{fig:nuc_cand_observed_mags} shows the range of observed
magnitudes of our secondary nucleus candidates, while
Fig.~\ref{fig:cand_photometry} shows the distribution of the candidate
absolute magnitudes.
The PHZ processing function \citep{Q1-TP005} has computed stellar masses for
MER-detected objects. As a consequence of the errors in redshift, the
PHZ-derived stellar mass estimates are also associated with large uncertainties.
As we argue in Sect.~\ref{sec:chance_alignments}, the lower spatial resolution
in \Euclid's near-infrared channels prevents us from obtaining mass estimates
from fits to the spectral energy distributions.

We employ the following method to compute stellar mass estimates for our sample
of candidates: First, we compute extinction-corrected absolute magnitudes for
all objects in our early-type sample that also have spectroscopic redshifts. We
then further limit this sample to objects for which the PHZ-derived redshift
falls within 5\% of the spectroscopic redshift. For this subsample, we compute
the biweight location \citep{Beers1990} of the ratio of PHZ stellar mass to
luminosity and obtain

\begin{equation}
{\rm{biweight} }\left( \frac{{\rm{log}}_{10} \left(M^{\rm{PHZ}}\,/\rm{M}_{\rm{*}}\right)}{M\left(I_{\rm{E}}^{\mbox{Sérsic}}\right)} \right) = 0.50.
\end{equation}

\begin{sidewaystable*}
\small
    \centering
    \caption{Catalogue of secondary nuclei candidates. A random selection of 20 galaxies from our sample that hold potential secondary (or tertiary) nuclei is shown here as an example. The full catalogue is available in electronic form. 
    (\emph{1}) MER catalog ID of the host,
    (\emph{2}) host right ascension,
    (\emph{3}) host declination,
    (\emph{4}) host redshift,
    (\emph{5}) type of redshift, photometric or spectroscopic,
    (\emph{6}) source of redshift:
NEPs (\citealp{Ortiz2023}, Balzer in prep.), 
ENIA25 \citep{Q1-SP031},
2dFGRS \citep{TwodFGRS},
2dFLenS \citep{2dflens},
6dFGS \citep{6dFGS},
DESI\_DR1 \citep{DESIDR1},
OzDES \citep{OzDES},
PRIMUS \citep{PRIMUS1, PRIMUS2},
PHZ \citep{Q1-TP005},
    (\emph{7}) secondary nucleus candidate identifier.
    (\emph{8}) secondary nucleus candidate right ascension,
    (\emph{9}) secondary nucleus candidate declination.
   (\emph{10}) Kron radius.
    (\emph{11}) Angular separation from host center.
   (\emph{12}) nuclear \IE magnitude. 
    } 
    \begin{tabular}{rrrrllcrrrcc}
   ID$_{\rm MER,host}$          & RA         & Dec         & $z$           & $z_{\mbox{type}}$ & $z_{\mbox{src}}$  & ID$_{\mbox{cand.}}$  & RA$_{\mbox{cand.}}$  & Dec$_{\mbox{cand.}}$  & $s_{\mbox{cand}}$  & $d_{\mbox{p,cand}}$  & \IE \\
                  & $\deg$ & $\deg$ &   & & &  &  $\deg$ & $\deg$ &  arcsec & arcsec & \\      
                        & [J2000] & [J2000] &   & & &  &  [J2000] & [J2000] & & & \\      
   (1)          & (2)         & (3)         & (4)           & (5) & (6)  & (7)  & (8)  & (9)  & (10) & (11) & (12) \\
         \hline
         \hline
$ -519174801271206772$ & $   51.917480$ & $  -27.120677$ & $    0.2364$ & phot       & ENIA25       & $         1$  & $ 51.917621$  & $-27.120206$  & $      1.49$  & $      1.77$  & $     25.19$\\
$ 2715627779684561341$ & $  271.562778$ & $   68.456134$ & $    0.2656$ & phot       & ENIA25       & $         1$  & $271.563754$  & $ 68.456179$  & $      1.55$  & $      1.33$  & $     23.76$\\
                     &              &              &            &            &            &          2  & $271.563608$  & $ 68.456003$  & $      0.19$  & $      1.30$  & $     24.12$\\
                     &              &              &            &            &            &          3  & $271.562475$  & $ 68.455623$  & $      0.31$  & $      2.02$  & $     21.71$\\
$ 2642175996655430565$ & $  264.217600$ & $   65.543057$ & $    0.0562$ & phot       & PHZ        & $         1$  & $264.216547$  & $ 65.543334$  & $      1.08$  & $      1.90$  & $     25.10$\\
$ -519580090290970892$ & $   51.958009$ & $  -29.097089$ & $    0.1777$ & spec       & 2dFGRS     & $         1$  & $ 51.958214$  & $-29.096622$  & $      3.77$  & $      1.85$  & $     20.90$\\
$ 2676693273656463650$ & $  267.669327$ & $   65.646365$ & $    0.0884$ & spec       & DESI\_DR1   & $         1$  & $267.668138$  & $ 65.646415$  & $      4.60$  & $      2.04$  & $     19.38$\\
                     &              &              &            &            &            &          2  & $267.669545$  & $ 65.647207$  & $      0.12$  & $      3.12$  & $     25.77$\\
$ 2733043435663082954$ & $  273.304344$ & $   66.308295$ & $    0.2179$ & spec       & NEP        & $         1$  & $273.303725$  & $ 66.308265$  & $      1.46$  & $      0.93$  & $     21.79$\\
$ -608649072504762971$ & $   60.864907$ & $  -50.476297$ & $    0.3228$ & phot       & ENIA25       & $         1$  & $ 60.865472$  & $-50.476692$  & $      1.22$  & $      1.91$  & $     25.53$\\
$ -609670363469236320$ & $   60.967036$ & $  -46.923632$ & $    0.3252$ & phot       & ENIA25       & $         1$  & $ 60.966307$  & $-46.923550$  & $      1.72$  & $      1.30$  & $     21.46$\\
                     &              &              &            &            &            &          2  & $ 60.968506$  & $-46.924572$  & $      0.24$  & $      5.69$  & $     22.96$\\
                     &              &              &            &            &            &          3  & $ 60.968729$  & $-46.922580$  & $      0.17$  & $      5.70$  & $     25.74$\\
                     &              &              &            &            &            &          4  & $ 60.967934$  & $-46.921987$  & $      0.23$  & $      6.04$  & $     23.50$\\
                     &              &              &            &            &            &          5  & $ 60.969446$  & $-46.923248$  & $      0.15$  & $      6.50$  & $     24.58$\\
                     &              &              &            &            &            &          6  & $ 60.969268$  & $-46.924968$  & $      0.24$  & $      8.04$  & $     25.32$\\
$ 2679144071652882920$ & $  267.914407$ & $   65.288292$ & $    0.0429$ & spec       & NEP        & $         1$  & $267.914277$  & $ 65.287953$  & $      1.29$  & $      1.23$  & $     25.46$\\
$ 2658766329646760654$ & $  265.876633$ & $   64.676065$ & $    0.1712$ & phot       & ENIA25       & $         1$  & $265.876914$  & $ 64.677490$  & $      3.20$  & $      5.52$  & $     21.13$\\
                     &              &              &            &            &            &          2  & $265.880439$  & $ 64.676299$  & $      0.27$  & $      6.79$  & $     22.36$\\
                     &              &              &            &            &            &          3  & $265.877218$  & $ 64.674594$  & $      0.70$  & $      5.35$  & $     19.15$\\
$ 2749253759662933221$ & $  274.925376$ & $   66.293322$ & $    0.0562$ & phot       & PHZ        & $         1$  & $274.924932$  & $ 66.293615$  & $      1.34$  & $      1.25$  & $     22.69$\\
                     &              &              &            &            &            &          2  & $274.924343$  & $ 66.293833$  & $      0.13$  & $      2.38$  & $     26.13$\\
                     &              &              &            &            &            &          3  & $274.923485$  & $ 66.293652$  & $      0.13$  & $      2.97$  & $     26.29$\\
                     &              &              &            &            &            &          4  & $274.926953$  & $ 66.294033$  & $      0.22$  & $      3.47$  & $--$\\
                     &              &              &            &            &            &          5  & $274.923228$  & $ 66.294023$  & $      0.11$  & $      4.00$  & $--$\\
$ 2740201515661233824$ & $  274.020152$ & $   66.123382$ & $    0.4724$ & phot       & ENIA25       & $         1$  & $274.019620$  & $ 66.123275$  & $      0.90$  & $      0.82$  & $     25.89$\\
                     &              &              &            &            &            &          2  & $274.020375$  & $ 66.123301$  & $      0.04$  & $      0.45$  & $     27.21$\\
$ -609670363469236320$ & $   60.967036$ & $  -46.923632$ & $    0.3252$ & phot       & ENIA25       & $         1$  & $ 60.966307$  & $-46.923550$  & $      1.72$  & $      1.30$  & $     21.46$\\
                     &              &              &            &            &            &          2  & $ 60.968506$  & $-46.924572$  & $      0.24$  & $      5.69$  & $     22.96$\\
                     &              &              &            &            &            &          3  & $ 60.968729$  & $-46.922580$  & $      0.17$  & $      5.70$  & $     25.74$\\
                     &              &              &            &            &            &          4  & $ 60.967934$  & $-46.921987$  & $      0.23$  & $      6.04$  & $     23.50$\\
                     &              &              &            &            &            &          5  & $ 60.969446$  & $-46.923248$  & $      0.15$  & $      6.50$  & $     24.58$\\
                     &              &              &            &            &            &          6  & $ 60.969268$  & $-46.924968$  & $      0.24$  & $      8.04$  & $     25.32$\\
$ -627320338487824217$ & $   62.732034$ & $  -48.782422$ & $    0.3077$ & phot       & ENIA25       & $         1$  & $ 62.732398$  & $-48.782407$  & $      2.75$  & $      0.92$  & $     22.31$\\
$ -521607871267149403$ & $   52.160787$ & $  -26.714940$ & $    0.2040$ & phot       & ENIA25       & $         1$  & $ 52.161098$  & $-26.714886$  & $      1.91$  & $      1.09$  & $     23.27$\\

\hline
\hline 
    \end{tabular}
    \label{tab:catalog}
\end{sidewaystable*}

\begin{figure}
    \centering
    \includegraphics[width=\linewidth]{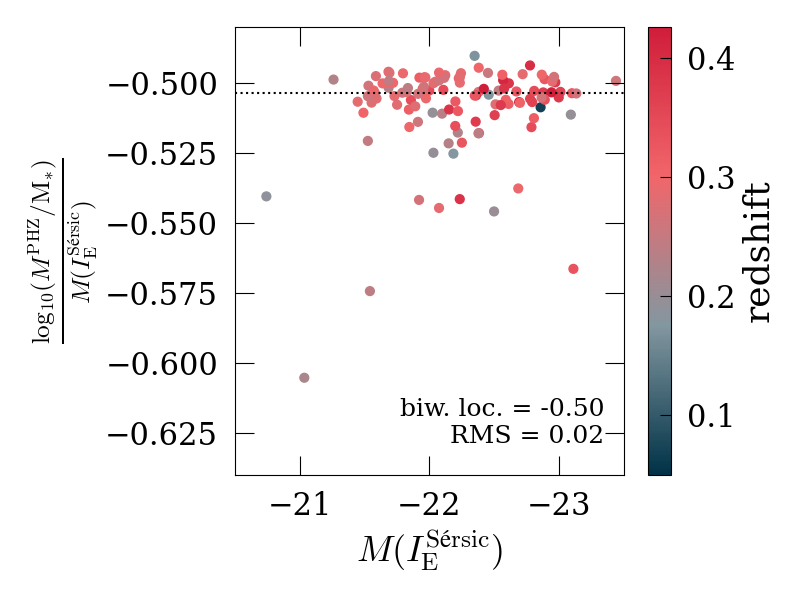}
    \caption{Ratio of stellar mass to (extinction-corrected) \IE
    for all early-type objects in our sample that have a spectroscopic
    redshift.
    }
    \label{fig:mag_to_stellarmass}
\end{figure}

While a few outliers
exist, the core of the distribution is tight, with an RMS scatter
of less than 0.02. We observe no obvious trend with redshift
(see Fig.~\ref{fig:mag_to_stellarmass}).

\begin{figure}
    \centering
    \includegraphics[width=\myfigurewidth\linewidth]{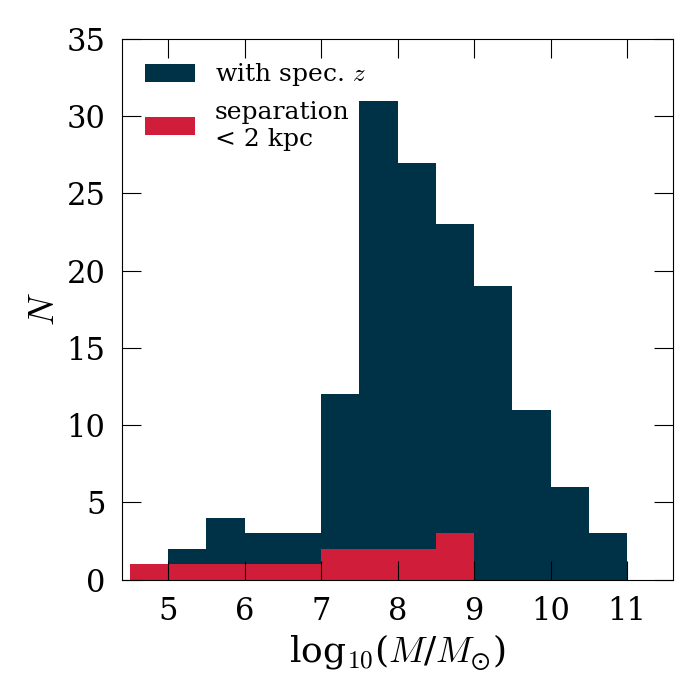}
    \caption{
    Distribution of stellar masses for the secondary nucleus
    candidates in our sample that have spectroscopic redshifts.
    \numcandlttwokpcwithspecz~objects have separations of less than
    2~kpc.
    }
    \label{fig:cand_stellar_mass}
\end{figure}

Figure~\ref{fig:cand_stellar_mass} shows the distribution of candidate stellar
masses for all objects for which we know spectroscopic redshifts for their
hosts. Masses range from $10^{4.1}\,M_\odot$ to $10^{10.6}\,M_\odot$, with a
median of $10^{8.2}\,M_\odot$. If we restrict our sample to those
\numcandlttwokpcwithspecz~systems with projected physical separations of less
than 2\,kpc, the masses range from $10^{4.6}\,M_\odot$ to $10^{8.7}\,M_\odot$,
with a median of $10^{7.2}$\,$M_\odot$.

\section{Discussion}
\label{sec:discussion}
The aim of the presented work is to open the avenue for the detection
of secondary nuclei in future \Euclid data releases. We seek to
establish a basis for the expected number of observed secondary
nuclei and to define the methodology for detecting them.

A number of caveats remain, of course. Visual inspection by a single or a few
experts does not allow us to make statements about completeness. Consequently,
any statement about the frequency of occurrence of secondary nuclei remains
tentative. The nature of this work should therefore be understood as
exploratory.

\subsection{Chance alignments}
\label{sec:chance_alignments}
Furthermore, the fraction of objects that appear superimposed on
galaxy images and in fact are not genuine secondary nuclei is not clearly
known. Ideally, we would have spectroscopic information on the candidates.
The slitless nature of the \Euclid{} spectroscopic channel, however,
makes it very difficult to disentangle the host light contribution
from that of the nucleus. Also, unfortunately, the spatial resolution
of the near-infrared channels is considerably lower than that of VIS.
Our tests showed that virtually all interesting close-separation candidates
only appear as nuclear asymmetries in the near-infrared bands.
Deriving sufficiently accurate photometry, including modelling and
subtraction of the host light, to compute photometric redshifts
for the candidate secondary nuclei is very challenging.

\begin{figure}
    \centering
    \includegraphics[width=\myfigurewidth\linewidth]
    {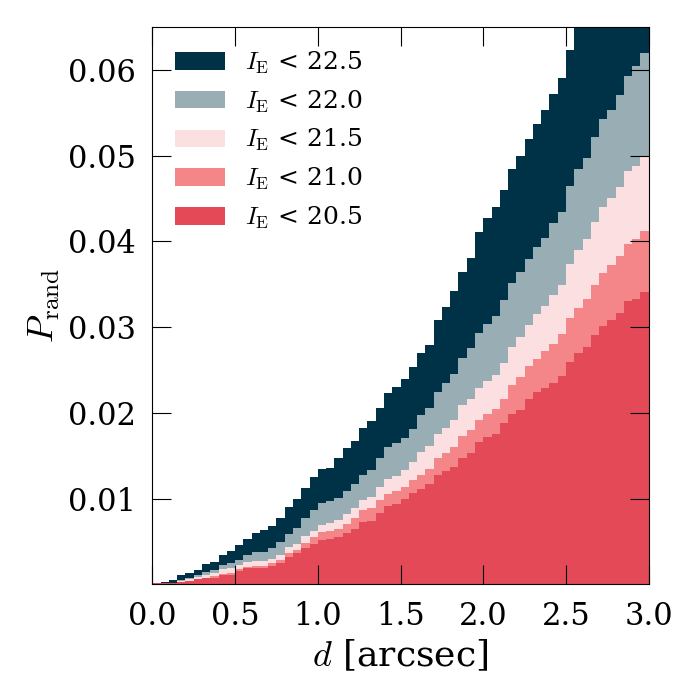}
    \caption{Probability of a secondary light peak being the
    consequence of a chance alignment with a foreground or background
    object, as determined through random sampling in the EDF-N
    region. The different coloured histograms are cumulative;
    that is, they include all objects up to the stated magnitude limit.
    } \label{fig:random_alignment}
\end{figure}

At least regarding potential chance alignments with foreground
or background objects, we can gain some insight from the data
themselves. To this end, we randomly and uniformly select 10\,000
locations on the sky that fall within the EDF-N.
At each location we then ask: What is the chance of finding a source
of at least a given magnitude within a certain radius? This is shown
in Fig.~\ref{fig:random_alignment}. Even in the faintest magnitude
bin, we find that within one arcsecond of angular separation, the
probability of finding an object is less than one percent. At a two
arcsecond separation, this number increases to about three percent.

\begin{figure}
    \centering
    \includegraphics[width=\myfigurewidth\linewidth]{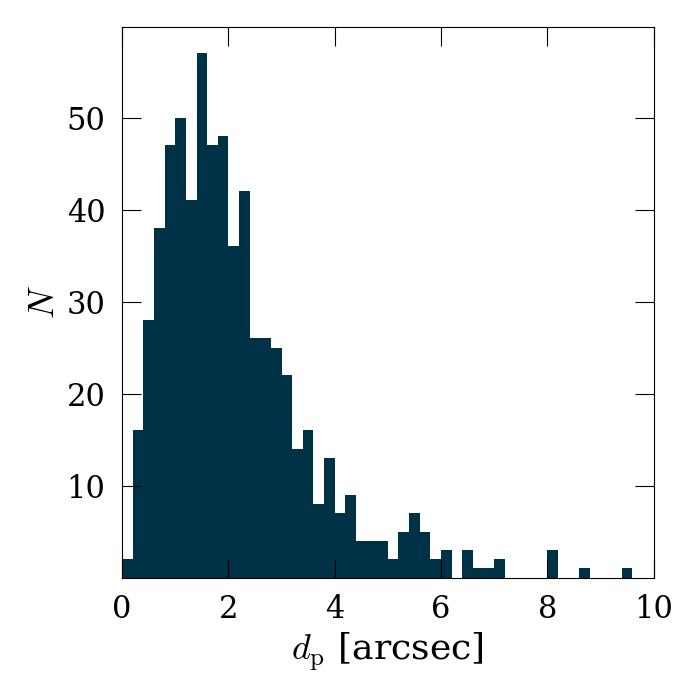}
    \caption{Distribution of angular separations.
    We estimate angular separations of the secondary light peak centre
    from the centre of the host as determined by the Sérsic model fit.
    The centres of the candidate nuclei are estimated using a simple
    centre-of-mass approach within a manually placed aperture on the
    residual image.}
    \label{fig:angular_separation}
\end{figure}

The bulk of our candidate hosts have redshifts of 0.4 or less. At this redshift,
an angular separation of \ang{;;2} -- and the majority of our secondary our candidates
is separated by less than \mbox{\ang{;;0.5}} (see Fig.~\ref{fig:angular_separation}) --
translates to a projected physical separation of approximately 11\,kpc, a range
at which tidal effects during a merger become significant. A separation of
\ang{;;0.36} corresponds to 2\,kpc which is of particular interest in the
context of this work, as explained above.

\begin{figure}
    \centering
    \includegraphics[width=\myfigurewidth\linewidth]{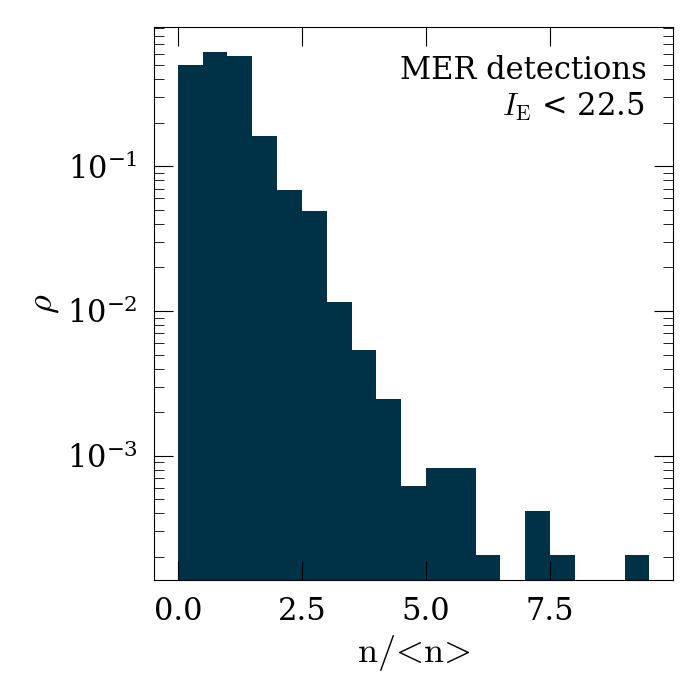}
    \caption{Normalized density distribution of spatial overdensities (over the mean density) of the MER catalogue
    for $\IE < 22.5$.
    } \label{fig:mer_overdensities}
\end{figure}

The random sampling of field locations would only yield chance-alignment
probabilities in the case of a perfectly homogeneous distribution of MER
detections. To estimate how much the probability of a chance alignment can be
boosted in regions of high cosmic density, we place 1\,arcminute-wide circular
apertures at 10\,000 random locations within the EDF-N and count the
number of
MER detections with $\IE < 22.5$ within these apertures. We then plot the
distribution of counts normalised to the mean of all apertures
(see Fig.~\ref{fig:mer_overdensities}). This indicates
that the frequency of chance alignments can be boosted by up to a factor of
approximately seven in very rare cases (probability $< 0.1\%$). We thus conclude
that the majority of closely separated secondary objects will not be due to
chance alignment.

\subsection{Nature of secondary nuclei}
\begin{figure}
    \centering
    \includegraphics[width=\myfigurewidth\linewidth]{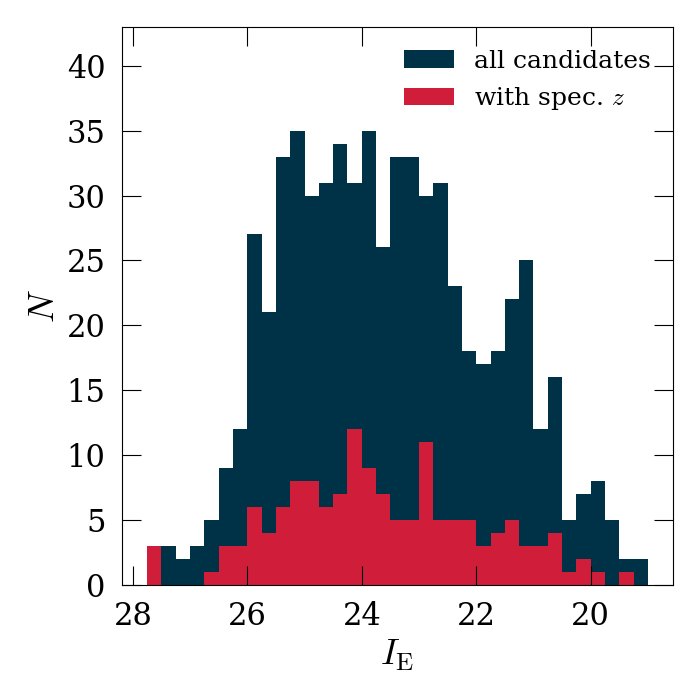}
    \caption{Distribution of observed magnitudes of the
    secondary nucleus candidates in our sample. No extinction correction
    has been applied here.
    }
    \label{fig:nuc_cand_observed_mags}
\end{figure}
\begin{figure}
    \centering
    \includegraphics[width=\myfigurewidth\linewidth]{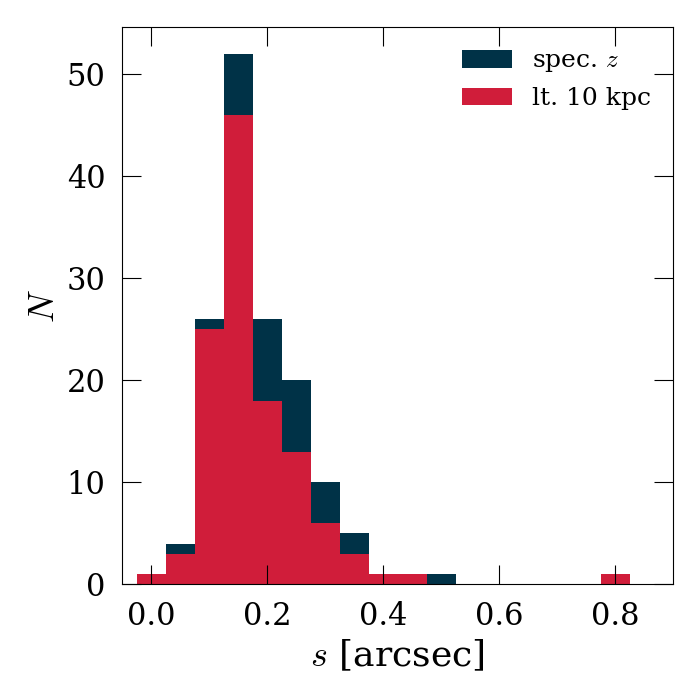}
    \caption{Angular Kron radii, $s$, for the candidate nuclei. The blue histogram
    shows all candidates with spectroscopic redshifts and the red histogram
    shows only those candidates that are closer than 10~kpc to the host galaxy
    centre. }
    \label{fig:angular_sizes}
\end{figure}

\begin{figure}
    \centering
    \includegraphics[width=\myfigurewidth\linewidth]{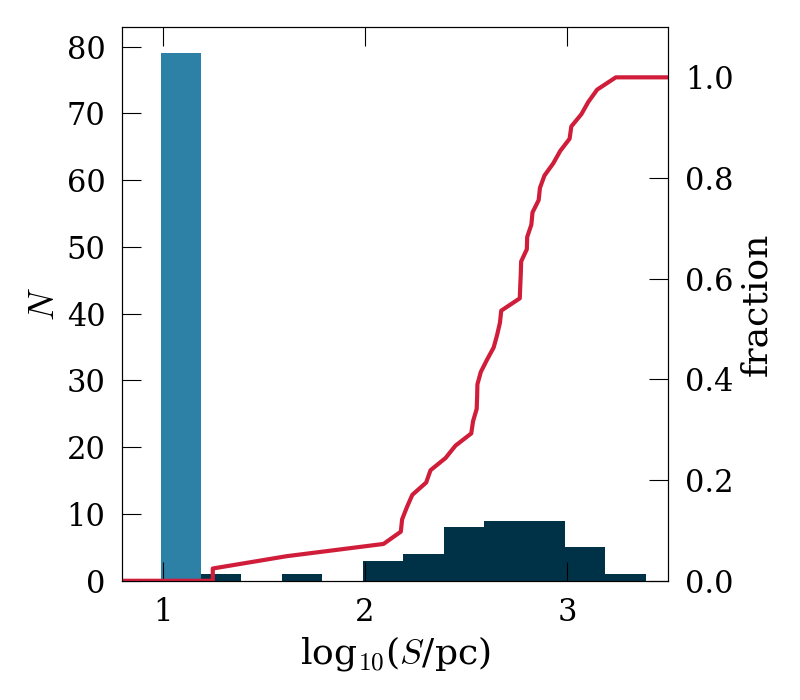}
    \caption{Histogram of physical sizes (Kron radii, $S$). Unresolved objects are
    accounted for in the light blue bar. The red curve shows
    the normalized cumulative distribution for all resolved object
    with projected physical separations of less than \SI{10}{\kilo\parsec}.}
    \label{fig:physical_sizes}
\end{figure}

For the following, we will focus on the group of candidates that have
spectroscopic redshifts, to enable us to derive precise estimates of projected
physical dimensions.
Now that we know that many of our close separation candidates are not the
consequence of a chance alignment, it is instructive to have a look at their
structural parameters. Their small sizes prevent us from fitting them with
Sérsic models, and we therefore resort to computing Kron radii using the data within
the previously drawn apertures, on the residual images. In
Fig.~\ref{fig:angular_sizes}, we show the angular sizes that we obtain this way.
The Kron radii were derived from the residual images, ignoring the convolution of
the true images by the VIS PSF. To first order we corrected for this making
use of Gaussian deconvolution, that is by subtracting in
quadrature \ang{;;0.18} (VIS spatial resolution; \citealp{EuclidSkyVIS}).
The Kron radius measurement yields a value of less than \ang{;;0.18} in 72
cases. We regard these objects as unresolved by \Euclid. In
Fig.~\ref{fig:physical_sizes} we translated the Kron radii to physical sizes
using our best available redshift.
The distribution of sizes stretches from 60\,pc to 1750\,pc with a median of
450\,pc, not accounting for the unresolved 72 objects.

Kron radii are not effective radii, however. By generating S\'ersic models with
different indices $n$ we find that Kron radii are equal to the effective
radius for a circular Gaussian light distribution ($n = 0.5$) and 2.5 times
larger for a S\'ersic light distribution with $n = 5$
(\citealp{GrahamDriver2005} and Appendix~\ref{sec:kron_to_re}).

Of our resolved objects, 90\% have Kron radii of less than 1\,kpc,
which is well within the regime of dwarf galaxies (e.g.~ \citealp{KFCB2009,
Zoeller2024, EROPerseusDGs, Q1-SP001}), even more so when the just mentioned
caveat about the Kron radius to effective radius relation is taken into account.

\begin{figure}
    \centering
    \includegraphics[width=\myfigurewidth\linewidth]{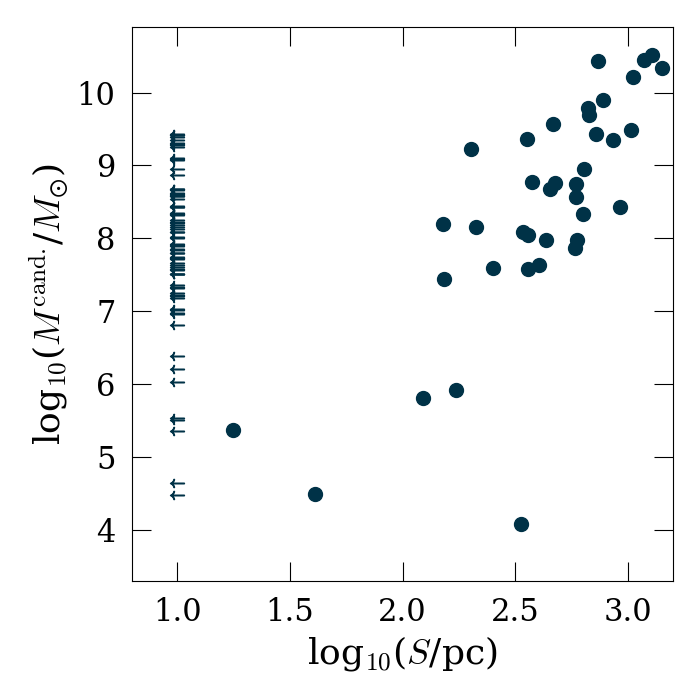}
    \caption{Stellar mass to physical size relation for the candidate nuclei We
    only show candidate nuclei with projected physical separation from the host
    centre of less than 10~kpc and available spectroscopic redshifts. The arrows
    indicate unresolved candidate nuclei.}
    \label{fig:mass_size_relation}
\end{figure}

In Fig.~\ref{fig:mass_size_relation}, we compare the physical sizes to
the stellar mass estimates. These appear roughly correlated. A simple
linear fit yields
\begin{equation*}
\logten\paren{M/{\rm M}_*} = 3.35\,\logten\paren{S/\text{pc}}\,,
\end{equation*}
albeit with large scatter.

\begin{figure}
    \centering
    \includegraphics[width=\myfigurewidth\linewidth]{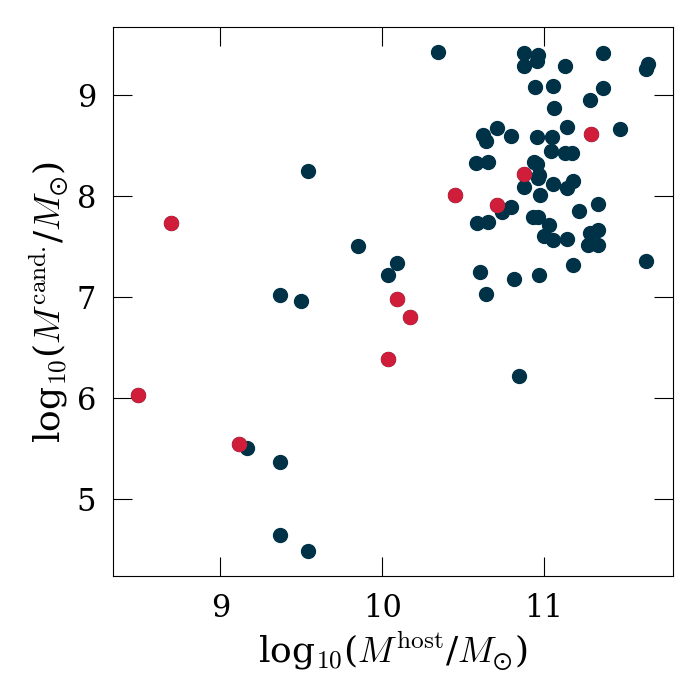}
    \caption{Stellar mass to host stellar mass relation
    for the unresolved secondary nuclei
    candidates.
    The red points show candidates that lie
    at projected physical separations of less
    than \SI{2}{\kilo\parsec}.}
    \label{fig:mass_hostmass_relation}
\end{figure}

Finally, in Fig.~\ref{fig:mass_hostmass_relation} we plot the candidate mass
estimates versus the host mass estimates of all unresolved sources.
Both the stellar masses of the nuclei candidates and the hosts
are computed using our $M(\IE)$ to stellar mass conversion. In
the case of the nuclei, we use our own photometry; in the case of the
hosts we use the VIS Sérsic fitting photometry after the application of
the extinction correction.
One
can see that a majority of the candidates at masses larger than $10^7M_{\odot}$
occur in hosts with masses larger than $10^{10.5}M_{\odot}$. Only a few of the candidate
masses of these unresolved candidates are compatible with the typical stellar
mass range of globular clusters in ETGs. In fact, only two
objects fall below $2 \times 10^{5} M_\odot$, which is the turnover mass of the
globular cluster stellar mass function (e.g.~\citealp{Jordan2007}) in ETGs.
Thus, a good number of these objects may indeed result from a prior
merger: the central, highest stellar density cusp of the lower mass progenitor
now sinking towards the centre of the newly formed system.
Two of these have masses around $10^8 M_{\odot}$
and are found in hosts with stellar masses of about $10^{11} M_{\odot}$
and are therefore good candidates to become binary SMBH, generating a cored light profile, and a tangentially anisotropic orbital structure.

Not previously considered in this paper is the
possibility that some of our secondary nuclei may result from a SMBH
recoil event following the coalescence of SMBH binaries.
During the final stages of SMBH mergers, anisotropic gravitational wave emission
carries away linear momentum (e.g. \citealp{Gonzalez2007a}), resulting in a
recoil of the remnant SMBH in the opposite direction  with some velocity $v_\mathrm{kick}$ (e.g.
\citealp{Bekenstein1973}). Recoiling SMBHs
are expected to carry a stellar envelope with them
\citep{Komossa2008a,Merritt2009,Rawlings2025} and may thus be observable as
offset stellar nuclei which have indeed been detected in HST observations
(e.g.\ \citealp{Turner2012}).
Observational signatures of SMBH recoils include velocity
offsets between the SMBH and the galaxy nucleus, with candidate
systems identified at both moderate (\SI{\sim 100}{\kms};
\citealp{Kim2017}) and extreme velocities (\SI{>2600}{\kms}, e.g.
\citealp{Komossa2008}). Recoiling SMBHs may also be identified
through a spatial offset from the host centre, where the offset can
range from several parsecs \citep{Batcheldor2010, Lena2014,
Barrows2016} to more than a kiloparsec \citep{Koss2014, SkipperBrowne2018}.

In our analysis of \Euclid Q1 data, it is therefore plausible that
in a subset of galaxies exhibiting multiple or offset nuclei we may be
witnessing the aftermath of such SMBH recoil events.
High spatial resolution,
spectroscopic follow-up observations with confirmed velocity offsets
on the order of \SI{400}{\kms} would be required to establish
these as viable recoil candidates.
The potential detection of recoiling SMBHs offers a compelling
observational window into the final stages of galaxy mergers and
the dynamics of black hole coalescence.

\subsection{Towards DR1 and automated detections}
\label{sec:towards_dr1_automated_detections}
\Euclid DR1 will cover a sky area of approximately 1900\,$\rm{deg}^2$. It is reasonable to assume that we can scale our number
of candidates by the increase in sky coverage. From this, we estimate
that more than 10\,000 secondary objects will be detected at projected
physical separations of less than 10\,kpc. We also expect to detect
about 2000 secondary nuclei at projected separations of less than
2\,kpc. These numbers are substantial. With DR1, we will be able
to divide the future sample by redshift and/or environment and assess
whether the occurrence of secondary nuclei is consistent with our
picture of core formation.

Visual inspections, while straightforward to conduct, have obvious
limitations in terms of objectivity and scalability. While inspecting
a sample of approximately 15\,000 VIS images is feasible for a
single person over the course of a few weeks, this approach will
become impractical for the 30 times larger dataset expected from
DR1.

We plan to use the sample presented herein, including both the
candidates for secondary nuclei and the early-type galaxy sample
(objects with smooth light distributions), as a testbed for developing
and training automated detection routines.
Today, convolutional neural networks (CNNs) are readily available and have been
successfully applied to various problems such as the automatic detection of
gravitational strong lenses \citep{Q1-SP053, Q1-SP048, Q1-SP054, Q1-SP059,
Q1-SP013} and the morphological classification of galaxies \citep{Q1-SP047} in
\Euclid's Q1.

As a test case, we construct a generic sequential CNN composed of three
convolutional blocks followed by a fully connected classifier. The input layer
has a size of $50\,{\rm pixels}\times 50\,{\rm pixels}$, and we present only the
central $50\,{\rm pixels}\times 50\,{\rm pixels}$ of each VIS cutout to the
network. Each block consists of a convolutional layer with rectified linear unit
(ReLU) activation and a $2 \times 2$ max-pooling layer. The convolutional layers
use $3 \times 3$ kernels with increasing filter counts (32, 64, 128). After
feature extraction, the output is flattened and passed through a dense layer
with 128 units and ReLU activation, followed by a dropout layer (30\%). The
final output layer uses a sigmoid activation function for binary classification.

We use 20\% of our complete sample (including all non-multiple-nucleated
early-type galaxies) as a test sample during training. After 50
epochs, the model achieves an accuracy of 98.8\%.
Using this CNN, we estimate the completeness that we might achieve when applying it
to \Euclid DR1. For this purpose, we generate a series of mock images by
implanting a secondary nucleus into one of the non-multiple early-type VIS
images from this work. The secondary nucleus is modelled as a Gaussian with a
standard deviation of 1\,pixel (\ang{;;0.1}, i.e. unresolved). We convolve the
mock nucleus with the VIS PSF before adding it to the VIS
image. We generate these mock images for a number of angular separations from
the centre, ranging from \ang{;;0.1} to \ang{;;1.5}, and for a range of
observed brightnesses from $\IE = 24.1$ to $\IE = 20.5$, resulting in 435
distinct realisations. We randomly select 500 of the
\numearlytypes~non-multiple-nucleated early-type input images and insert into
each of those, the 435 mock nuclei one after the other generating 217\,500 mock
images. The CNN is then applied to classify this set.

\begin{figure}
    \centering
    \includegraphics[width=1.\linewidth]{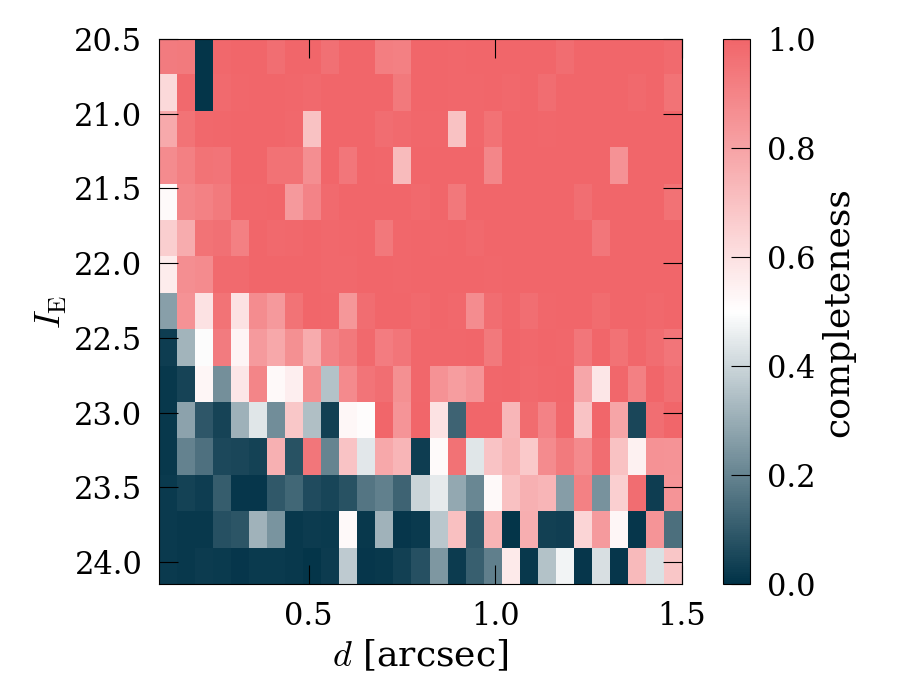}
    \caption{
    Completeness as a function of separation and magnitude using a
    convolutional neural network approach.
    }
    \label{fig:cnn_completeness}
\end{figure}

Figure~\ref{fig:cnn_completeness} shows the result of this exercise.
This two-dimensional histogram displays the fraction of successfully
detected secondary nuclei at each combination of input brightness
of the mock nucleus and its angular separation from the host centre.
At a brightness of $\IE = 23$ and separations greater than
\ang{;;0.2}, the CNN achieves approximately 50\% completeness.
Of course at larger relative brightness the detection probability
generally increases.

A comparison with Fig.~\ref{fig:nuc_cand_observed_mags} indicates that the CNN
still does not detect the full range of brightnesses that appear accessible via
visual inspection; thus, there is certainly room for improvement. Nevertheless,
CNNs may provide a viable approach to studying the occurrence of bright
secondary nuclei across the hundreds of thousands of early-type galaxies
expected in DR1.

Towards DR1, we plan to further improve our detection methodology through the
use of a more optimal interpolation scheme during the stacking of the individual
VIS frames. The VIS stacked frames provided by the MER processing function are
currently generated using bilinear interpolation. Initial tests indicate that
this degrades the spatial resolution compared to a Lanczos interpolation scheme
\citep{Duchon1979, Bertin1996}: We fit bivariate Gaussian models to stars in a
single Q1 MER VIS stacked frame. We then rerun the MER stacking pipeline on the
same tile using a Lanczos interpolation. We find that the latter yields about
20\% smaller full width half maxima of the Gaussian models potentially enabling
the detection of secondary nuclei at even smaller separations.

\section{Conclusions}
\label{sec:conclusions}
We conduct an exploratory investigation into the incidence and
properties of multiple nuclei in ETGs, using
the high-resolution VIS imaging data from the \Euclid Q1 data
release. Starting from the MER detection catalogue, we employ a
systematic selection procedure which combines photometric filtering,
stellar contamination rejection using \textit{Gaia} DR3, and
structural modelling of stellar light profiles. The resulting sample
enables us to visually identify and characterise sub-galactic central
structures suggestive of secondary nuclei.

Our methodology involves a two-stage modelling approach: an initial
multi-Gaussian expansion to describe the global light
distribution, followed by refined Sérsic profile fits using
\texttt{imfit}, incorporating position-dependent VIS PSFs.
Residual images generated from these fits are effective
in highlighting features that are inconsistent with the
smooth Sérsic host profile.

Through careful visual inspection of \numearlytypes~ETGs, we classify a
significant subset of \numetwithdnuclei~as hosting plausible candidate secondary
nuclei. Out of these, \numhavemultiple~systems host multiple secondary nuclei candidate. We list a total
of \numcanddnuclei~candidates.
The majority of our detections fall at projected physical separations of more than
\SI{2}{\kilo\parsec} (93\% of all) and up to \SI{15}{\kilo\parsec} (85\%). We consider these to be normal merger candidates. The dynamical effects on the host will be
significant at this stage, however they have not reached a stage at which an
inset of core flattening will occur.
In our sample \numcandlttwokpc~fall at physical
separations of less than \SI{2}{\kilo\parsec}.
We argue that for candidates with very
small angular separations from the host centre, a chance alignment
with foreground or background sources is unlikely.

We measure an aperture-based photometry for the objects in our catalogue. For
objects that have spectroscopic redshift determinations, we find that the range
of absolute magnitudes stretches from $M(\IE) = -8.1$ to $M(\IE) = -21.1$. This
indicates that, at the bright end, we are picking up other galaxies (mergers),
while at the faint end the measured values are compatible with bright globular
clusters. Using a Kron radius as a proxy, we compute physical sizes and find a
typical value of about \SI{600}{\parsec} for the objects that are resolved at
the VIS spatial resolution, while 72 objects are unresolved. Of particular
interest are objects of close physical separation to the host centre. Focusing
on all objects at projected separations of less than \SI{10}{\kilo\parsec}, we
see that the vast majority, 90\%, have Kron radii of less than
\SI{1}{\kilo\parsec}. Using a simple empirical translation from absolute
magnitude to stellar mass, we also compute stellar masses for the candidates,
and find that, for this subset, masses range from
$10^{4.1} M_\odot$ to $10^{10.6} M_\odot$. Restricting this to the spatially
resolved objects, sizes stretch from \SI{62}{\parsec} to \SI{1750}{\parsec}.
Both in terms of size and stellar mass, these are compatible with the masses and
effective radii that are found for dwarf galaxies in the local universe.

Within the sample of unresolved objects, only two have masses of less than
$10^{5} M_\odot$, the turnover mass of typical globular cluster
populations in early-type galaxies. These close-separation and with up to
$10^{9.4} M_\odot$ relatively massive but compact objects, are good candidates
for the remnants of mergers: the largest stellar density centres of the
progenitors, likely hosting an SMBH, that are sinking into the newly formed
central regions of the gravitational potentials. In fact, their masses fall into the
range of UCD galaxies, that are also suspected to be merger remnants, whilst at
larger radial offsets. Restricting the sample further to physical separations of
\SI{2}{\kilo\parsec}, which puts them close to, or within, the range of core radii
in massive elliptical galaxies, we have 14 galaxies with spectroscopic redshifts.
Some of these are our most interesting candidates to develop cored light profiles and
tangential anisotropic structures, or even to host a SMBH recoil event.

Looking ahead, the methodology developed here establishes a framework
for extending this analysis to the full \Euclid DR1. We demonstrate
that convolutional neural networks (CNNs) offer a viable approach
for detecting secondary nuclei in the 30 times larger DR1 dataset.
The anticipated increase in sample size and depth will enhance the
statistical significance and more robustly constrains the
frequency, morphology, and physical origins of multiple nuclei in
massive galaxies.
Our findings underscore the power of \Euclid's high spatial resolution
VIS imaging over large sky areas in probing the complex assembly
history of galaxies and highlight its potential for revealing rare
structural features indicative of past interaction events.

\begin{acknowledgements}
\AckEC
\AckQone
MF, RS and RB acknowledge support by the Deutsches Zentrum fur Luft- und
Raumfahrt (DLR) grant 50 QE 1101.  This research makes use of ESA Datalabs
(\url{datalabs.esa.int}), an initiative by ESA’s Data Science and Archives
Division in the Science and Operations Department, Directorate of Science. This
work has made use of data from the European Space Agency (ESA) mission
\textit{Gaia} (\url{https://www.cosmos.esa.int/gaia}), processed by the
\textit{Gaia} Data Processing and Analysis Consortium (DPAC,
\url{https://www.cosmos.esa.int/web/gaia/dpac/consortium}). Funding for the DPAC
has been provided by national institutions, in particular the institutions
participating in the \textit{Gaia} Multilateral Agreement. Based on observations
obtained with the Hobby--Eberly Telescope (HET), which is a joint project of the
University of Texas at Austin, the Pennsylvania State University,
Ludwig-Maximillians-Universit\"at M\"unchen, and Georg-August Universit\"at
G\"ottingen. The HET is named in honour of its principal benefactors, William P.
Hobby and Robert E. Eberly. This research has made use of the Astrophysics Data
System, funded by NASA under Cooperative Agreement 80NSSC21M00561.
\end{acknowledgements}

\section*{Acknowledgement of AI Tool Usage }
The text in this manuscript has been proofread, checked for grammar and
consistency with the \Euclid style guide using OpenAI’s ChatGPT (version GPT-4,
accessed May 2025).  ChatGPT has also been used for prototyping various sections
of analysis code. AI tools, specifically Google's Tensorflow have been used to
generate CNN models for testing automated detection schemes.
scienceOS\footnote{https://app.scienceos.ai/} has been used for literature
discovery and addition to the usual databases like Astrophysics Data
System\footnote{https://ui.adsabs.harvard.edu/}. The authors take full
responsibility of content, data interpretation, and conclusions.

\bibliography{dnuclei, Euclid, ext_specz, Q1}

\begin{thebibliography}{102}
\expandafter\ifx\csname natexlab\endcsname\relax\def\natexlab#1{#1}\fi

\bibitem[{{Agazie} {et~al.}(2023){Agazie}, {Anumarlapudi}, {Archibald},
  {Baker}, {B{\'e}csy}, {Blecha}, {Bonilla}, {Brazier}, {Brook},
  {Burke-Spolaor}, {Burnette}, {Case}, {Casey-Clyde}, {Charisi}, {Chatterjee},
  {Chatziioannou}, {Cheeseboro}, {Chen}, {Cohen}, {Cordes}, {Cornish},
  {Crawford}, {Cromartie}, {Crowter}, {Cutler}, {D'Orazio}, {Decesar}, {Degan},
  {Demorest}, {Deng}, {Dolch}, {Drachler}, {Ferrara}, {Fiore}, {Fonseca},
  {Freedman}, {Gardiner}, {Garver-Daniels}, {Gentile}, {Gersbach}, {Glaser},
  {Good}, {G{\"u}ltekin}, {Hazboun}, {Hourihane}, {Islo}, {Jennings},
  {Johnson}, {Jones}, {Kaiser}, {Kaplan}, {Kelley}, {Kerr}, {Key}, {Laal},
  {Lam}, {Lamb}, {Lazio}, {Lewandowska}, {Littenberg}, {Liu}, {Luo}, {Lynch},
  {Ma}, {Madison}, {McEwen}, {McKee}, {McLaughlin}, {McMann}, {Meyers},
  {Meyers}, {Mingarelli}, {Mitridate}, {Natarajan}, {Ng}, {Nice}, {Ocker},
  {Olum}, {Pennucci}, {Perera}, {Petrov}, {Pol}, {Radovan}, {Ransom}, {Ray},
  {Romano}, {Runnoe}, {Sardesai}, {Schmiedekamp}, {Schmiedekamp}, {Schmitz},
  {Schult}, {Shapiro-Albert}, {Siemens}, {Simon}, {Siwek}, {Stairs},
  {Stinebring}, {Stovall}, {Sun}, {Susobhanan}, {Swiggum}, {Taylor}, {Taylor},
  {Turner}, {Unal}, {Vallisneri}, {Vigeland}, {Wachter}, {Wahl}, {Wang},
  {Witt}, {Wright}, {Young}, \& {Nanograv Collaboration}}]{Agazie2023}
{Agazie}, G., {Anumarlapudi}, A., {Archibald}, A.~M., {et~al.} 2023, \apjl,
  952, L37

\bibitem[{{Babusiaux} {et~al.}(2023){Babusiaux}, {Fabricius}, {Khanna},
  {Muraveva}, {Reyl{\'e}}, {Spoto}, {Vallenari}, {Luri}, {Arenou},
  {{\'A}lvarez}, {Anders}, {Antoja}, {Balbinot}, {Barache}, {Bauchet},
  {Bossini}, {Busonero}, {Cantat-Gaudin}, {Carrasco}, {Dafonte}, {Diakit{\'e}},
  {Figueras}, {Garcia-Gutierrez}, {Garofalo}, {Helmi}, {Jim{\'e}nez-Arranz},
  {Jordi}, {Kervella}, {Kostrzewa-Rutkowska}, {Leclerc}, {Licata}, {Manteiga},
  {Masip}, {Mongui{\'o}}, {Ramos}, {Robichon}, {Robin}, {Romero-G{\'o}mez},
  {S{\'a}ez}, {Santove{\\textasciitilde n}a}, {Spina}, {Torralba Elipe}, \&
  {Weiler}}]{BabusiauxFabricius2023}
{Babusiaux}, C., {Fabricius}, C., {Khanna}, S., {et~al.} 2023, \aap, 674, A32

\bibitem[{Barrows {et~al.}(2016)Barrows, Comerford, Greene, \&
  Pooley}]{Barrows2016}
Barrows, R.~S., Comerford, J.~M., Greene, J.~E., \& Pooley, D. 2016, ApJ, 829,
  37

\bibitem[{Batcheldor {et~al.}(2010)Batcheldor, Robinson, Axon, Perlman, \&
  Merritt}]{Batcheldor2010}
Batcheldor, D., Robinson, A., Axon, D.~J., Perlman, E.~S., \& Merritt, D. 2010,
  ApJL, 717, L6

\bibitem[{Beers {et~al.}(1990)Beers, Flynn, \& Gebhardt}]{Beers1990}
Beers, T.~C., Flynn, K., \& Gebhardt, K. 1990, AJ, 100, 32

\bibitem[{{Bekenstein}(1973)}]{Bekenstein1973}
{Bekenstein}, J.~D. 1973, \apj, 183, 657

\bibitem[{{Bender}(1988)}]{Bender1988}
{Bender}, R. 1988, \aap, 202, L5

\bibitem[{{Bender} \& {Moellenhoff}(1987)}]{BenderMoellenhoff1987}
{Bender}, R. \& {Moellenhoff}, C. 1987, \aap, 177, 71

\bibitem[{{Bertin} \& {Arnouts}(1996)}]{Bertin1996}
{Bertin}, E. \& {Arnouts}, S. 1996, \aaps, 117, 393

\bibitem[{{Bhattacharya} {et~al.}(2023){Bhattacharya}, {Nehal}, {Das},
  {Paswan}, {Saha}, \& {Combes}}]{Bhattacharya2023}
{Bhattacharya}, A., {Nehal}, C.~P., {Das}, M., {et~al.} 2023, \mnras, 524, 4482

\bibitem[{{Blake} {et~al.}(2016){Blake}, {Amon}, {Childress}, {Erben},
  {Glazebrook}, {Harnois-Deraps}, {Heymans}, {Hildebrandt}, {Hinton},
  {Janssens}, {Johnson}, {Joudaki}, {Klaes}, {Kuijken}, {Lidman}, {Marin},
  {Parkinson}, {Poole}, \& {Wolf}}]{2dflens}
{Blake}, C., {Amon}, A., {Childress}, M., {et~al.} 2016, \mnras, 462, 4240

\bibitem[{{Capetti} {et~al.}(2005){Capetti}, {Verdoes Kleijn}, \&
  {Chiaberge}}]{Capetti2005}
{Capetti}, A., {Verdoes Kleijn}, G., \& {Chiaberge}, M. 2005, \aap, 439, 935

\bibitem[{{Cappellari}(2002)}]{Cappellari2002}
{Cappellari}, M. 2002, \mnras, 333, 400

\bibitem[{{Carollo} {et~al.}(1997){Carollo}, {Franx}, {Illingworth}, \&
  {Forbes}}]{Carollo1997}
{Carollo}, C.~M., {Franx}, M., {Illingworth}, G.~D., \& {Forbes}, D.~A. 1997,
  \apj, 481, 710

\bibitem[{{Ch{\'a}vez Ortiz} {et~al.}(2023){Ch{\'a}vez Ortiz}, {Finkelstein},
  {Davis}, {Leung}, {Mentuch Cooper}, {Bagley}, {Larson}, {Casey}, {McCarron},
  {Gebhardt}, {Guo}, {Liu}, {Laseter}, {Rhodes}, {Bender}, {Fabricius},
  {S{\'a}nchez}, {Scarlata}, {Capak}, {Zalesky}, {Sanders}, {Szapudi},
  {Baxter}, {McPartland}, {Weaver}, {Toft}, {Mobasher}, {Suzuki}, \&
  {Chartab}}]{Ortiz2023}
{Ch{\'a}vez Ortiz}, {\'O}.~A., {Finkelstein}, S.~L., {Davis}, D., {et~al.}
  2023, \apj, 952, 110

\bibitem[{{Coil} {et~al.}(2011){Coil}, {Blanton}, {Burles}, {Cool},
  {Eisenstein}, {Moustakas}, {Wong}, {Zhu}, {Aird}, {Bernstein}, {Bolton}, \&
  {Hogg}}]{PRIMUS1}
{Coil}, A.~L., {Blanton}, M.~R., {Burles}, S.~M., {et~al.} 2011, \apj, 741, 8

\bibitem[{{Colless} {et~al.}(2003){Colless}, {Peterson}, {Jackson}, {Peacock},
  {Cole}, {Norberg}, {Baldry}, {Baugh}, {Bland-Hawthorn}, {Bridges}, {Cannon},
  {Collins}, {Couch}, {Cross}, {Dalton}, {De Propris}, {Driver}, {Efstathiou},
  {Ellis}, {Frenk}, {Glazebrook}, {Lahav}, {Lewis}, {Lumsden}, {Maddox},
  {Madgwick}, {Sutherland}, \& {Taylor}}]{TwodFGRS}
{Colless}, M., {Peterson}, B.~A., {Jackson}, C., {et~al.} 2003,
  arXiv:astro-ph/0306581

\bibitem[{{Cool} {et~al.}(2013){Cool}, {Moustakas}, {Blanton}, {Burles},
  {Coil}, {Eisenstein}, {Wong}, {Zhu}, {Aird}, {Bernstein}, {Bolton}, {Hogg},
  \& {Mendez}}]{PRIMUS2}
{Cool}, R.~J., {Moustakas}, J., {Blanton}, M.~R., {et~al.} 2013, \apj, 767, 118

\bibitem[{{Cropper} {et~al.}(2016){Cropper}, {Pottinger}, {Niemi}, {Azzollini},
  {Denniston}, {Szafraniec}, {Awan}, {Mellier}, {Berthe}, {Martignac}, {Cara},
  {Di Giorgio}, {Sciortino}, {Bozzo}, {Genolet}, {Cole}, {Philippon}, {Hailey},
  {Hunt}, {Swindells}, {Holland}, {Gow}, {Murray}, {Hall}, {Skottfelt},
  {Amiaux}, {Laureijs}, {Racca}, {Salvignol}, {Short}, {Lorenzo Alvarez},
  {Kitching}, {Hoekstra}, {Massey}, \& {Israel}}]{Cropper16}
{Cropper}, M., {Pottinger}, S., {Niemi}, S., {et~al.} 2016, in Society of
  Photo-Optical Instrumentation Engineers (SPIE) Conference Series, Vol. 9904,
  Space Telescopes and Instrumentation 2016: Optical, Infrared, and Millimeter
  Wave, ed. H.~A. {MacEwen}, G.~G. {Fazio}, M.~{Lystrup}, N.~{Batalha},
  N.~{Siegler}, \& E.~C. {Tong}, 99040Q

\bibitem[{{Cuillandre} {et~al.}(2025){Cuillandre}, {Bolzonella}, {Boselli},
  {Marleau}, {Mondelin}, {Sorce}, {Stone}, {Buitrago}, {Cantiello}, {George},
  {Hatch}, {Quilley}, {Mannucci}, {Saifollahi}, {S{\'a}nchez-Janssen},
  {Tarsitano}, {Tortora}, {Xu}, {Bouy}, {Gwyn}, {Kluge}, {Lan{\c{c}}on},
  {Laureijs}, {Schirmer}, {Abdurro'uf}, {Awad}, {Baes}, {Bournaud}, {Carollo},
  {Codis}, {Conselice}, {De Lapparent}, {Duc}, {Ferr{\'e}-Mateu}, {Gillard},
  {Golden-Marx}, {Jablonka}, {Habas}, {Hunt}, {Mei}, {Miville-Desch{\^e}nes},
  {Montes}, {Nersesian}, {Peletier}, {Poulain}, {Scaramella}, {Scialpi},
  {Sola}, {Stephan}, {Ulivi}, {Urbano}, {Z{\"o}ller}, {Aghanim}, {Altieri},
  {Amara}, {Andreon}, {Auricchio}, {Baldi}, {Balestra}, {Bardelli}, {Bender},
  {Biviano}, {Bodendorf}, {Bonino}, {Branchini}, {Brescia}, {Brinchmann},
  {Camera}, {Capobianco}, {Carbone}, {Carretero}, {Casas}, {Castander},
  {Castellano}, {Castignani}, {Cavuoti}, {Cimatti}, {Congedo}, {Conversi},
  {Copin}, {Courbin}, {Courtois}, {Cropper}, {Da Silva}, {Degaudenzi}, {De
  Lucia}, {Di Giorgio}, {Dinis}, {Douspis}, {Dubath}, {Duncan}, {Dupac},
  {Dusini}, {Farina}, {Farrens}, {Ferriol}, {Fotopoulou}, {Frailis},
  {Franceschi}, {Galeotta}, {Gillis}, {Giocoli}, {G{\'o}mez-Alvarez},
  {Grazian}, {Grupp}, {Guzzo}, {Haugan}, {Hoar}, {Hoekstra}, {Holmes}, {Hook},
  {Hormuth}, {Hornstrup}, {Hudelot}, {Jahnke}, {Jhabvala}, {Keih{\"a}nen},
  {Kermiche}, {Kiessling}, {Kilbinger}, {Kitching}, {Kohley}, {Kubik},
  {Kuijken}, {K{\"u}mmel}, {Kunz}, {Kurki-Suonio}, {Lahav}, {Le Mignant},
  {Ligori}, {Lilje}, {Lindholm}, {Lloro}, {Maino}, {Maiorano}, {Mansutti},
  {Marggraf}, {Markovic}, {Martinet}, {Marulli}, {Massey}, {Maurogordato},
  {McCracken}, {Medinaceli}, {Melchior}, {Mellier}, {Meneghetti}, {Merlin},
  {Meylan}, {Mohr}, {Mora}, {Moresco}, {Moscardini}, {Nakajima}, {Nichol},
  {Niemi}, {Padilla}, {Paltani}, {Pasian}, {Pedersen}, {Percival}, {Pettorino},
  {Pires}, {Polenta}, {Poncet}, {Popa}, {Pozzetti}, {Raison}, {Renzi},
  {Rhodes}, {Riccio}, {Romelli}, {Roncarelli}, {Saglia}, {Sapone}, {Schneider},
  {Schrabback}, {Secroun}, {Seidel}, {Serrano}, {Simon}, {Sirignano}, {Sirri},
  {Skottfelt}, {Stanco}, {Tallada-Cresp{\'\i}}, {Taylor}, {Teplitz}, {Tereno},
  {Toledo-Moreo}, {Tutusaus}, {Valentijn}, {Valenziano}, {Vassallo}, {Verdoes
  Kleijn}, {Wang}, {Weller}, {Zucca}, {Burigana}, \&
  {Scottez}}]{Cuillandre2025}
{Cuillandre}, J.~C., {Bolzonella}, M., {Boselli}, A., {et~al.} 2025, \aap, 697,
  A11

\bibitem[{{DESI Collaboration: Abdul-Karim} {et~al.}(2025){DESI Collaboration:
  Abdul-Karim}, {Adame}, {Aguado}, {Aguilar}, {Ahlen}, {Alam}, {Aldering},
  {Alexander}, {Alfarsy}, {Allen}, {Allende Prieto}, {Alves}, {Anand},
  {Andrade}, {Armengaud}, {Avila}, {Aviles}, {Awan}, {Bailey}, {Baleato
  Lizancos}, {Ballester}, {Bault}, {Bautista}, {BenZvi}, {Beraldo e Silva},
  {Bermejo-Climent}, {Beutler}, {Bianchi}, {Blake}, {Blum}, {Bolton}, {Bonici},
  {Brieden}, {Brodzeller}, {Brooks}, {Buckley-Geer}, {Burtin}, {Canning},
  {Carnero Rosell}, {Carr}, {Carrilho}, {Casas}, {Castander}, {Cereskaite},
  {Cervantes-Cota}, {Chaussidon}, {Chaves-Montero}, {Chen}, {Chen},
  {Claybaugh}, {Cole}, {Cooper}, {Cousinou}, {Cuceu}, {Davis}, {Dawson}, {de
  Belsunce}, {de la Cruz}, {de la Macorra}, {de Mattia}, {Deiosso}, {Della
  Costa}, {Demina}, {Demirbozan}, {DeRose}, {Dey}, {Dey}, {Ding}, {Ding},
  {Doel}, {Douglass}, {Dowicz}, {Ebina}, {Edelstein}, {Eisenstein}, {Elbers},
  {Emas}, {Escoffier}, {Fagrelius}, {Fan}, {Fanning}, {Fawcett},
  {Fern\'andez-Garc\'ia}, {Ferraro}, {Findlay}, {Font-Ribera}, {Forero-Romero},
  {Forero-S\'anchez}, {Frenk}, {G\''ansicke}, {Galbany}, {Garc\'ia-Bellido},
  {Garcia-Quintero}, {Garrison}, {Gazta\\~naga}, {Gil-Mar\'in}, {Gnedin},
  {Gontcho}, {Gonzalez-Morales}, {Gonzalez-Perez}, {Gordon}, {Graur}, {Green},
  {Gruen}, {Gsponer}, {Guandalin}, {Gutierrez}, {Guy}, {Hahn}, {Han}, {Han},
  {He}, {Herrera-Alcantar}, {Honscheid}, {Hou}, {Howlett}, {Huterer},
  {Ir\v\{s\}i\v\{c\}}, {Ishak}, {Jacques}, {Jimenez}, {Jing}, {Joachimi},
  {Joudaki}, {Joyce}, {Jullo}, {Juneau}, {Kara\c\{c\}ayl\{\i\}}, {Karim},
  {Kehoe}, {Kent}, {Khederlarian}, {Kirkby}, {Kisner}, {Kitaura},
  {Kizhuprakkat}, {Kong}, {Koposov}, {Kremin}, {Krolewski}, {Lahav}, {Lai},
  {Lamman}, {Lan}, {Landriau}, {Lang}, {Lange}, {Lasker}, {Le Goff}, {Le
  Guillou}, {Leauthaud}, {Levi}, {Li}, {Li}, {Lodha}, {Lokken}, {Luo},
  {Magneville}, {Manera}, {Manser}, {Margala}, {Martini}, {Maus}, {McCullough},
  {McDonald}, {Medina}, {Medina-Varela}, {Meisner}, {Mena-Fern\'andez},
  {Menegas}, {Mezcua}, {Miquel}, {Montero-Camacho}, {Moon}, {Moustakas},
  {Mu\~noz-Guti\'errez}, {Mu\~noz-Santos}, {Myers}, {Myles}, {Nadathur},
  {Najita}, {Napolitano}, {Newman}, {Nikakhtar}, {Nikutta}, {Niz}, {Noriega},
  {Padmanabhan}, {Paillas}, {Palanque-Delabrouille}, {Palmese}, {Pan}, {Pan},
  {Parkinson}, {Peacock}, {Percival}, {P\'erez-Fern\'andez},
  {P\'erez-R\`afols}, \& {Peterson}}]{DESIDR1}
{DESI Collaboration: Abdul-Karim}, M., {Adame}, A.~G., {Aguado}, D., {et~al.}
  2025, \aj, submitted, arXiv:2503.14745

\bibitem[{{Duchon}(1979)}]{Duchon1979}
{Duchon}, C.~E. 1979, Journal of Applied Meteorology, 18, 1016

\bibitem[{{{EPTA Collaboration} and {InPTA Collaboration}: Antoniadis}
  {et~al.}(2023){{EPTA Collaboration} and {InPTA Collaboration}: Antoniadis},
  {Arumugam}, {Arumugam}, {Babak}, {Bagchi}, {Bak Nielsen}, {Bassa}, {Bathula},
  {Berthereau}, {Bonetti}, {Bortolas}, {Brook}, {Burgay}, {Caballero},
  {Chalumeau}, {Champion}, {Chanlaridis}, {Chen}, {Cognard}, {Dandapat}, {Deb},
  {Desai}, {Desvignes}, {Dhanda-Batra}, {Dwivedi}, {Falxa}, {Ferdman},
  {Franchini}, {Gair}, {Goncharov}, {Gopakumar}, {Graikou}, {Grie{\ss}meier},
  {Guillemot}, {Guo}, {Gupta}, {Hisano}, {Hu}, {Iraci}, {Izquierdo-Villalba},
  {Jang}, {Jawor}, {Janssen}, {Jessner}, {Joshi}, {Kareem}, {Karuppusamy},
  {Keane}, {Keith}, {Kharbanda}, {Kikunaga}, {Kolhe}, {Kramer}, {Krishnakumar},
  {Lackeos}, {Lee}, {Liu}, {Liu}, {Lyne}, {McKee}, {Maan}, {Main},
  {Mickaliger}, {Ni{\c{t}}u}, {Nobleson}, {Paladi}, {Parthasarathy}, {Perera},
  {Perrodin}, {Petiteau}, {Porayko}, {Possenti}, {Prabu}, {Quelquejay Leclere},
  {Rana}, {Samajdar}, {Sanidas}, {Sesana}, {Shaifullah}, {Singha}, {Speri},
  {Spiewak}, {Srivastava}, {Stappers}, {Surnis}, {Susarla}, {Susobhanan},
  {Takahashi}, {Tarafdar}, {Theureau}, {Tiburzi}, {van der Wateren}, {Vecchio},
  {Venkatraman Krishnan}, {Verbiest}, {Wang}, {Wang}, \& {Wu}}]{EPTA_InPTA2023}
{{EPTA Collaboration} and {InPTA Collaboration}: Antoniadis}, J., {Arumugam},
  P., {Arumugam}, S., {et~al.} 2023, \aap, 678, A50

\bibitem[{{Erwin}(2015)}]{Erwin2015}
{Erwin}, P. 2015, \apj, 799, 226

\bibitem[{{Euclid Collaboration: Aussel} {et~al.}(2025){Euclid Collaboration:
  Aussel}, {Tereno}, {Schirmer}, {et~al.}}]{Q1-TP001}
{Euclid Collaboration: Aussel}, H., {Tereno}, I., {Schirmer}, M., {et~al.}
  2025, A\&A, submitted (Euclid Q1 SI), arXiv:2503.15302

\bibitem[{{Euclid Collaboration: Cropper} {et~al.}(2025){Euclid Collaboration:
  Cropper}, {Al-Bahlawan}, {Amiaux}, {et~al.}}]{EuclidSkyVIS}
{Euclid Collaboration: Cropper}, M., {Al-Bahlawan}, A., {Amiaux}, J., {et~al.}
  2025, A\&A, 697, A2

\bibitem[{{Euclid Collaboration: Ecker} {et~al.}(2025){Euclid Collaboration:
  Ecker}, {Fabricius}, {Seitz}, {et~al.}}]{Q1-SP082}
{Euclid Collaboration: Ecker}, L., {Fabricius}, M., {Seitz}, S., {et~al.} 2025,
  A\&A, hopefully submitted

\bibitem[{{Euclid Collaboration: Enia} {et~al.}(2025){Euclid Collaboration:
  Enia}, {Pozzetti}, {Bolzonella}, {et~al.}}]{Q1-SP031}
{Euclid Collaboration: Enia}, A., {Pozzetti}, L., {Bolzonella}, M., {et~al.}
  2025, A\&A, in press (Euclid Q1 SI),
  \url{https://doi.org/10.1051/0004-6361/202554576}, arXiv:2503.15314

\bibitem[{{Euclid Collaboration: Holloway} {et~al.}(2025){Euclid Collaboration:
  Holloway}, {Verma}, {Walmsley}, {et~al.}}]{Q1-SP059}
{Euclid Collaboration: Holloway}, P., {Verma}, A., {Walmsley}, M., {et~al.}
  2025, A\&A, submitted (Euclid Q1 SI), arXiv:2503.15328

\bibitem[{{Euclid Collaboration: Jahnke} {et~al.}(2025){Euclid Collaboration:
  Jahnke}, {Gillard}, {Schirmer}, {et~al.}}]{EuclidSkyNISP}
{Euclid Collaboration: Jahnke}, K., {Gillard}, W., {Schirmer}, M., {et~al.}
  2025, A\&A, 697, A3

\bibitem[{{Euclid Collaboration: La Marca} {et~al.}(2025){Euclid Collaboration:
  La Marca}, {Wang}, {Margalef-Bentabol}, {et~al.}}]{Q1-SP013}
{Euclid Collaboration: La Marca}, A., {Wang}, L., {Margalef-Bentabol}, B.,
  {et~al.} 2025, A\&A, in press (Euclid Q1 SI),
  \url{https://doi.org/10.1051/0004-6361/202554579}, arXiv:2503.15317

\bibitem[{{Euclid Collaboration: Li} {et~al.}(2025){Euclid Collaboration: Li},
  {Collett}, {Walmsley}, {et~al.}}]{Q1-SP054}
{Euclid Collaboration: Li}, T., {Collett}, T.~E., {Walmsley}, M., {et~al.}
  2025, A\&A, in press (Euclid Q1 SI),
  \url{https://doi.org/10.1051/0004-6361/202554543}, arXiv:2503.15327

\bibitem[{{Euclid Collaboration: Lines} {et~al.}(2025){Euclid Collaboration:
  Lines}, {Collett}, {Walmsley}, {et~al.}}]{Q1-SP053}
{Euclid Collaboration: Lines}, N.~E.~P., {Collett}, T.~E., {Walmsley}, M.,
  {et~al.} 2025, A\&A, in press (Euclid Q1 SI),
  \url{https://doi.org/10.1051/0004-6361/202554542}, arXiv:2503.15326

\bibitem[{{Euclid Collaboration: Marleau} {et~al.}(2025){Euclid Collaboration:
  Marleau}, {Habas}, {Carollo}, {et~al.}}]{Q1-SP001}
{Euclid Collaboration: Marleau}, F.~R., {Habas}, R., {Carollo}, D., {et~al.}
  2025, A\&A, accepted (Euclid Q1 SI), arXiv:2503.15335

\bibitem[{{Euclid Collaboration: McCracken} {et~al.}(2025){Euclid
  Collaboration: McCracken}, {Benson}, {Dolding}, {et~al.}}]{Q1-TP002}
{Euclid Collaboration: McCracken}, H.~J., {Benson}, K., {Dolding}, C., {et~al.}
  2025, A\&A, in press (Euclid Q1 SI),
  \url{https://doi.org/10.1051/0004-6361/202554594}, arXiv:2503.15303

\bibitem[{{Euclid Collaboration: Mellier} {et~al.}(2025){Euclid Collaboration:
  Mellier}, {Abdurro'uf}, {Acevedo~Barroso}, {et~al.}}]{EuclidSkyOverview}
{Euclid Collaboration: Mellier}, Y., {Abdurro'uf}, {Acevedo~Barroso}, J.,
  {et~al.} 2025, A\&A, 697, A1

\bibitem[{{Euclid Collaboration: Quilley} {et~al.}(2025){Euclid Collaboration:
  Quilley}, {Damjanov}, {de Lapparent}, {et~al.}}]{Q1-SP040}
{Euclid Collaboration: Quilley}, L., {Damjanov}, I., {de Lapparent}, V.,
  {et~al.} 2025, A\&A, in press (Euclid Q1 SI),
  \url{https://doi.org/10.1051/0004-6361/202554585}, arXiv:2503.15309

\bibitem[{{Euclid Collaboration: Romelli} {et~al.}(2025){Euclid Collaboration:
  Romelli}, {K\"ummel}, {Dole}, {et~al.}}]{Q1-TP004}
{Euclid Collaboration: Romelli}, E., {K\"ummel}, M., {Dole}, H., {et~al.} 2025,
  A\&A, in press (Euclid Q1 SI),
  \url{https://doi.org/10.1051/0004-6361/202554586}, arXiv:2503.15305

\bibitem[{{Euclid Collaboration: Scaramella} {et~al.}(2022){Euclid
  Collaboration: Scaramella}, {Amiaux}, {Mellier}, {et~al.}}]{Scaramella-EP1}
{Euclid Collaboration: Scaramella}, R., {Amiaux}, J., {Mellier}, Y., {et~al.}
  2022, \aap, 662, A112

\bibitem[{{Euclid Collaboration: Tucci} {et~al.}(2025){Euclid Collaboration:
  Tucci}, {Paltani}, {Hartley}, {et~al.}}]{Q1-TP005}
{Euclid Collaboration: Tucci}, M., {Paltani}, S., {Hartley}, W.~G., {et~al.}
  2025, A\&A, in press (Euclid Q1 SI),
  \url{https://doi.org/10.1051/0004-6361/202554588}, arXiv:2503.15306

\bibitem[{{Euclid Collaboration: Walmsley} {et~al.}(2025{\natexlab{a}}){Euclid
  Collaboration: Walmsley}, {Holloway}, {Lines}, {et~al.}}]{Q1-SP048}
{Euclid Collaboration: Walmsley}, M., {Holloway}, P., {Lines}, N.~E.~P.,
  {et~al.} 2025{\natexlab{a}}, A\&A, submitted (Euclid Q1 SI), arXiv:2503.15324

\bibitem[{{Euclid Collaboration: Walmsley} {et~al.}(2025{\natexlab{b}}){Euclid
  Collaboration: Walmsley}, {Huertas-Company}, {Quilley}, {et~al.}}]{Q1-SP047}
{Euclid Collaboration: Walmsley}, M., {Huertas-Company}, M., {Quilley}, L.,
  {et~al.} 2025{\natexlab{b}}, A\&A, accepted (Euclid Q1 SI), arXiv:2503.15310

\bibitem[{{Euclid Quick Release Q1}(2025)}]{Q1cite}
{Euclid Quick Release Q1}. 2025, \url{https://doi.org/10.57780/esa-2853f3b}

\bibitem[{{Fu} {et~al.}(2012){Fu}, {Yan}, {Myers}, {Stockton}, {Djorgovski},
  {Aldering}, \& {Rich}}]{Fu2012}
{Fu}, H., {Yan}, L., {Myers}, A.~D., {et~al.} 2012, \apj, 745, 67

\bibitem[{{Gaia Collaboration: Brown} {et~al.}(2021){Gaia Collaboration:
  Brown}, {Vallenari}, {Prusti}, {de Bruijne}, {Babusiaux}, {Biermann},
  {Creevey}, {Evans}, {Eyer}, {Hutton}, {Jansen}, {Jordi}, {Klioner},
  {Lammers}, {Lindegren}, {Luri}, {Mignard}, {Panem}, {Pourbaix}, {Randich},
  {Sartoretti}, {Soubiran}, {Walton}, {Arenou}, {Bailer-Jones}, {Bastian},
  {Cropper}, {Drimmel}, {Katz}, {Lattanzi}, {van Leeuwen}, {Bakker},
  {Cacciari}, {Casta{\\textasciitilde n}eda}, {De Angeli}, {Ducourant},
  {Fabricius}, {Fouesneau}, {Fr{\'e}mat}, {Guerra}, {Guerrier}, {Guiraud},
  {Jean-Antoine Piccolo}, {Masana}, {Messineo}, {Mowlavi}, {Nicolas},
  {Nienartowicz}, {Pailler}, {Panuzzo}, {Riclet}, {Roux}, {Seabroke}, {Sordo},
  {Tanga}, {Th{\'e}venin}, {Gracia-Abril}, {Portell}, {Teyssier}, {Altmann},
  {Andrae}, {Bellas-Velidis}, {Benson}, {Berthier}, {Blomme}, {Brugaletta},
  {Burgess}, {Busso}, {Carry}, {Cellino}, {Cheek}, {Clementini}, {Damerdji},
  {Davidson}, {Delchambre}, {Dell'Oro}, {Fern{\'a}ndez-Hern{\'a}ndez},
  {Galluccio}, {Garc{\'\i}a-Lario}, {Garcia-Reinaldos},
  {Gonz{\'a}lez-N{\'u}{\\textasciitilde n}ez}, {Gosset}, {Haigron},
  {Halbwachs}, {Hambly}, {Harrison}, {Hatzidimitriou}, {Heiter},
  {Hern{\'a}ndez}, {Hestroffer}, {Hodgkin}, {Holl}, {Jan{\ss}en}, {Jevardat de
  Fombelle}, {Jordan}, {Krone-Martins}, {Lanzafame}, {L{\"o}ffler}, {Lorca},
  {Manteiga}, {Marchal}, {Marrese}, {Moitinho}, {Mora}, {Muinonen}, {Osborne},
  {Pancino}, {Pauwels}, {Petit}, {Recio-Blanco}, {Richards}, {Riello},
  {Rimoldini}, {Robin}, {Roegiers}, {Rybizki}, {Sarro}, {Siopis}, {Smith},
  {Sozzetti}, {Ulla}, {Utrilla}, {van Leeuwen}, {van Reeven}, {Abbas}, {Abreu
  Aramburu}, {Accart}, {Aerts}, {Aguado}, {Ajaj}, {Altavilla}, {{\'A}lvarez},
  {{\'A}lvarez Cid-Fuentes}, {Alves}, {Anderson}, {Anglada Varela}, {Antoja},
  {Audard}, {Baines}, {Baker}, {Balaguer-N{\'u}{\\textasciitilde n}ez},
  {Balbinot}, {Balog}, {Barache}, {Barbato}, {Barros}, {Barstow},
  {Bartolom{\'e}}, {Bassilana}, {Bauchet}, {Baudesson-Stella}, {Becciani},
  {Bellazzini}, {Bernet}, {Bertone}, {Bianchi}, {Blanco-Cuaresma}, {Boch},
  {Bombrun}, {Bossini}, {Bouquillon}, {Bragaglia}, {Bramante}, {Breedt},
  {Bressan}, {Brouillet}, {Bucciarelli}, {Burlacu}, {Busonero}, {Butkevich},
  {Buzzi}, {Caffau}, {Cancelliere}, {C{\'a}novas}, {Cantat-Gaudin}, {Carballo},
  {Carlucci}, {Carnerero}, {Carrasco}, {Casamiquela}, {Castellani},
  {Castro-Ginard}, {Castro Sampol}, {Chaoul}, {Charlot}, {Chemin}, {Chiavassa},
  {Cioni}, {Comoretto}, {Cooper}, {Cornez}, {Cowell}, {Crifo}, {Crosta},
  {Crowley}, {Dafonte}, {Dapergolas}, {David}, \&
  {David}}]{Gaia-CollaborationBrown2021}
{Gaia Collaboration: Brown}, A. G.~A., {Vallenari}, A., {Prusti}, T., {et~al.}
  2021, \aap, 649, A1

\bibitem[{{Gaia Collaboration: Prusti} {et~al.}(2016){Gaia Collaboration:
  Prusti}, {de Bruijne}, {Brown}, {Vallenari}, {Babusiaux}, {Bailer-Jones},
  {Bastian}, {Biermann}, {Evans}, {Eyer}, {Jansen}, {Jordi}, {Klioner},
  {Lammers}, {Lindegren}, {Luri}, {Mignard}, {Milligan}, {Panem}, {Poinsignon},
  {Pourbaix}, {Randich}, {Sarri}, {Sartoretti}, {Siddiqui}, {Soubiran},
  {Valette}, {van Leeuwen}, {Walton}, {Aerts}, {Arenou}, {Cropper}, {Drimmel},
  {H{\o}g}, {Katz}, {Lattanzi}, {O'Mullane}, {Grebel}, {Holland}, {Huc},
  {Passot}, {Bramante}, {Cacciari}, {Casta{\\textasciitilde n}eda}, {Chaoul},
  {Cheek}, {De Angeli}, {Fabricius}, {Guerra}, {Hern{\'a}ndez},
  {Jean-Antoine-Piccolo}, {Masana}, {Messineo}, {Mowlavi}, {Nienartowicz},
  {Ord{\'o}{\\textasciitilde n}ez-Blanco}, {Panuzzo}, {Portell}, {Richards},
  {Riello}, {Seabroke}, {Tanga}, {Th{\'e}venin}, {Torra}, {Els},
  {Gracia-Abril}, {Comoretto}, {Garcia-Reinaldos}, {Lock}, {Mercier},
  {Altmann}, {Andrae}, {Astraatmadja}, {Bellas-Velidis}, {Benson}, {Berthier},
  {Blomme}, {Busso}, {Carry}, {Cellino}, {Clementini}, {Cowell}, {Creevey},
  {Cuypers}, {Davidson}, {De Ridder}, {de Torres}, {Delchambre}, {Dell'Oro},
  {Ducourant}, {Fr{\'e}mat}, {Garc{\'\i}a-Torres}, {Gosset}, {Halbwachs},
  {Hambly}, {Harrison}, {Hauser}, {Hestroffer}, {Hodgkin}, {Huckle}, {Hutton},
  {Jasniewicz}, {Jordan}, {Kontizas}, {Korn}, {Lanzafame}, {Manteiga},
  {Moitinho}, {Muinonen}, {Osinde}, {Pancino}, {Pauwels}, {Petit},
  {Recio-Blanco}, {Robin}, {Sarro}, {Siopis}, {Smith}, {Smith}, {Sozzetti},
  {Thuillot}, {van Reeven}, {Viala}, {Abbas}, {Abreu Aramburu}, {Accart},
  {Aguado}, {Allan}, {Allasia}, {Altavilla}, {{\'A}lvarez}, {Alves},
  {Anderson}, {Andrei}, {Anglada Varela}, {Antiche}, {Antoja}, {Ant{\'o}n},
  {Arcay}, {Atzei}, {Ayache}, {Bach}, {Baker},
  {Balaguer-N{\'u}{\\textasciitilde n}ez}, {Barache}, {Barata}, {Barbier},
  {Barblan}, {Baroni}, {Barrado y Navascu{\'e}s}, {Barros}, {Barstow},
  {Becciani}, {Bellazzini}, {Bellei}, {Bello Garc{\'\i}a}, {Belokurov},
  {Bendjoya}, {Berihuete}, {Bianchi}, {Bienaym{\'e}}, {Billebaud},
  {Blagorodnova}, {Blanco-Cuaresma}, {Boch}, {Bombrun}, {Borrachero},
  {Bouquillon}, {Bourda}, {Bouy}, {Bragaglia}, {Breddels}, {Brouillet},
  {Br{\"u}semeister}, {Bucciarelli}, {Budnik}, {Burgess}, {Burgon}, {Burlacu},
  {Busonero}, {Buzzi}, {Caffau}, {Cambras}, {Campbell}, {Cancelliere},
  {Cantat-Gaudin}, {Carlucci}, {Carrasco}, {Castellani}, {Charlot}, {Charnas},
  {Charvet}, {Chassat}, {Chiavassa}, {Clotet}, {Cocozza}, {Collins}, {Collins},
  \& {Costigan}}]{Gaia-CollaborationPrusti2016}
{Gaia Collaboration: Prusti}, T., {de Bruijne}, J. H.~J., {Brown}, A. G.~A.,
  {et~al.} 2016, \aap, 595, A1

\bibitem[{{Gaia Collaboration: Vallenari} {et~al.}(2023){Gaia Collaboration:
  Vallenari}, {Brown}, {Prusti}, {de Bruijne}, {Arenou}, {Babusiaux},
  {Biermann}, {Creevey}, {Ducourant}, {Evans}, {Eyer}, {Guerra}, {Hutton},
  {Jordi}, {Klioner}, {Lammers}, {Lindegren}, {Luri}, {Mignard}, {Panem},
  {Pourbaix}, {Randich}, {Sartoretti}, {Soubiran}, {Tanga}, {Walton},
  {Bailer-Jones}, {Bastian}, {Drimmel}, {Jansen}, {Katz}, {Lattanzi}, {van
  Leeuwen}, {Bakker}, {Cacciari}, {Casta{\\textasciitilde n}eda}, {De Angeli},
  {Fabricius}, {Fouesneau}, {Fr{\'e}mat}, {Galluccio}, {Guerrier}, {Heiter},
  {Masana}, {Messineo}, {Mowlavi}, {Nicolas}, {Nienartowicz}, {Pailler},
  {Panuzzo}, {Riclet}, {Roux}, {Seabroke}, {Sordo}, {Th{\'e}venin},
  {Gracia-Abril}, {Portell}, {Teyssier}, {Altmann}, {Andrae}, {Audard},
  {Bellas-Velidis}, {Benson}, {Berthier}, {Blomme}, {Burgess}, {Busonero},
  {Busso}, {C{\'a}novas}, {Carry}, {Cellino}, {Cheek}, {Clementini},
  {Damerdji}, {Davidson}, {de Teodoro}, {Nu{\\textasciitilde n}ez Campos},
  {Delchambre}, {Dell'Oro}, {Esquej}, {Fern{\'a}ndez-Hern{\'a}ndez}, {Fraile},
  {Garabato}, {Garc{\'\i}a-Lario}, {Gosset}, {Haigron}, {Halbwachs}, {Hambly},
  {Harrison}, {Hern{\'a}ndez}, {Hestroffer}, {Hodgkin}, {Holl}, {Jan{\ss}en},
  {Jevardat de Fombelle}, {Jordan}, {Krone-Martins}, {Lanzafame},
  {L{\"o}ffler}, {Marchal}, {Marrese}, {Moitinho}, {Muinonen}, {Osborne},
  {Pancino}, {Pauwels}, {Recio-Blanco}, {Reyl{\'e}}, {Riello}, {Rimoldini},
  {Roegiers}, {Rybizki}, {Sarro}, {Siopis}, {Smith}, {Sozzetti}, {Utrilla},
  {van Leeuwen}, {Abbas}, {{\'A}brah{\'a}m}, {Abreu Aramburu}, {Aerts},
  {Aguado}, {Ajaj}, {Aldea-Montero}, {Altavilla}, {{\'A}lvarez}, {Alves},
  {Anders}, {Anderson}, {Anglada Varela}, {Antoja}, {Baines}, {Baker},
  {Balaguer-N{\'u}{\\textasciitilde n}ez}, {Balbinot}, {Balog}, {Barache},
  {Barbato}, {Barros}, {Barstow}, {Bartolom{\'e}}, {Bassilana}, {Bauchet},
  {Becciani}, {Bellazzini}, {Berihuete}, {Bernet}, {Bertone}, {Bianchi},
  {Binnenfeld}, {Blanco-Cuaresma}, {Blazere}, {Boch}, {Bombrun}, {Bossini},
  {Bouquillon}, {Bragaglia}, {Bramante}, {Breedt}, {Bressan}, {Brouillet},
  {Brugaletta}, {Bucciarelli}, {Burlacu}, {Butkevich}, {Buzzi}, {Caffau},
  {Cancelliere}, {Cantat-Gaudin}, {Carballo}, {Carlucci}, {Carnerero},
  {Carrasco}, {Casamiquela}, {Castellani}, {Castro-Ginard}, {Chaoul},
  {Charlot}, {Chemin}, {Chiaramida}, {Chiavassa}, {Chornay}, {Comoretto},
  {Contursi}, {Cooper}, {Cornez}, {Cowell}, {Crifo}, {Cropper}, {Crosta},
  {Crowley}, {Dafonte}, {Dapergolas}, {David}, {David}, {de Laverny}, {De
  Luise}, \& {De March}}]{Gaia-CollaborationVallenari2023}
{Gaia Collaboration: Vallenari}, A., {Brown}, A. G.~A., {Prusti}, T., {et~al.}
  2023, \aap, 674, A1

\bibitem[{{Gebhardt} {et~al.}(2021){Gebhardt}, {Mentuch Cooper}, {Ciardullo},
  {Acquaviva}, {Bender}, {Bowman}, {Castanheira}, {Dalton}, {Davis}, {de Jong},
  {DePoy}, {Devarakonda}, {Dongsheng}, {Drory}, {Fabricius}, {Farrow},
  {Feldmeier}, {Finkelstein}, {Froning}, {Gawiser}, {Gronwall}, {Herold},
  {Hill}, {Hopp}, {House}, {Janowiecki}, {Jarvis}, {Jeong}, {Jogee}, {Kakuma},
  {Kelz}, {Kollatschny}, {Komatsu}, {Krumpe}, {Landriau}, {Liu}, {Niemeyer},
  {MacQueen}, {Marshall}, {Mawatari}, {McLinden}, {Mukae}, {Nagaraj}, {Ono},
  {Ouchi}, {Papovich}, {Sakai}, {Saito}, {Schneider}, {Schulze},
  {Shanmugasundararaj}, {Shetrone}, {Sneden}, {Snigula}, {Steinmetz}, {Thomas},
  {Thomas}, {Tuttle}, {Urrutia}, {Wisotzki}, {Wold}, {Zeimann}, \&
  {Zhang}}]{Gebhardt2021}
{Gebhardt}, K., {Mentuch Cooper}, E., {Ciardullo}, R., {et~al.} 2021, \apj,
  923, 217

\bibitem[{{Gonz{\'a}lez} {et~al.}(2007){Gonz{\'a}lez}, {Sperhake},
  {Br{\"u}gmann}, {Hannam}, \& {Husa}}]{Gonzalez2007a}
{Gonz{\'a}lez}, J.~A., {Sperhake}, U., {Br{\"u}gmann}, B., {Hannam}, M., \&
  {Husa}, S. 2007, \prl, 98, 091101

\bibitem[{{Goulding} {et~al.}(2019){Goulding}, {Pardo}, {Greene}, {Mingarelli},
  {Nyland}, \& {Strauss}}]{Goulding2019}
{Goulding}, A.~D., {Pardo}, K., {Greene}, J.~E., {et~al.} 2019, \apjl, 879, L21

\bibitem[{{Goullaud} {et~al.}(2018){Goullaud}, {Jensen}, {Blakeslee}, {Ma},
  {Greene}, \& {Thomas}}]{Goullaud2018}
{Goullaud}, C.~F., {Jensen}, J.~B., {Blakeslee}, J.~P., {et~al.} 2018, \apj,
  856, 11

\bibitem[{{Graham} \& {Driver}(2005)}]{GrahamDriver2005}
{Graham}, A.~W. \& {Driver}, S.~P. 2005, \pasa, 22, 118

\bibitem[{{Hill} {et~al.}(2021){Hill}, {Lee}, {MacQueen}, {Kelz}, {Drory},
  {Vattiat}, {Good}, {Ramsey}, {Kriel}, {Peterson}, {DePoy}, {Gebhardt},
  {Marshall}, {Tuttle}, {Bauer}, {Chonis}, {Fabricius}, {Froning},
  {H{\"a}user}, {Indahl}, {Jahn}, {Landriau}, {Leck}, {Montesano}, {Prochaska},
  {Snigula}, {Zeimann}, {Bryant}, {Damm}, {Fowler}, {Janowiecki}, {Martin},
  {Mrozinski}, {Odewahn}, {Rostopchin}, {Shetrone}, {Spencer}, {Mentuch
  Cooper}, {Armandroff}, {Bender}, {Dalton}, {Hopp}, {Komatsu}, {Nicklas},
  {Ramsey}, {Roth}, {Schneider}, {Sneden}, \& {Steinmetz}}]{Hill2021}
{Hill}, G.~J., {Lee}, H., {MacQueen}, P.~J., {et~al.} 2021, \aj, 162, 298

\bibitem[{{Hoaglin} {et~al.}(1983){Hoaglin}, {Mosteller}, \&
  {Tukey}}]{Hoaglin1983}
{Hoaglin}, D.~C., {Mosteller}, F., \& {Tukey}, J.~W. 1983, {Understanding
  robust and exploratory data anlysis}

\bibitem[{{Hoessel}(1980)}]{Hoessel1980}
{Hoessel}, J.~G. 1980, \apj, 241, 493

\bibitem[{{Jones} {et~al.}(2009){Jones}, {Read}, {Saunders}, {Colless},
  {Jarrett}, {Parker}, {Fairall}, {Mauch}, {Sadler}, {Watson}, {Burton},
  {Campbell}, {Cass}, {Croom}, {Dawe}, {Fiegert}, {Frankcombe}, {Hartley},
  {Huchra}, {James}, {Kirby}, {Lahav}, {Lucey}, {Mamon}, {Moore}, {Peterson},
  {Prior}, {Proust}, {Russell}, {Safouris}, {Wakamatsu}, {Westra}, \&
  {Williams}}]{6dFGS}
{Jones}, D.~H., {Read}, M.~A., {Saunders}, W., {et~al.} 2009, \mnras, 399, 683

\bibitem[{{Jord{\'a}n} {et~al.}(2007){Jord{\'a}n}, {McLaughlin},
  {C{\^o}t{\'e}}, {Ferrarese}, {Peng}, {Mei}, {Villegas}, {Merritt}, {Tonry},
  \& {West}}]{Jordan2007}
{Jord{\'a}n}, A., {McLaughlin}, D.~E., {C{\^o}t{\'e}}, P., {et~al.} 2007,
  \apjs, 171, 101

\bibitem[{Khonji {et~al.}(2024)Khonji, Gualandris, Read, \&
  Dehnen}]{Khonji2024}
Khonji, N., Gualandris, A., Read, J.~I., \& Dehnen, W. 2024, \apj, 974, 204

\bibitem[{{Kim} {et~al.}(2017){Kim}, {Yoon}, {Privon}, {Evans}, {Harvey},
  {Stierwalt}, \& {Kim}}]{Kim2017}
{Kim}, D.~C., {Yoon}, I., {Privon}, G.~C., {et~al.} 2017, \apj, 840, 71

\bibitem[{{Komossa} \& {Merritt}(2008)}]{Komossa2008a}
{Komossa}, S. \& {Merritt}, D. 2008, \apjl, 683, L21

\bibitem[{{Komossa} {et~al.}(2008){Komossa}, {Zhou}, \& {Lu}}]{Komossa2008}
{Komossa}, S., {Zhou}, H., \& {Lu}, H. 2008, \apjl, 678, L81

\bibitem[{{Kormendy} {et~al.}(2009){Kormendy}, {Fisher}, {Cornell}, \&
  {Bender}}]{KFCB2009}
{Kormendy}, J., {Fisher}, D.~B., {Cornell}, M.~E., \& {Bender}, R. 2009, \apjs,
  182, 216

\bibitem[{{Koss} {et~al.}(2014){Koss}, {Blecha}, {Mushotzky}, {Veilleux},
  {Hung}, {Man}, \& {Li}}]{Koss2014}
{Koss}, M., {Blecha}, L., {Mushotzky}, R., {et~al.} 2014, in American
  Astronomical Society Meeting Abstracts, Vol. 223, American Astronomical
  Society Meeting Abstracts 223, 251.20

\bibitem[{{Kron}(1980)}]{Kron1980}
{Kron}, R.~G. 1980, \apjs, 43, 305

\bibitem[{{Lauer} {et~al.}(2005){Lauer}, {Faber}, {Gebhardt}, {Richstone},
  {Tremaine}, {Ajhar}, {Aller}, {Bender}, {Dressler}, {Filippenko}, {Green},
  {Grillmair}, {Ho}, {Kormendy}, {Magorrian}, {Pinkney}, \&
  {Siopis}}]{Lauer2005}
{Lauer}, T.~R., {Faber}, S.~M., {Gebhardt}, K., {et~al.} 2005, \aj, 129, 2138

\bibitem[{{Lauer} {et~al.}(2007){Lauer}, {Gebhardt}, {Faber}, {Richstone},
  {Tremaine}, {Kormendy}, {Aller}, {Bender}, {Dressler}, {Filippenko}, {Green},
  \& {Ho}}]{Lauer2007}
{Lauer}, T.~R., {Gebhardt}, K., {Faber}, S.~M., {et~al.} 2007, \apj, 664, 226

\bibitem[{Lena {et~al.}(2014)Lena, Robinson, Marconi, Axon, Capetti, Merritt,
  \& Batcheldor}]{Lena2014}
Lena, D., Robinson, A., Marconi, A., {et~al.} 2014, ApJ, 795, 146

\bibitem[{{Lidman} {et~al.}(2020){Lidman}, {Tucker}, {Davis}, {Uddin},
  {Asorey}, {Bolejko}, {Brout}, {Calcino}, {Carollo}, {Carr}, {Childress},
  {Hoormann}, {Foley}, {Galbany}, {Glazebrook}, {Hinton}, {Kessler}, {Kim},
  {King}, {Kremin}, {Kuehn}, {Lagattuta}, {Lewis}, {Macaulay}, {Malik},
  {March}, {Martini}, {M{\"o}ller}, {Mudd}, {Nichol}, {Panther}, {Parkinson},
  {Pursiainen}, {Sako}, {Swann}, {Scalzo}, {Scolnic}, {Sharp}, {Smith},
  {Sommer}, {Sullivan}, {Webb}, {Wiseman}, {Yu}, {Yuan}, {Zhang}, {Abbott},
  {Aguena}, {Allam}, {Annis}, {Avila}, {Bertin}, {Bhargava}, {Brooks}, {Carnero
  Rosell}, {Carrasco Kind}, {Carretero}, {Castander}, {Costanzi}, {da Costa},
  {De Vicente}, {Doel}, {Eifler}, {Everett}, {Fosalba}, {Frieman},
  {Garc{\'\i}a-Bellido}, {Gaztanaga}, {Gruen}, {Gruendl}, {Gschwend},
  {Gutierrez}, {Hartley}, {Hollowood}, {Honscheid}, {James}, {Kuropatkin},
  {Li}, {Lima}, {Lin}, {Maia}, {Marshall}, {Melchior}, {Menanteau}, {Miquel},
  {Palmese}, {Paz-Chinch{\'o}n}, {Plazas}, {Roodman}, {Rykoff}, {Sanchez},
  {Santiago}, {Scarpine}, {Schubnell}, {Serrano}, {Sevilla-Noarbe}, {Suchyta},
  {Swanson}, {Tarle}, {Tucker}, {Varga}, {Walker}, {Wester}, {Wilkinson}, \&
  {DES Collaboration}}]{OzDES}
{Lidman}, C., {Tucker}, B.~E., {Davis}, T.~M., {et~al.} 2020, \mnras, 496, 19

\bibitem[{{Liepold} \& {Ma}(2024)}]{Liepold2024}
{Liepold}, E.~R. \& {Ma}, C.-P. 2024, \apjl, 971, L29

\bibitem[{{Mannucci} {et~al.}(2022){Mannucci}, {Pancino}, {Belfiore}, {Cicone},
  {Ciurlo}, {Cresci}, {Lusso}, {Marasco}, {Marconi}, {Nardini}, {Pinna},
  {Severgnini}, {Saracco}, {Tozzi}, \& {Yeh}}]{Mannucci2022}
{Mannucci}, F., {Pancino}, E., {Belfiore}, F., {et~al.} 2022, Nature Astronomy,
  6, 1185

\bibitem[{{Mannucci} {et~al.}(2023){Mannucci}, {Scialpi}, {Ciurlo}, {Yeh},
  {Marconcini}, {Tozzi}, {Cresci}, {Marconi}, {Amiri}, {Belfiore}, {Carniani},
  {Cicone}, {Nardini}, {Pancino}, {Rubinur}, {Severgnini}, {Ulivi}, {Venturi},
  {Vignali}, {Volonteri}, {Pinna}, {Rossi}, {Puglisi}, {Agapito}, {Plantet},
  {Ghose}, {Carbonaro}, {Xompero}, {Grani}, {Esposito}, {Power}, {Guerra
  Ramon}, {Lefebvre}, {Cavallaro}, {Davies}, {Riccardi}, {Macintosh}, {Taylor},
  {Dolci}, {Baruffolo}, {Feuchtgruber}, {Kravchenko}, {Rau}, {Sturm},
  {Wiezorrek}, {Dallilar}, \& {Kenworthy}}]{Mannucci2023}
{Mannucci}, F., {Scialpi}, M., {Ciurlo}, A., {et~al.} 2023, \aap, 680, A53

\bibitem[{{Marleau} {et~al.}(2025){Marleau}, {Cuillandre}, {Cantiello},
  {et~al.}}]{EROPerseusDGs}
{Marleau}, F., {Cuillandre}, J.-C., {Cantiello}, M., {et~al.} 2025, A\&A, 697,
  A12

\bibitem[{{Mazzalay} {et~al.}(2016){Mazzalay}, {Thomas}, {Saglia}, {Wegner},
  {Bender}, {Erwin}, {Fabricius}, \& {Rusli}}]{Mazzalay2016}
{Mazzalay}, X., {Thomas}, J., {Saglia}, R.~P., {et~al.} 2016, \mnras, 462, 2847

\bibitem[{{Mehrgan} {et~al.}(2019){Mehrgan}, {Thomas}, {Saglia}, {Mazzalay},
  {Erwin}, {Bender}, {Kluge}, \& {Fabricius}}]{Mehrgan2019}
{Mehrgan}, K., {Thomas}, J., {Saglia}, R., {et~al.} 2019, \apj, 887, 195

\bibitem[{{Merritt} {et~al.}(2009){Merritt}, {Schnittman}, \&
  {Komossa}}]{Merritt2009}
{Merritt}, D., {Schnittman}, J.~D., \& {Komossa}, S. 2009, \apj, 699, 1690

\bibitem[{{Mieske} {et~al.}(2013){Mieske}, {Frank}, {Baumgardt},
  {L{\"u}tzgendorf}, {Neumayer}, \& {Hilker}}]{Mieske2013}
{Mieske}, S., {Frank}, M.~J., {Baumgardt}, H., {et~al.} 2013, \aap, 558, A14

\bibitem[{{Navarro} {et~al.}(2024){Navarro}, {del Rio}, {Angel Diego},
  {Lopez-Caniego}, {Marinic}, {Kruk}, {Reerink}, \& {Arviset}}]{Navarro2024}
{Navarro}, V., {del Rio}, S., {Angel Diego}, M., {et~al.} 2024, in Space Data
  Management. Studies in Big Data, Vol. 141, 1

\bibitem[{{Neumayer} \& {Walcher}(2012)}]{Neumeyer2012}
{Neumayer}, N. \& {Walcher}, C.~J. 2012, Advances in Astronomy, 2012, 709038

\bibitem[{{Neureiter} {et~al.}(2023){Neureiter}, {Thomas}, {Rantala}, {Naab},
  {Mehrgan}, {Saglia}, {de Nicola}, \& {Bender}}]{Neureiter2023}
{Neureiter}, B., {Thomas}, J., {Rantala}, A., {et~al.} 2023, \apj, 950, 15

\bibitem[{{Planck Collaboration: Aghanim} {et~al.}(2020){Planck Collaboration:
  Aghanim}, {Akrami}, {Ashdown}, {Aumont}, {Baccigalupi}, {Ballardini},
  {Banday}, {Barreiro}, {Bartolo}, {Basak}, {Battye}, {Benabed}, {Bernard},
  {Bersanelli}, {Bielewicz}, {Bock}, {Bond}, {Borrill}, {Bouchet}, {Boulanger},
  {Bucher}, {Burigana}, {Butler}, {Calabrese}, {Cardoso}, {Carron},
  {Challinor}, {Chiang}, {Chluba}, {Colombo}, {Combet}, {Contreras}, {Crill},
  {Cuttaia}, {de Bernardis}, {de Zotti}, {Delabrouille}, {Delouis}, {Di
  Valentino}, {Diego}, {Dor{\'e}}, {Douspis}, {Ducout}, {Dupac}, {Dusini},
  {Efstathiou}, {Elsner}, {En{\ss}lin}, {Eriksen}, {Fantaye}, {Farhang},
  {Fergusson}, {Fernandez-Cobos}, {Finelli}, {Forastieri}, {Frailis},
  {Fraisse}, {Franceschi}, {Frolov}, {Galeotta}, {Galli}, {Ganga},
  {G{\'e}nova-Santos}, {Gerbino}, {Ghosh}, {Gonz{\'a}lez-Nuevo}, {G{\'o}rski},
  {Gratton}, {Gruppuso}, {Gudmundsson}, {Hamann}, {Handley}, {Hansen},
  {Herranz}, {Hildebrandt}, {Hivon}, {Huang}, {Jaffe}, {Jones}, {Karakci},
  {Keih{\"a}nen}, {Keskitalo}, {Kiiveri}, {Kim}, {Kisner}, {Knox},
  {Krachmalnicoff}, {Kunz}, {Kurki-Suonio}, {Lagache}, {Lamarre}, {Lasenby},
  {Lattanzi}, {Lawrence}, {Le Jeune}, {Lemos}, {Lesgourgues}, {Levrier},
  {Lewis}, {Liguori}, {Lilje}, {Lilley}, {Lindholm}, {L{\'o}pez-Caniego},
  {Lubin}, {Ma}, {Mac{\'\i}as-P{\'e}rez}, {Maggio}, {Maino}, {Mandolesi},
  {Mangilli}, {Marcos-Caballero}, {Maris}, {Martin}, {Martinelli},
  {Mart{\'\i}nez-Gonz{\'a}lez}, {Matarrese}, {Mauri}, {McEwen}, {Meinhold},
  {Melchiorri}, {Mennella}, {Migliaccio}, {Millea}, {Mitra},
  {Miville-Desch{\^e}nes}, {Molinari}, {Montier}, {Morgante}, {Moss}, {Natoli},
  {N{\o}rgaard-Nielsen}, {Pagano}, {Paoletti}, {Partridge}, {Patanchon},
  {Peiris}, {Perrotta}, {Pettorino}, {Piacentini}, {Polastri}, {Polenta},
  {Puget}, {Rachen}, {Reinecke}, {Remazeilles}, {Renzi}, {Rocha}, {Rosset},
  {Roudier}, {Rubi{\\textasciitilde n}o-Mart{\'\i}n}, {Ruiz-Granados},
  {Salvati}, {Sandri}, {Savelainen}, {Scott}, {Shellard}, {Sirignano}, {Sirri},
  {Spencer}, {Sunyaev}, {Suur-Uski}, {Tauber}, {Tavagnacco}, {Tenti},
  {Toffolatti}, {Tomasi}, {Trombetti}, {Valenziano}, {Valiviita}, {Van Tent},
  {Vibert}, {Vielva}, {Villa}, {Vittorio}, {Wandelt}, {Wehus}, {White},
  {White}, {Zacchei}, \& {Zonca}}]{Planck2018}
{Planck Collaboration: Aghanim}, N., {Akrami}, Y., {Ashdown}, M., {et~al.}
  2020, \aap, 641, A6

\bibitem[{{Ramsey} {et~al.}(1998){Ramsey}, {Adams}, {Barnes}, {Booth},
  {Cornell}, {Fowler}, {Gaffney}, {Glaspey}, {Good}, {Hill}, {Kelton},
  {Krabbendam}, {Long}, {MacQueen}, {Ray}, {Ricklefs}, {Sage}, {Sebring},
  {Spiesman}, \& {Steiner}}]{Ramsey1998}
{Ramsey}, L.~W., {Adams}, M.~T., {Barnes}, T.~G., {et~al.} 1998, in Society of
  Photo-Optical Instrumentation Engineers (SPIE) Conference Series, Vol. 3352,
  Advanced Technology Optical/IR Telescopes VI, ed. L.~M. {Stepp}, 34

\bibitem[{{Rantala} {et~al.}(2018){Rantala}, {Johansson}, {Naab}, {Thomas}, \&
  {Frigo}}]{Rantala2018}
{Rantala}, A., {Johansson}, P.~H., {Naab}, T., {Thomas}, J., \& {Frigo}, M.
  2018, \apj, 864, 113

\bibitem[{{Rawlings} {et~al.}(2025){Rawlings}, {Johansson}, {Naab}, {Rantala},
  {Thomas}, \& {Neureiter}}]{Rawlings2025}
{Rawlings}, A., {Johansson}, P.~H., {Naab}, T., {et~al.} 2025, arXiv:2505.17183

\bibitem[{{Reardon} {et~al.}(2023){Reardon}, {Zic}, {Shannon}, {Hobbs},
  {Bailes}, {Di Marco}, {Kapur}, {Rogers}, {Thrane}, {Askew}, {Bhat},
  {Cameron}, {Cury{\l}o}, {Coles}, {Dai}, {Goncharov}, {Kerr}, {Kulkarni},
  {Levin}, {Lower}, {Manchester}, {Mandow}, {Miles}, {Nathan}, {Os{\l}owski},
  {Russell}, {Spiewak}, {Zhang}, \& {Zhu}}]{Reardon2023}
{Reardon}, D.~J., {Zic}, A., {Shannon}, R.~M., {et~al.} 2023, \apjl, 951, L6

\bibitem[{{Rowell} {et~al.}(2021){Rowell}, {Davidson}, {Lindegren}, {van
  Leeuwen}, {Casta{\\textasciitilde n}eda}, {Fabricius}, {Bastian}, {Hambly},
  {Hern{\'a}ndez}, {Bombrun}, {Evans}, {De Angeli}, {Riello}, {Busonero},
  {Crowley}, {Mora}, {Lammers}, {Gracia}, {Portell}, {Biermann}, \&
  {Brown}}]{RowellDavidson2021}
{Rowell}, N., {Davidson}, M., {Lindegren}, L., {et~al.} 2021, \aap, 649, A11

\bibitem[{{Rusli} {et~al.}(2013){Rusli}, {Erwin}, {Saglia}, {Thomas},
  {Fabricius}, {Bender}, \& {Nowak}}]{Rusli2013}
{Rusli}, S.~P., {Erwin}, P., {Saglia}, R.~P., {et~al.} 2013, \aj, 146, 160

\bibitem[{{Saglia} {et~al.}(2024){Saglia}, {Mehrgan}, {de Nicola},
  {et~al.}}]{Saglia24}
{Saglia}, R., {Mehrgan}, K., {de Nicola}, S., {et~al.} 2024, \aap, 692, A124

\bibitem[{{Schlafly} \& {Finkbeiner}(2011)}]{Schlafly2011}
{Schlafly}, E.~F. \& {Finkbeiner}, D.~P. 2011, \apj, 737, 103

\bibitem[{{Schneider} {et~al.}(1983){Schneider}, {Gunn}, \&
  {Hoessel}}]{Schneider1983}
{Schneider}, D.~P., {Gunn}, J.~E., \& {Hoessel}, J.~G. 1983, \apj, 264, 337

\bibitem[{{Seth} {et~al.}(2014){Seth}, {van den Bosch}, {Mieske}, {Baumgardt},
  {Brok}, {Strader}, {Neumayer}, {Chilingarian}, {Hilker}, {McDermid},
  {Spitler}, {Brodie}, {Frank}, \& {Walsh}}]{Seth2014}
{Seth}, A.~C., {van den Bosch}, R., {Mieske}, S., {et~al.} 2014, \nat, 513, 398

\bibitem[{{Skipper} \& {Browne}(2018)}]{SkipperBrowne2018}
{Skipper}, C.~J. \& {Browne}, I. W.~A. 2018, \mnras, 475, 5179

\bibitem[{{Springel} {et~al.}(2005){Springel}, {Di Matteo}, \&
  {Hernquist}}]{Springel2005}
{Springel}, V., {Di Matteo}, T., \& {Hernquist}, L. 2005, \apjl, 620, L79

\bibitem[{{Thomas} {et~al.}(2016){Thomas}, {Ma}, {McConnell}, {Greene},
  {Blakeslee}, \& {Janish}}]{Thomas2016}
{Thomas}, J., {Ma}, C.-P., {McConnell}, N.~J., {et~al.} 2016, \nat, 532, 340

\bibitem[{{Thomas} {et~al.}(2014){Thomas}, {Saglia}, {Bender}, {Erwin}, \&
  {Fabricius}}]{Thomas2014}
{Thomas}, J., {Saglia}, R.~P., {Bender}, R., {Erwin}, P., \& {Fabricius}, M.
  2014, \apj, 782, 39

\bibitem[{{Tremmel} {et~al.}(2015){Tremmel}, {Governato}, {Volonteri}, \&
  {Quinn}}]{Tremmel2015}
{Tremmel}, M., {Governato}, F., {Volonteri}, M., \& {Quinn}, T.~R. 2015,
  \mnras, 451, 1868

\bibitem[{{Turner} {et~al.}(2012){Turner}, {C{\^o}t{\'e}}, {Ferrarese},
  {Jord{\'a}n}, {Blakeslee}, {Mei}, {Peng}, \& {West}}]{Turner2012}
{Turner}, M.~L., {C{\^o}t{\'e}}, P., {Ferrarese}, L., {et~al.} 2012, \apjs,
  203, 5

\bibitem[{{Ulivi} {et~al.}(2025){Ulivi}, {Mannucci}, {Scialpi},
  {et~al.}}]{Q1-SP072}
{Ulivi}, L., {Mannucci}, F., {Scialpi}, M., {et~al.} 2025, A\&A, submitted,
  arXiv:2508.19494

\bibitem[{{Valtonen}(1996)}]{Valtonen1996}
{Valtonen}, M.~J. 1996, \mnras, 278, 186

\bibitem[{{Volonteri} {et~al.}(2003{\natexlab{a}}){Volonteri}, {Haardt}, \&
  {Madau}}]{Volonteri2003a}
{Volonteri}, M., {Haardt}, F., \& {Madau}, P. 2003{\natexlab{a}}, \apj, 582,
  559

\bibitem[{{Volonteri} {et~al.}(2003{\natexlab{b}}){Volonteri}, {Madau}, \&
  {Haardt}}]{Volonteri2003b}
{Volonteri}, M., {Madau}, P., \& {Haardt}, F. 2003{\natexlab{b}}, \apj, 593,
  661

\bibitem[{{Xu} \& {Komossa}(2009)}]{Xu2009}
{Xu}, D. \& {Komossa}, S. 2009, \apjl, 705, L20

\bibitem[{{Z{\"o}ller} {et~al.}(2024){Z{\"o}ller}, {Kluge}, {Staiger}, \&
  {Bender}}]{Zoeller2024}
{Z{\"o}ller}, R., {Kluge}, M., {Staiger}, B., \& {Bender}, R. 2024, \apjs, 271,
  52

\end{thebibliography}

\begin{appendix}

\section{Effect of the selection criteria on sample
characteristics}
\label{sec:cut_study}

Here we study how the specific selection criteria that we apply to the MER
catalogue, affect the distribution of selected galaxies compared to the
underlying MER catalogue. The MER catalogue provides detections and includes
several columns to exclude point-like sources, artefacts, and spurious
detections. We follow the criteria of \citet[Q25 hereafter]{Q1-SP040}, to
select -- as far as possible -- a representative set of all galaxies within
\Euclid{}'s Q1:

\begin{itemize}
\item {\tt VIS\_DET} = 1: galaxies must be detected in the VIS filter.
\item {\tt spurious\_flags = 0}: to exclude spurious detections from the sample.
\item {\tt point\_like\_prob} $\leq$ 0.1: to restrict the probability of a
source being point-like.
\item \IE $< 23$
\item $|${\tt PHZ\_PP\_MEDIAN\_Z} – {\tt PHZ\_MEDIAN}$|$ < 0.2: the difference
between the median redshift from SED-fitting (for galaxy properties) and the
photometric redshift to select for reliable redshift estimates.
\item |{\tt PHZ\_PP\_MEDIAN\_Z} – {\tt PHZ\_PP\_MODE\_Z}| < 0.2: the difference
between the median and mode of the SED-fitting redshift posterior must be less
than 0.2.
\item {\tt phys\_param\_flags} = 0: to ensure reliable physical parameter estimates.
\end{itemize}

As we explain in the main text, our sample selection
only cuts on the size of the segmentation area and the detected flux
within a 1-FWHM aperture.
The specific astronomical
data language (ADQL) {\tt SELECT} statement is as follows:

\begin{center}
\begin{minipage}{1.0\linewidth}
\lstset{basicstyle=\sffamily\ttfamily}
\begin{lstlisting}[language=SQL]
SELECT * FROM catalogue.mer_catalogue
WHERE
segmentation_area > 5500 AND flux_vis_1fwhm_aper > 20
\end{lstlisting}
\end{minipage}
\end{center}

\begin{figure}
    \centering
    \includegraphics[width=.9\linewidth]{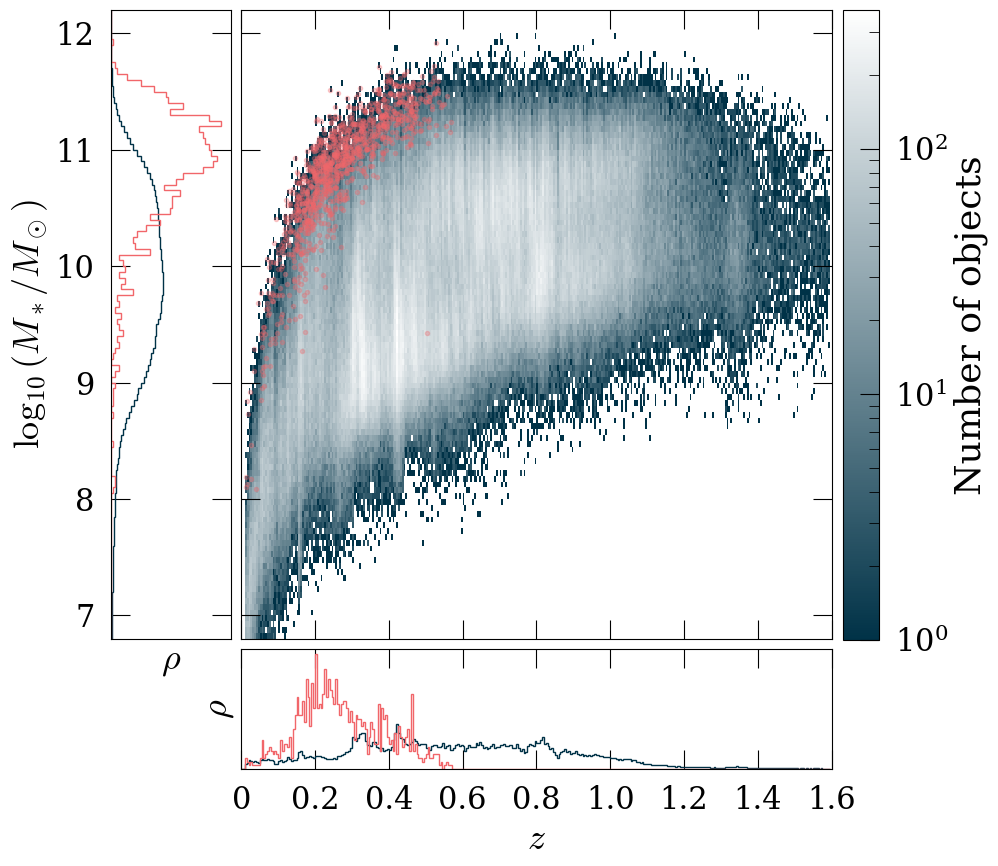}
    \caption{Sample redshift stellar-mass distribution.
    The main panel of the diagrams shows in the background a
    2D histogram of all MER detection following the selection criteria
    from Q25. The over-plotted points show what remains from
    this sample after applying the cuts in segmentation area and
    flux as we did for our sample. The 1D histograms in the left and
    the bottom panels compare the two different samples in
    stellar mass space and redshift space. The scaling of these histograms
    is arbitrary and chosen for both distributions separately to enhance visibility.
    \label{fig:cut_study}}
\end{figure}

In Fig.~\ref{fig:cut_study} we plot histograms of all galaxies
that follow the selection of Q25 and then to those
that also fall within our cut in segmentation area and detected flux.
Unsurprisingly, the requirements for a specific number of
segmentation pixels and flux, prefer galaxies at lower redshift --
median$(z) = 0.26$ in our sample vs.\ median$(z) = 0.57$ in Q25 -- and
higher stellar mass. The median stellar mass in our sample is
$\log_{10}(M_*/{\rm M}_\odot) = 10.92 $ vs.\ $\log_{10}(M_*/{\rm M}_\odot) = 9.89$ in Q25.

\begin{figure}
    \centering
    \includegraphics[width=.9\linewidth]{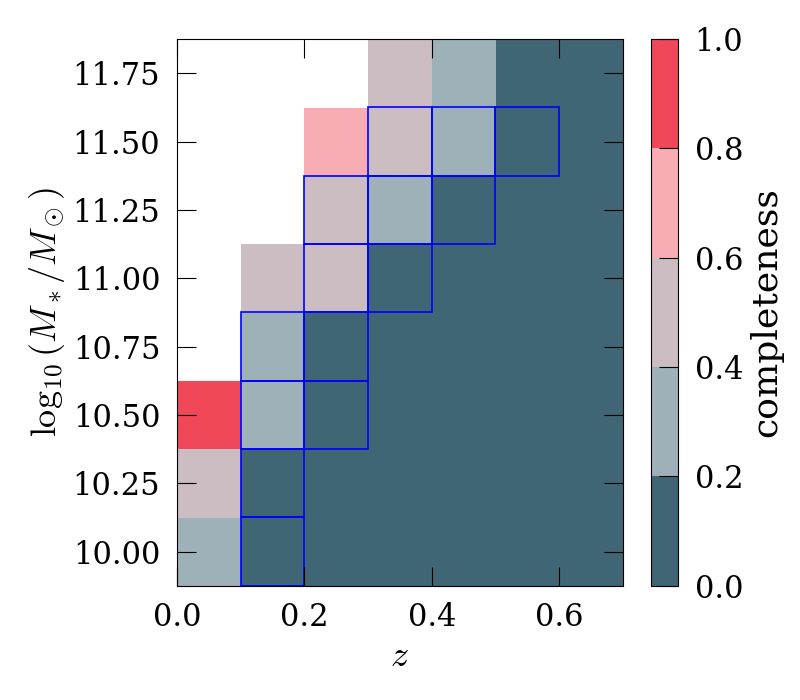}
    \caption{
    Selection function. This 2D histogram shows how our cuts in flux and segmentation
    area select targets compared to the sample of Q25.
    Every bin shows the ratio of the number of objects that remain after our cuts,
    to all objects following their selection. Where the relative
    error on the ratio from Poisson statistics falls below 30\% we also draw a blue
    box around the bin.
    \label{fig:cut_study_2Dhist}}
\end{figure}

Fig.~\ref{fig:cut_study_2Dhist} shows in more detail the completenesses that we reach
in specific redshift and stellar mass bins. The completeness falls close to or above 60\% for objects
at the highest stellar mass bins up to a redshift of 0.4.
Closer investigation of the completeness as function of segmentation area
shows only variations of up to 10\%.

\section{Conversion of Kron radius to effective radius}
\label{sec:kron_to_re}
\begin{figure}
    \centering
    \includegraphics[width=\myfigurewidth\linewidth]{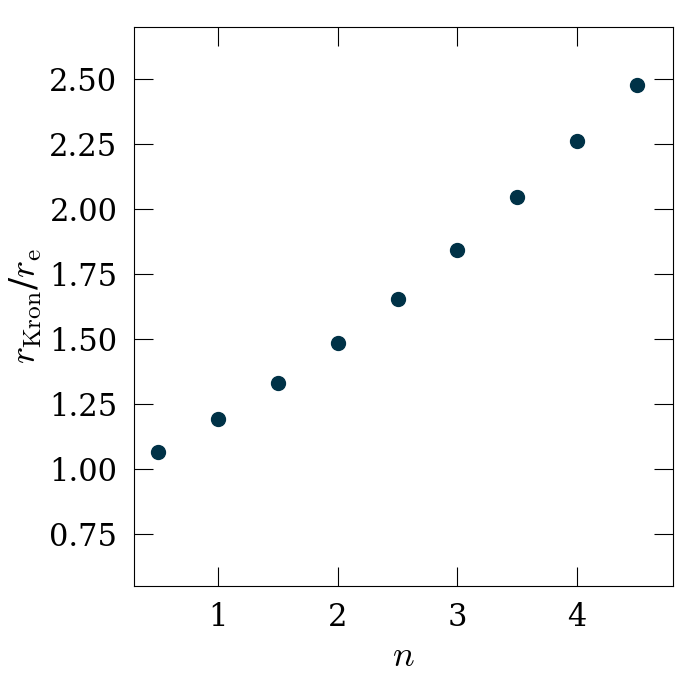}
    \caption{
    Conversion of Kron radius to effective radius for different values of the
    Sérsic index.
    }
    \label{fig:kron_to_re}
\end{figure}

We compute the relationship between Kron radii and effective radii for Sérsic
light profiles with varying Sérsic indices \( n \), ranging from 0.5 to 2.5. For
each value of \( n \), we generate a two-dimensional Sérsic light distribution using
\texttt{imfits}'s \texttt{makeimage} tool and save it as a FITS file. We then
calculate the Kron radius by first computing the image’s centre of mass and
then computing the Kron radius according to
\begin{equation*}
R_{\text{Kron}} = \frac{\sum\limits_{i} r_i I_i}{\sum\limits_{i} I_i},
\end{equation*}
where $I_i$ and $r_i$ are the fluxes and the radial distances for each
pixel respectively. Figure~\ref{fig:kron_to_re} shows the result. At
Sérsic indices of 0.5, the Kron radius equals the effective radius
(which is also straightforward to derive analytically). At a Sérsic index
of 2.5, the Kron radius is five times larger than the effective radius.

\end{appendix}

\label{LastPage}
\end{document}